\documentclass[aps,pra,10pt,
        twocolumn,superscriptaddress,
        floatfix,nofootinbib,
        showpacs,longbibliography]{revtex4-1}

\usepackage[utf8]{inputenc}  
\usepackage[T1]{fontenc}     
\usepackage[british]{babel}  
\usepackage[sc,osf]{mathpazo}
 
\usepackage[colorlinks=true, 
            citecolor=blue, 
            urlcolor=blue,
            breaklinks]{hyperref}  
\usepackage{graphicx} 
\usepackage{amsmath, amssymb,
            amsthm, bm,
            amsfonts, mathrsfs} 
\usepackage{physics}

\usepackage{algpseudocode}
\usepackage{algorithm}

\makeatletter
\def\bibsection{%
   \par
   \begingroup
    \baselineskip26\p@
    \bib@device{\hsize}{72\p@}%
   \endgroup
   \nobreak\@nobreaktrue
   \addvspace{19\p@}%
  }%
\makeatother

\begin{document}

\title{Disti-Mator: an entanglement {\it disti}llation-based state esti{\it mator}}

\author{Joshua Carlo A. Casapao$^{\dagger}$}
\affiliation{Networked Quantum Devices Unit, Okinawa Institute of Science and Technology Graduate University, Onna-son, Okinawa 904-0495, Japan}

\author{Ananda G. Maity$^{\dagger}$}
\affiliation{Networked Quantum Devices Unit, Okinawa Institute of Science and Technology Graduate University, Onna-son, Okinawa 904-0495, Japan}
\def\thefootnote{$\dagger$}\footnotetext{These authors contributed equally}

\author{Naphan Benchasattabuse}
\affiliation{Keio University, Shonan Fujisawa Campus, 5322 Endo, Fujisawa, Kanagawa 252-0882, Japan}

\author{Michal Hajdu\v{s}ek}
\affiliation{Keio University, Shonan Fujisawa Campus, 5322 Endo, Fujisawa, Kanagawa 252-0882, Japan}

\author{Rodney Van Meter}
\affiliation{Keio University, Shonan Fujisawa Campus, 5322 Endo, Fujisawa, Kanagawa 252-0882, Japan}

\author{David Elkouss}
\affiliation{Networked Quantum Devices Unit, Okinawa Institute of Science and Technology Graduate University, Onna-son, Okinawa 904-0495, Japan}

\begin{abstract}

Minimizing both experimental effort and consumption of valuable quantum resources in state estimation is vital in practical quantum information processing. Here, we explore characterizing states as an additional benefit of the entanglement distillation protocols.
We show that the Bell-diagonal parameters of any undistilled state can be efficiently estimated solely from the measurement statistics of probabilistic distillation protocols. We further introduce the state estimator `Disti-Mator' designed specifically for a realistic experimental setting, and exhibit its robustness through numerical simulations. Our results demonstrate that a separate estimation protocol can be circumvented whenever distillation is an indispensable communication-based task.
\end{abstract}
\maketitle

\section{Introduction}\label{chap:introduction}

Quantum networks, and ultimately the quantum Internet, hold the promise of enabling a wide range of distributed information processing and secure communication tasks that can outperform any classical network \cite{Wehner18,VanMeter22,rfc9340}.
Among the potentially impactful applications are quantum key distribution \cite{Bennet84,Ekert91,Bauml20}, enhanced interferometry for telescopes \cite{Gottesman12}, higher-precision clock synchronization \cite{Kmr14,IloOkeke18}, and distributed quantum computing \cite{Buhrman03,Broadbent09,Fitzsimons17,Cuomo20}.

But, these very applications are concomitant with a sustained distribution of high-fidelity entanglement between remote network nodes \cite{Chakraborty20,Azuma21,Coopmans21,Satoh22}. Entanglement distribution either in free space or through optical fibers remains an architectural bottleneck since the optical attenuation worsens as the communication distance increases.
While optical signals can be easily amplified in classical networks to effectively nullify this attenuation, a classical-like amplification in quantum networks is impossible as stipulated by the no-cloning theorem \cite{dieks1982communication,wootters1982single}.
To overcome this problem, quantum networks can be supplemented with quantum repeater nodes that perform splicing of short-ranged entangled states via entanglement swapping~\cite{azuma2023quantum,Goodenough21}.

Unfortunately, repeated entanglement swapping operations needed for establishing long-distance communication exponentially degrade the end-to-end fidelity. Entanglement distillation protocols can be used to suppress this fidelity loss by transforming a number of noisy entangled states into a smaller number of high-fidelity states using only local operations and classical communication \cite{Bennett96(1),Bennett96(2),Deutsch96,Yamamoto01,Pan01}.

Still, realizing these accompanying communication-based tasks often require fidelity or some other measure of quality to pass a certain threshold in order to work in practice. The guarantee of having a complete or sufficient knowledge of the quality of the network resources and devices can help in efficiently implementing these tasks. 
On this account, it is desirable to develop characterization methods compatible with emerging quantum communication technologies in the noisy intermediate-scale quantum (NISQ) era and beyond. Some of the approaches include quantum tomography \cite{nielsen_chuang_2010,Altepeter03,Poyatos97,Chuang97,Fujiwara01,OBrien04}, randomized benchmarking \cite{Emerson05,Knill08,Dankert09,Magesan11,Magesan12,Erhard19,Helsen22}, self-testing \cite{Supic2020}, and quantum gate set tomography \cite{Merkel13,Nielsen21}. In recent years, other state estimation and quantum certification methods have also been developed~\cite{Yuan2017,Qi2017,Yu2022,Xu2019,Meyer2021,Pereira2022,Xiao2022,Nolan2021,Conlon2021,Hou2016,Grinko2021,Helsen2023,Andrade22,Andrade23,Andrade24}. We direct the readers to Ref.~\cite{Eisert2020} for a brief overview.

\begin{figure*}[htbp]
    \centering
    \includegraphics[width=\textwidth]{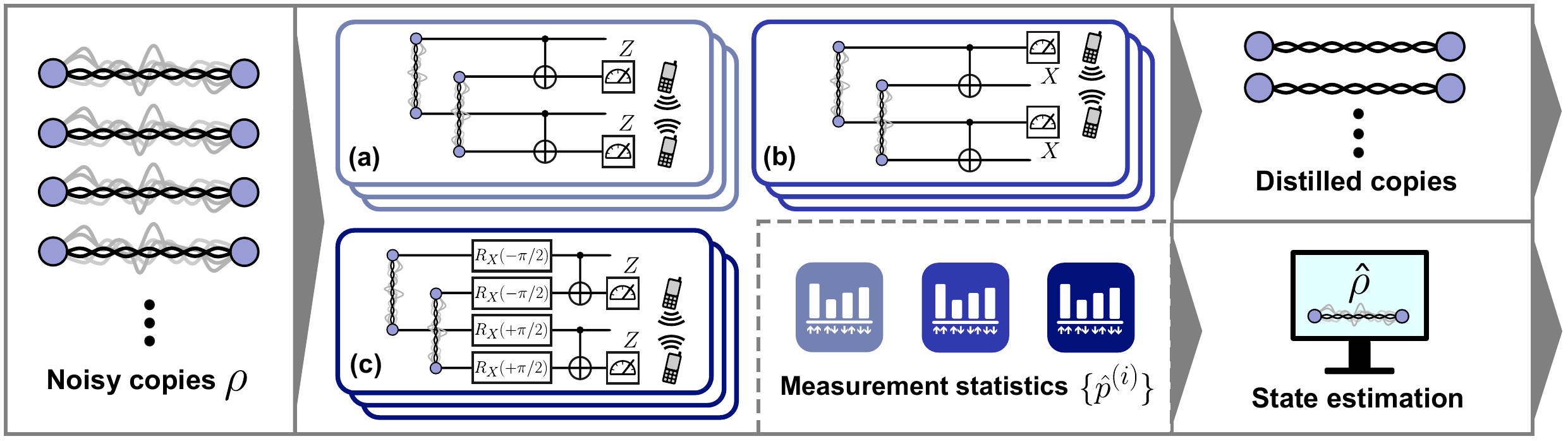}
    \caption{
    \textbf{The distillation-based estimation.} Several copies of noisy entangled state $\rho$ are prepared and shared between two parties. Both parties jointly perform one of the shown entanglement distillation protocols, which, given some probability of success, results to a higher-quality distilled state: \textbf{(a)} the distillation protocol proposed by Bennett {\it et al.} \cite{Bennett96(1)} (which we call \texttt{Distillation-(a)}); \textbf{(b)} a modification of Bennett {\it et al.}'s protocol, where the first copy is locally measured in the $X$ basis (\texttt{Distillation-(b)}); \textbf{(c)} the distillation protocol proposed in Deutsch {\it et al.} \cite{Deutsch96} (\texttt{Distillation-(c)}). After repeated applications of the protocols, we obtain a set of measurement statistics $\{\smash[t]{\hat{p}^{(i)}}\}$ which can be post-processed via our proposed state estimator to generate an estimation $\hat{\rho}$ of the prepared noisy state. 
    }
    \label{fig:distillation}
\end{figure*}

The feasibility of these characterization methods depends on a compromise between the complexity of the protocol and the information gain for a meaningful estimation \cite{Eisert2020}.
For small-scale systems, quantum tomography is a standard tool for estimation since it directly probes an unknown quantum state or process.
However, a complete tomographic reconstruction needs a substantial amount of resources that will later be unusable for further communication-based tasks. 

In contrast to the resource-intensive tomography, it is sometimes possible to obtain a partial characterization as a by-product of a quantum information processing task. In the case of learning Pauli noise parameters, the syndrome statistics of an error-correcting code can be used to efficiently characterize such noise without consuming additional resources beyond what is needed for error correction~\cite{Wagner21,Wagner22}. 

Motivated by the above, we propose a new state estimator that solely uses the measurement statistics obtained from probabilistic entanglement distillation protocols to efficiently characterize the Bell-diagonal elements of the states prepared for distillation. We also demonstrate that this estimator is robust to imperfections in realistic distillation protocols.
While the resulting estimate is generally tomographically incomplete, the protocol can be beneficial for entanglement-based networks, where the information about a state's fidelity with respect to the Bell states is sufficient in practice. Moreover, whenever distillation is unavoidable, our approach can simplify the network management by eliminating the need for a separate estimation task.


\section{Results}\label{chap:results}

\subsection{Distillation protocols}\label{sec:overview}

We begin with general two-way probabilistic distillation protocols~\cite{Bennett96(1),Bennett96(2),Deutsch96,Yamamoto01,Pan01}. Consider two spatially separated parties, Alice and Bob, sharing several copies of a noisy entangled state $\rho$ that are prepared for distillation. The goal is to distill these copies towards the maximally entangled state $\ket{\Phi^+}=(\ket{00}+\ket{11})/\sqrt{2}$, which Alice and Bob achieve by performing bilateral operations on their respective halves, followed by local measurements on a subset of the shared states in an agreed upon basis. Subsequent exchange of classical information about the measurement statistics enables Alice and Bob to detect whether the unmeasured copies are afflicted with errors. When this succeeds, the process transforms the noisy entangled states $\rho$ into a smaller number of entangled states with higher fidelity to $\ket{\Phi^+}$. 

For this work, we focus on the distillation protocols proposed by Bennett {\it et al.} \cite{Bennett96(1)} in Fig.~\ref{fig:distillation}(a) and its variation in (b), and the protocol proposed by Deutsch {\it et al.} \cite{Deutsch96} in (c), which we refer to as \texttt{Distillation-(a)}, \texttt{-(b)}, \texttt{-(c)}, respectively. These protocols herald a successful distillation whenever the local measurements coincide, which happens probabilistically. A detailed discussion of the distillation protocols is provided in Methods-\ref{sec:methods_distillation}.

In this regard, we ask whether the measurement statistics from these distillation protocols can be further used to characterize the undistilled $\rho$ as an additional benefit. For our analysis, we split $\rho$ into two,
\begin{equation}\label{eqn:arbitrary-state}
    \rho := \overline{\rho}(\mathbf{q}) + \rho_{\text{off-diagonal}},
\end{equation}
so that $\overline{\rho}(\mathbf{q})$ is a Bell-diagonal state:
\begin{align}\label{eqn:Bell-diagonal}
    \overline{\rho}(\mathbf{q}) =& \,\,q_1 \ketbra{\Phi^+}{\Phi^+} + q_2 \ketbra{\Phi^-}{\Phi^-}\nonumber\\ 
    &+ q_3 \ketbra{\Psi^+}{\Psi^+} + q_4 \ketbra{\Psi^-}{\Psi^-},
\end{align}
Here, the coefficients $\mathbf{q}:=(q_1,q_2,q_3,q_4)$ satisfy $q_1+q_2+q_3+q_4 =1$, and the pure states in Eq.~\eqref{eqn:Bell-diagonal} transform from $\ket{\Phi^+}$ via the Pauli operations $\{I,Z,X,ZX\}$, respectively. We show that, without assuming any symmetries of $\overline{\rho}$, the measurement statistics from the three protocols can accurately estimate $\mathbf{q}$ (with the fourth coefficient constrained by normalization). As we go through the discussion of our estimation scheme, we define the success probability $p^{(i)}$ of both measurement outcomes being ``up'' as half the coincidence probability of the $i$-th distillation protocol. A cartoon picture depicting our estimation scheme is shown in Fig. \ref{fig:distillation}.


\subsection{Werner parameter estimation}\label{sec:werner_estimation}

Before we outline the distillation-based estimation for arbitrary undistilled copies, a simple yet nontrivial example to begin with is to assume that the undistilled copies are Werner states, defined by $q_1=1-\frac{3w}{4}$ and $q_2= q_3=q_4=\frac{w}{4}$ for some noise parameter $w$:
\begin{align}\label{eqn:Werner_state}
    \rho_{w}=& (1-w)\ketbra{\Phi^+}{\Phi^+} + \frac{w}{4} \mathbb{I}.
\end{align}
The inherent symmetry of Werner states implies that it is sufficient to take the measurement statistics from a single distillation protocol. This estimation is first outlined in Ref.~\cite{Maity23}.\\

Suppose that Alice and Bob possess two copies of a Werner state (assuming these are independent and identically distributed (i.i.d.)) and choose to carry out \texttt{Distillation-(a)}. The success probability of \texttt{Distillation-(a)} is $p^{(1)} = \frac{1}{4}(2-2w+ w^2)$, which we can rearrange to obtain an expression for $w$:
\begin{equation}\label{eqn:wfromp1}
    w = 1- \sqrt{4p^{(1)}-1}. 
\end{equation}

Let us suppose that one repeats the distillation several times. One then obtains an empirical probability $\hat{p}^{(1)}$ of both parties observing ``up'' outcomes given $N^{(1)}$ Werner pairs. Using $\hat{p}^{(1)}$ and Eq.~\eqref{eqn:wfromp1}, we can obtain an estimate $\hat w$ of $w$ at a user-specified error threshold $\epsilon_w$.
Equivalently, if $\abs{\hat{w}- w} \geq \epsilon_w$ then $\hat{p}^{(1)}$ deviates from $p^{(1)}$ such that either $p^{(1)} -\hat{p}^{(1)} \geq \frac{1}{4} (\epsilon_w^{2} + 2\epsilon_w(1-\hat{w})) =:\epsilon^{(1)}_R(\hat{w}, \epsilon_w)$ or $\hat{p}^{(1)} - p^{(1)} \geq \frac{1}{4} (-\epsilon_w^{2} + 2\epsilon_w(1-\hat{w})) =:\epsilon^{(1)}_L(\hat{w}, \epsilon_w)$. Applying Hoeffding's inequality \cite{hoeffding1963prob} leads to a bound for the estimation's failure probability,
\begin{equation}\label{estimation_werner}
    \text{Pr}(|\hat{w}- w| \geq \epsilon_w) \leq \sum_{m=L,R}\exp{-2 N^{(1)} (\epsilon^{(1)}_m(\hat{w}, \epsilon_w))^2}.
\end{equation}
The estimator's sample complexity (see Supplementary Material) is given by
\begin{equation}
    N^{(1)}\leq \mathcal{O}\left(\frac{\log(1/\delta)}{\epsilon_w^{2}}\right),
\end{equation}
where $\delta$ is the failure probability.

As a first step towards applying this procedure in a realistic setting, we consider idling undistilled copies undergoing depolarization,
\begin{equation}\label{eqn:global_depolarizing}
    \Lambda (\rho) = (1-\lambda(t))\rho + \frac{\lambda(t)}{4} \mathbb{I},
\end{equation}
where $\lambda(t)=1-\exp(-t/T^{\text{dpo}})$ is the depolarizing parameter, and $T^{\text{dpo}}$ is the relaxation time that describes the rate at which the state decoheres to the maximally mixed state. This becomes relevant if we model the entanglement generation stage as a sequential process.
After the first copy is distributed, it depolarizes in the quantum memory for some time $t$ until the second copy is successfully distributed.
The second copy can be used immediately for distillation and is therefore unaffected by the depolarizing noise. In presence of this depolarization, the distillation statistics changes into $p_S^{(1)} = \frac{1}{4}(Sw^2-2S w+ S+1)$, where the average depolarization is $1-S= 1-(1/N^{(1)})\sum_{j=1}^{N^{(1)}} (1-\lambda_{j}(t))$. 
Following the same procedure as before, we can estimate $w$ with the estimator $\hat{w} = 1-\sqrt{1-\frac{S+1-4\hat{p}^{(1)}}{S}}$.

Explicit calculation (see Methods-\ref{M_b}) reveals that if we consider $|\hat{w}- w|  \geq \epsilon_w$, then either $p_S^{(1)} -\hat{p}^{(1)} \geq \frac{S}{4} (\epsilon_w^{2} + 2\epsilon_w(1-\hat{w}))=:\epsilon'^{(1)}_R(\hat{w}, \epsilon_w)$ or $\hat{p}^{(1)} - p_S^{(1)} \geq \frac{S}{4} (-\epsilon_w^{2} + 2\epsilon_w(1-\hat{w})) =:\epsilon'^{(1)}_L(\hat{w}, \epsilon_w)$. The corresponding failure probability for this estimator satisfies
\begin{equation}\label{eqn:bound-werner-depolarizing}
    \text{Pr}(|\hat{w}- w| \geq \epsilon_w) \leq \sum_{m=L,R}\exp{-2 N^{(1)} (\epsilon'^{(1)}_m(\hat{w}, \epsilon_w))^2}.
\end{equation}
A more general estimation protocol incorporating both state preparation and measurement (SPAM) errors and gate errors is considered in Results-\ref{sec:disti-mator_overview}.

\begin{figure}[ht] 
    \includegraphics[width=0.48\textwidth]{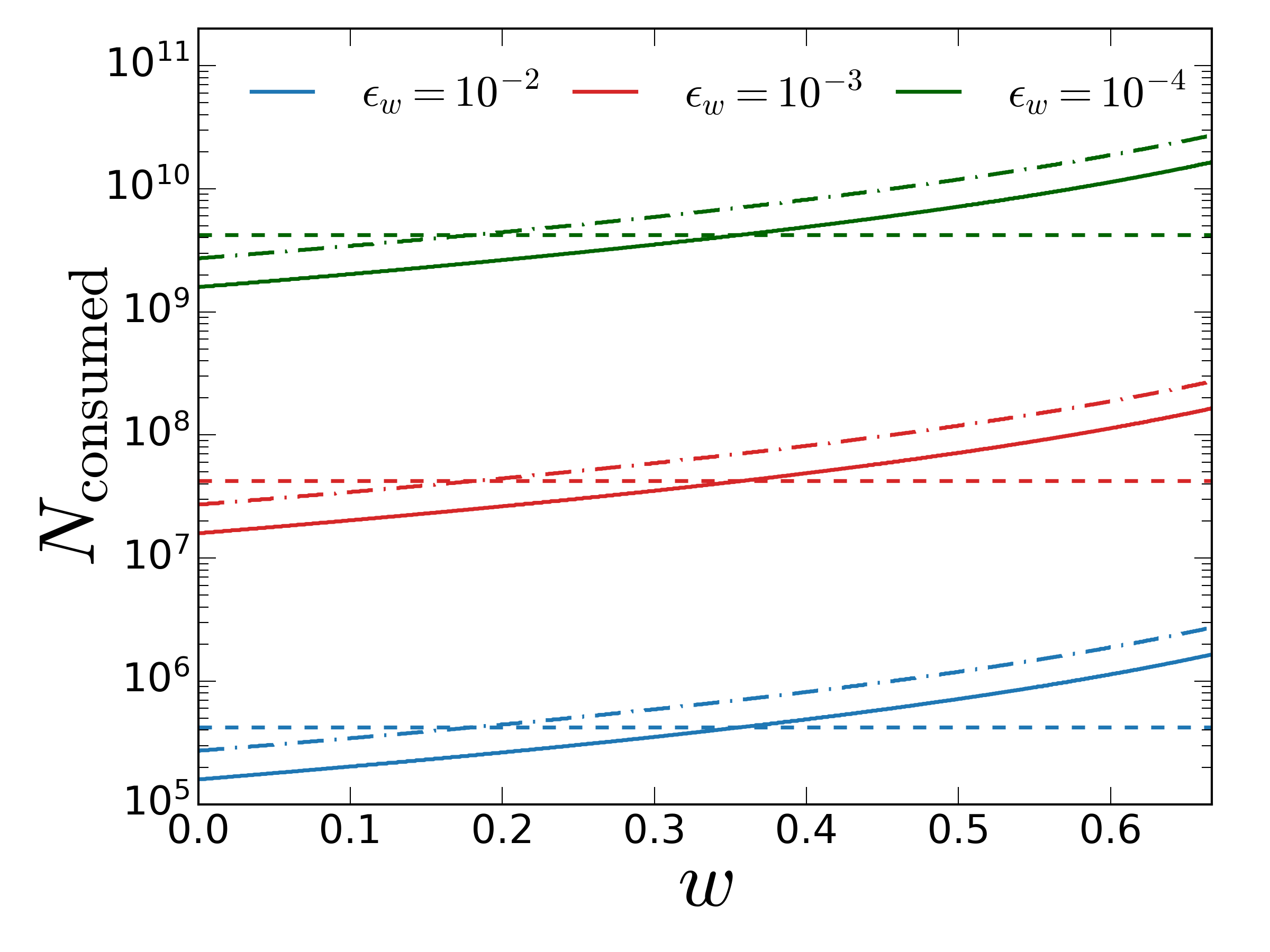}
    \caption{\textbf{Number of i.i.d. Werner states consumed for parameter estimation via a distillation experiment.} Here, $N^{(1)}$ is the number of i.i.d. Werner state pairs $\rho_w^{\otimes 2}$ required to estimate the Werner parameter $w$ with failure probability bound $\delta = 10^{-2}$, for different values of the error threshold $\epsilon_w$. The solid curves correspond to the expected total number of $\rho_w$ consumed with an estimation via a noiseless distillation protocol, $N_{\mathrm{consumed}}= (2-p^{(1)})N^{(1)}$, leaving $p^{(1)}N^{(1)}$ distilled states for further use. The dashed horizontal lines correspond to the total number of Werner states consumed to estimate using tomography. The dash-dotted curves correspond to the distillation protocol in the presence of depolarizing noise, where the average depolarization $1-S = 1-\exp(-1/4)$, and the expected $N_{\mathrm{consumed}}= (2 - p_S^{(1)})N^{(1)}$. Whenever $w$ is close to zero, we observe that an estimation via distillation consumes fewer overall resources than tomography.}
\label{fig:werner_samples}
\centering 
\end{figure}

We compare the complexity of our estimator with a standard tomographic protocol (see Methods-\ref{M_b} for a detailed analysis). Fig.~\ref{fig:werner_samples} shows the minimum number of i.i.d. Werner states to be consumed, $N_{\text{consumed}}$, for estimating $w$ with failure probability bound $\delta = 10^{-2}$ via a tomographic protocol, a noiseless distillation protocol, and a noisy distillation protocol with a severe depolarization of the first copy, where $S = \exp(-1/4)$. Whenever $w$ is close to zero, we find that the required number of samples for a successful distillation-based estimation can be less than that for tomography.

We note that we can successfully do estimation over the entire range of $w$ (irrespective of the distillability of the prepared states). This can be confirmed by observing that $p^{(1)}$ uniquely depends on $w$.

\subsection{Parameter estimation for Bell-diagonal states}
\label{sec:belldiagonal_estimation}

We now consider a situation where the copies shared between Alice and Bob are in Bell-diagonal form $\overline{\rho}(\mathbf{q})$, as in Eq.~\eqref{eqn:Bell-diagonal}. Unlike the Werner state, the number of unknown parameters to be estimated in a Bell-diagonal state is effectively three. To obtain an estimation $\hat{\mathbf{q}}$ for $\mathbf{q}$, we consider the empirical probabilities $\hat{p}^{(i)}$ obtained from all $i$-th distillation protocols, where $i\in \{1,2,3\}$ for \texttt{Distillation-(a)}, \texttt{-(b)}, and \texttt{-(c)}, respectively (see Methods-\ref{sec:methods_distillation}). For our analysis, we introduce the intermediate variables $x_i := q_1 + q_{i+1}$. We proceed in two steps. First, we obtain $\abs{\hat{x}_i- x_i} \leq \epsilon_i$ for all $i$ with high probability, given the collection of intermediate estimations $\hat{\mathbf{x}}:=(\hat{x}_1,\hat{x}_2,\hat{x}_3)$ and thresholds $\epsilon_i$. Second, we calculate $\hat{\mathbf{q}}$ (and ultimately the state $\hat{\rho}(\hat{\mathbf{q}})$) through $\hat{\mathbf{x}}$ with error $\abs{\hat{q}_i-q_i} \leq {\epsilon_T}/{2}$, where $\epsilon_T := \sum_{i\in\{1,2,3\}}\epsilon_i$. 

We first construct an estimator that assumes both i.i.d. input states and noiseless distillation protocols. In this case, we have
\begin{equation}\label{eqn:x_i}
    x_i = \frac{1}{2}\left(1+\sqrt{4p^{(i)} - 1} \right),
\end{equation}
for $i\in\{1,2,3\}$. The failure condition $\abs{\hat{x}_i- x_i} \geq \epsilon_i$ implies that $\hat{p}^{(i)}$ deviates from $p^{(i)}$ such that either $p^{(i)} -\hat{p}^{(i)} \geq  \epsilon_i^{2} + \epsilon_i(2\hat{x}_i-1) =:\epsilon^{(i)}_R (\hat{x}_i,\epsilon_i)$ or $\hat{p}^{(i)} - p^{(i)} \geq  -\epsilon_i^{2} + \epsilon_i(2\hat{x}_i-1) =:\epsilon^{(i)}_L(\hat{x}_i,\epsilon_i)$ for all $i$. 
We evaluate the failure probability of this estimator based on its trace distance $D(\hat{\rho}(\hat{\mathbf{q}}),\overline{\rho}(\mathbf{q})) :=\frac{1}{2}\|\hat{\mathbf{q}}-\mathbf{q}\|_1$ from the expected $\mathbf{q}$. Applying Hoeffding's inequality repeatedly, we obtain 
\begin{equation}
    \Pr[D(\hat{\rho}(\hat{\mathbf{q}}),\overline{\rho}(\mathbf{q}))\geq \epsilon_T] \leq \delta(\hat{\mathbf{x}}, \{N^{(i)}\}, \{\epsilon_i\}),
\end{equation}
where
\begin{align}
    &\delta(\hat{\mathbf{x}}, \{N^{(i)}\}, \{\epsilon_i\})\nonumber\\ &:= 1 -\!\!\! \prod_{i\in\{1,2,3\}}\left[ 1-\!\!\!\sum_{m=L,R}\exp(-2N^{(i)}\epsilon^{(i)}_m(\hat{x}_i,\epsilon_i)^2)\right].\label{eqn:delta_hoeffding_for_noiseless}
\end{align}
Here, $N^{(i)}$ is the number of pairs consumed for the $i$-th distillation protocol. A detailed derivation and analysis of this bound is provided in Methods-\ref{M_c}. To assess the estimator's sample complexity, suppose that $\epsilon_i=\epsilon$ and $N^{(i)} = \widetilde{N}$ for all $i$. Given the failure probability $\delta$, we can bound $\widetilde{N}$ with (see Supplementary Material)
\begin{equation}
    \widetilde{N} \leq \mathcal{O}\left(\frac{\log(1/\delta)}{\epsilon_T^{2}}\right).
\end{equation}

The protocol is also robust to noise. Similar to the Werner case, suppose that the first copy of the entangled state undergoes depolarization, represented by Eq.~\eqref{eqn:global_depolarizing}. In the presence of this depolarizing noise, Alice and Bob obtain the measurement statistics $p^{(i)} = S_ix_i^2-S_i x_i+ \frac{1}{4}(S_i+1)$ for $i\in \{1,2,3\}$, where $1-S_i= 1-({1}/{N^{(i)}})\sum_{j=1}^{N^{(i)}} (1-\lambda_{j}(t))$ describes the average depolarization.

One can now follow the same procedure as discussed earlier for the estimation of the Bell parameters in the noiseless scenario and obtain the bound 
\begin{equation}
    \Pr[D(\hat{\rho}(\hat{\mathbf{q}}),\overline{\rho}(\mathbf{q}))\geq \epsilon_T] \leq \delta'(\hat{\mathbf{x}}, \{N^{(i)}\}, \{\epsilon_i\}).
\end{equation}
Here, $\delta'$ is similar to $\delta$ in Eq.~\eqref{eqn:delta_hoeffding_for_noiseless}, but with $\epsilon^{(i)}_R(\hat{x}_i,\epsilon_i) :=  S_i(\epsilon_i^{2} + \epsilon_i(2\hat{x}_i-1))$ and $\epsilon^{(i)}_L(\hat{x}_i,\epsilon_i):= S_i(-\epsilon_i^{2} + \epsilon_i(2\hat{x}_i-1))$. Results-\ref{sec:disti-mator_overview} presents a general estimation protocol incorporating imperfections both in SPAM and in the protocol gates.

\begin{figure}[ht] 
    \centering
    \includegraphics[width=0.48\textwidth]{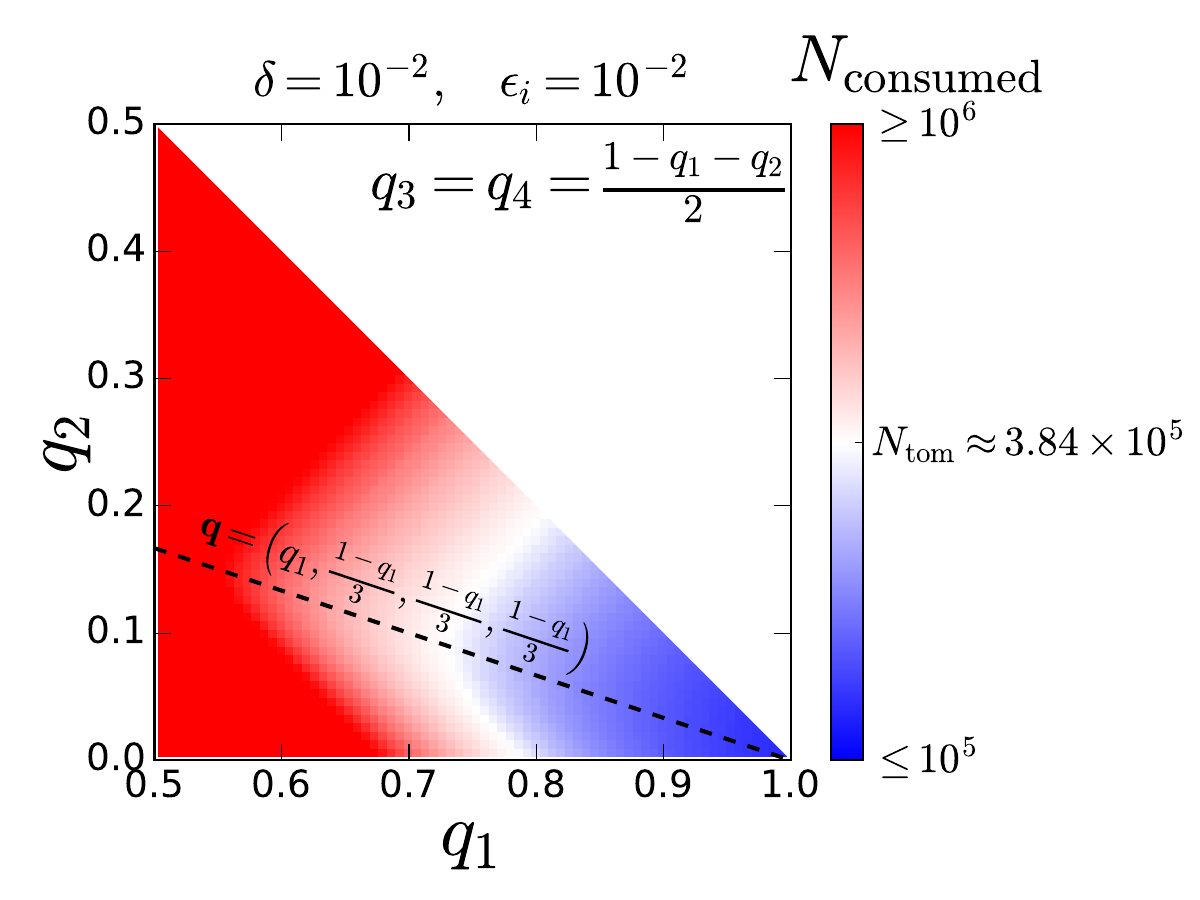}
    \caption{\textbf{Number of i.i.d. Bell-diagonal states consumed for parameter estimation via a noiseless distillation experiment.} Here, we assume that all distillation protocols are equally allocated with $N^{(i)}$ i.i.d. Bell-diagonal state pairs $\rho^{\otimes2}$. We compare this with conventional state tomography, where we assume that the joint measurements are also equally allocated with states $\rho$, totalling $N_{\mathrm{tom}}$ (see Methods~\ref{M_c}). The dashed line describes the collection of Werner states. We set the error bound $\epsilon_i= 10^{-2}$ for all $i$, and we impose $\delta=10^{-2}$ when estimating the trace distance $D(\hat{\rho}(\hat{\mathbf{q}}),\overline{\rho}(\mathbf{q}))$ either via distillation or tomography. While tomography consumes all $N_{\mathrm{tom}}$ of the states, the distillation-based estimation is expected to consume $N_{\mathrm{consumed}} =\sum_{i\in\{1,2,3\}} (2-p^{(i)})N^{(i)}$ of the states, leaving $\sum_{i\in\{1,2,3\}} p^{(i)}N^{(i)}$ distilled states for further use. Whenever $q_1$ is close to one, we observe that an estimation via distillation consumes fewer  resources than tomography (up to about $60\%$ fewer resources near $\ket*{\Phi^+}$). }
    \label{fig:belldiagonal_samples}
\end{figure}

We compare the complexity of our estimator with a noiseless tomographic protocol (see Methods-\ref{M_c}). Fig.~\ref{fig:belldiagonal_samples} shows the minimum number of i.i.d. Bell-diagonal states consumed, $N_{\text{consumed}}$, for a noiseless distillation-based estimation with the bound $\delta = 10^{-2}$ and error bound $\epsilon_i= 10^{-2}$, for all $i$. For $q_1$-values close to one, we find that utilizing a distillation-based estimation is again advantageous over tomography.

We also observe that the number of consumed Bell pairs $N_{\text{consumed}}$ increases as the parameter $q_2$ moves away from the Werner-state regime (for fixed $q_1$), represented by the dashed line in Fig.~\ref{fig:belldiagonal_samples}.
This can be understood by noting that the three distillation protocols extract different information about the noisy Bell pairs.
For example, \texttt{Distillation-(a)} can be used if the Bell pairs are affected by Pauli $X$ errors but is useless when the errors are Pauli $Z$.
In the case of Werner states, all three Pauli errors are equally likely.
Therefore, the optimal strategy in terms of the consumed number of Bell pairs is to split the Bell pairs equally between the three distillation protocols.


\subsection{Parameter estimation for arbitrary states}
\label{sec:arbitrary_estimation}

So far, we have found that a Bell-diagonal state can be characterized completely by our distillation-based estimation protocol. Our analysis for Bell-diagonal states can also be utilized to partially estimate the undistilled copies that deviate from the Bell-diagonal form, i.e., off-diagonal elements can be non-zero. Moreover, we can bound the precision of our estimation in this general case. 

Let $\rho$ be a state with nontrivial off-diagonal elements, and whose diagonal elements are $\mathbf{q}$ in the Bell basis. Suppose that by using our Bell-diagonal estimator we obtain an estimation $\hat{\rho}(\hat{\mathbf{q}})$. We can then find a Bell-diagonal state $\overline{\rho}(\mathbf{q})$ close to our estimation such that $D(\hat{\rho},\overline{\rho})\leq \epsilon_T$ with probability of at least $1-\delta(\hat{\mathbf{x}}, \{N^{(i)}\}, \{\epsilon_i\})$.  
Applying both the triangle inequality and the Fuchs-van de Graaf inequalities, we find that
\begin{equation}
    D(\hat{\rho}, \rho) \leq \epsilon_T + \sum_{k=1}^4 \sqrt{q_k^2(1- q_k^2)} =: \epsilon^{\ast},
\end{equation}
and the associated concentration bound satisfies
\begin{equation}
    \Pr[D(\hat{\rho}, \rho)\geq \epsilon^{\ast}] \leq  \delta(\hat{\mathbf{x}}, \{N^{(i)}\}, \{\epsilon_i\}).
\end{equation}
Notice that the order of magnitude of the summation can exceed $\epsilon_T$, rendering the bound $\epsilon^*$ too loose and therefore uninformative. As expected, an estimator for Bell-diagonal states performs poorly in fully estimating $\rho$, unless the state is already in the high-fidelity regime. 


\subsection{Disti-Mator: the state estimation toolbox}
\label{sec:disti-mator_overview}
In this section, we propose the state estimator `Disti-Mator' based on the measurement statistics of noisy distillation protocols, while retaining the assumption that resource states satisfy the i.i.d. property. Adopting this toolbox for characterizing entangled states generated between two network nodes can be efficient due to several reasons. For information processing tasks that necessitate performing a distillation protocol, our estimation method does not consume additional network resources than what is already required for the protocol. Moreover, our estimation method is more resource-efficient than conventional state tomography if the initially generated entangled states are already of high fidelity. 

The state estimator depends on a full characterization of the quantum devices used in distillation (see also the Supplementary Material for more details). This can be a reasonable assumption in the context of networks where the creation of entanglement is qualitatively more expensive than local operations.
We consider that each local quantum gate is afflicted with depolarizing noise that is independent of any noise acting on other quantum gates. We also assume that the SPAM is noisy. For state preparation, this means that each qubit of the first copy experiences both depolarization and dephasing over time $\Delta t$ while waiting in a quantum memory. These channels are parameterized by $\lambda_{A,B}(\Delta t) := 1-\exp(-\smash{{\Delta t}/{T_{A,B}^{\mathrm{dpo}}}})$ and $\zeta_{A,B}(\Delta t) := (1-\exp(\smash{-{\Delta t}/{T_{A,B}^{\mathrm{dph}}}}))/2$, respectively, where $T_{A,B}^{\mathrm{dpo}}$ and $T_{A,B}^{\mathrm{dph}}$ are their respective characteristic times. We also assume that the entire distillation experiment has a Markovian (i.e., memoryless) noise model: this assumption is usually seen in benchmarking procedures~\cite{Helsen2023}. 

The estimation process can be summarized with the following steps. The Disti-Mator initializes with the measurement statistics collected from a distillation experiment. This experiment may involve one or several distillation protocols in Fig.~\ref{fig:distillation}, depending on the number of unknown state parameters needed to be estimated. The estimator also requires a record of the noise parameters (as well as a timekeeping of the states generated) associated to the quantum devices and to SPAM during the distillation experiment. With this information, the estimator proceeds by inverting the measurement statistics to estimate the state parameters. Given that the distillation success probability $p^{(i)}$ can be effectively expressed as a monotonic function of a single variable (see Methods-\ref{M_d}), we can use a simple bisection search to invert from the empirical probabilities $\hat{p}^{(i)}$ towards an intermediate variable related to our desired estimation, or towards the estimation itself. Finally, the figure of merit for the Disti-Mator, which we choose to be the estimation's failure probability, is bounded using Hoeffding's inequality.


\subsection{Numerical simulations}\label{sec:disti-mator_simulation}

We now assess the utility and performance of the Disti-Mator by applying it in simulated noisy distillation experiments. With this aim, we discuss two versions of the state estimator: one for estimating Werner states and another for estimating Bell-diagonal states.

\textbf{Disti-Mator for Werner states:} For estimating unknown Werner states $\rho_w$, the empirical probability $\hat{p}^{(1)}$ obtained in \texttt{Distillation-(a)} is sufficient to estimate $\hat{w}$. We summarize this estimator in Algorithm~\ref{alg:wernerParamEstimationI} (see the Supplementary Material for more details).

The first input for the state estimator is the outcome statistics $\hat{p}^{(1)}:= n^{(1)}/N^{(1)}$ after $N^{(1)}$ repetitions of \texttt{Distillation-(a)} in the experiment, where $n^{(1)}$ counts the total number of successful distillation instances. We also set an error bound $\epsilon_w$ with respect to the expected Werner parameter $w$ for the state estimator. This will be used to calculate the corresponding error bounds $\epsilon_{L,R}^{(1)}$ for $\hat{p}^{(1)}$. Another input is $\mathcal{D}_1$ that contains the relevant empirical noise parameters in all $N^{(1)}$ repetitions of \texttt{Distillation-(a)}, and the timekeeping $\Delta t$ of the generated Werner states to track the noise effects during the state preparation. The notation $\mathcal{D}_1(\star)$ means that we calculate for $(p_s^{(1)}(\Delta t_1,\star) + \dots + p_s^{(1)}(\Delta t_{N^{(1)}},\star))/{N^{(1)}}$, with $\star$ as the Werner parameter for all states in all the $N^{(1)}$ instances, and with $p_s^{(1)}$ as the expected success probability of a single instance. We consider each instance separately since the success probability depends on $\Delta t$. We then define the expected success probability $p^{(1)} := \mathcal{D}_1(w)$ for the entire experiment. At the inversion step of the Disti-Mator, it is sufficient to use the bisection search algorithm to produce a unique estimation $\hat{w}$ since $p^{(1)}$ is a monotonically decreasing function of $w$ even in the presence of noise (see Methods-\ref{M_e}). Referring back to the calculations that resulted to Eq.~\eqref{estimation_werner}, we find that $\epsilon_{L,R}^{(1)}$ have $\mathcal{O}(\epsilon_w^2)$ subleading terms in the noiseless case. For the Disti-Mator, we choose the bisection search tolerance to be $o(\epsilon_w^2)$. Finally, using Hoeffding's inequality, the failure probability can be bounded from above by
\begin{equation}
    \delta= \sum_{m=L,R}\exp (-2N^{(1)}(\epsilon_m^{(1)})^2),
\end{equation}
where $\epsilon_L^{(1)} := \hat{p}^{(1)}-\mathcal{D}_1(\hat{w}+\epsilon_w)$ and 
$\epsilon_R^{(1)} := \mathcal{D}_1(\hat{w}-\epsilon_w)-\hat{p}^{(1)}$. 


\begin{figure}[htbp]
\begin{minipage}{\linewidth}
\begin{algorithm}[H]
{\small
\begin{algorithmic}[1]
\caption{\small Werner state estimation protocol}
\label{alg:wernerParamEstimationI}
\Require 
\Statex Total number of unknown $\rho_w\otimes\rho_w$: $N^{(1)}$ 
\Statex Distillation experiment: $\mathcal{D}_1$ \Comment{Via a noise model}
\Statex Outcome statistics: $\hat{p}^{(1)} := {n^{(1)}}/{N^{(1)}}$
\Statex Error bound for estimation: $\epsilon_w$

\Ensure 
\Statex Estimated Werner parameter: $\hat{w}$
\Statex Failure probability: $\delta$

\State $a_0 \leftarrow 0$; $b_0\leftarrow2/3$ \Comment{Initial endpoints of search}
\State $\hat{w} \leftarrow$ \Call{BisectionMethod}{$\mathcal{D}_1$, $a_0$, $b_0$, $\hat{p}^{(1)}$}
\Statex \Comment{see Supplementary Material}

\State $\epsilon_L^{(1)} \leftarrow \hat{p}^{(1)}-\mathcal{D}_1(\hat{w}+\epsilon_w)$
\State $\epsilon_R^{(1)} \leftarrow \mathcal{D}_1(\hat{w}-\epsilon_w)-\hat{p}^{(1)}$ \Comment{Monotonically decreasing}

\State $\delta \leftarrow \sum_{m=L,R}\exp (-2N^{(1)}(\epsilon_m^{(1)})^2)$ 
\Statex \Comment{Calculate failure probability}
\end{algorithmic}
}
\end{algorithm}
\end{minipage}
\end{figure}

\begin{figure*}[t]
    \centering
    \begin{minipage}{.32\textwidth}
        \includegraphics[width=\textwidth]{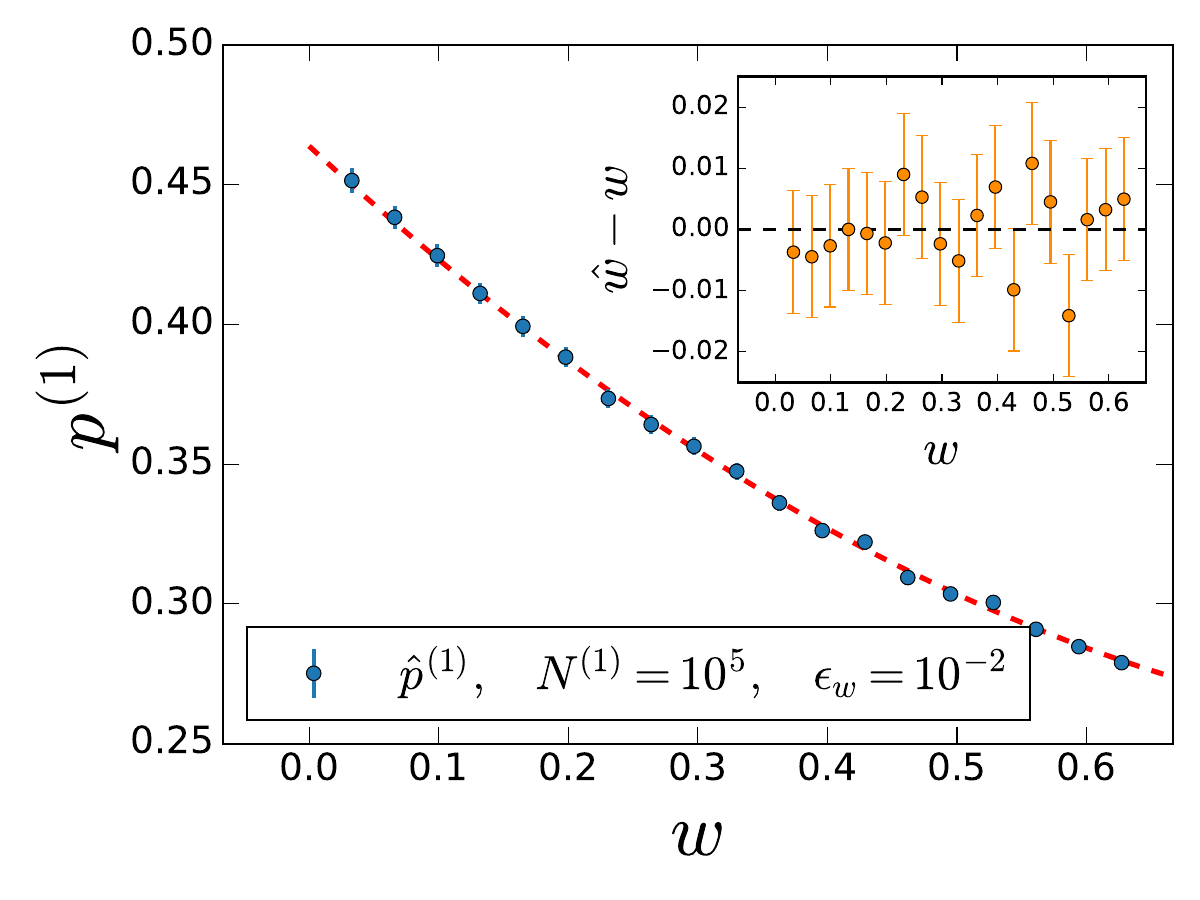}
        \textbf{(a)} \quad $N^{(1)} = 10^5$
    \end{minipage}%
    \quad
    \begin{minipage}{.32\textwidth}
        \centering
        \includegraphics[width=\textwidth]{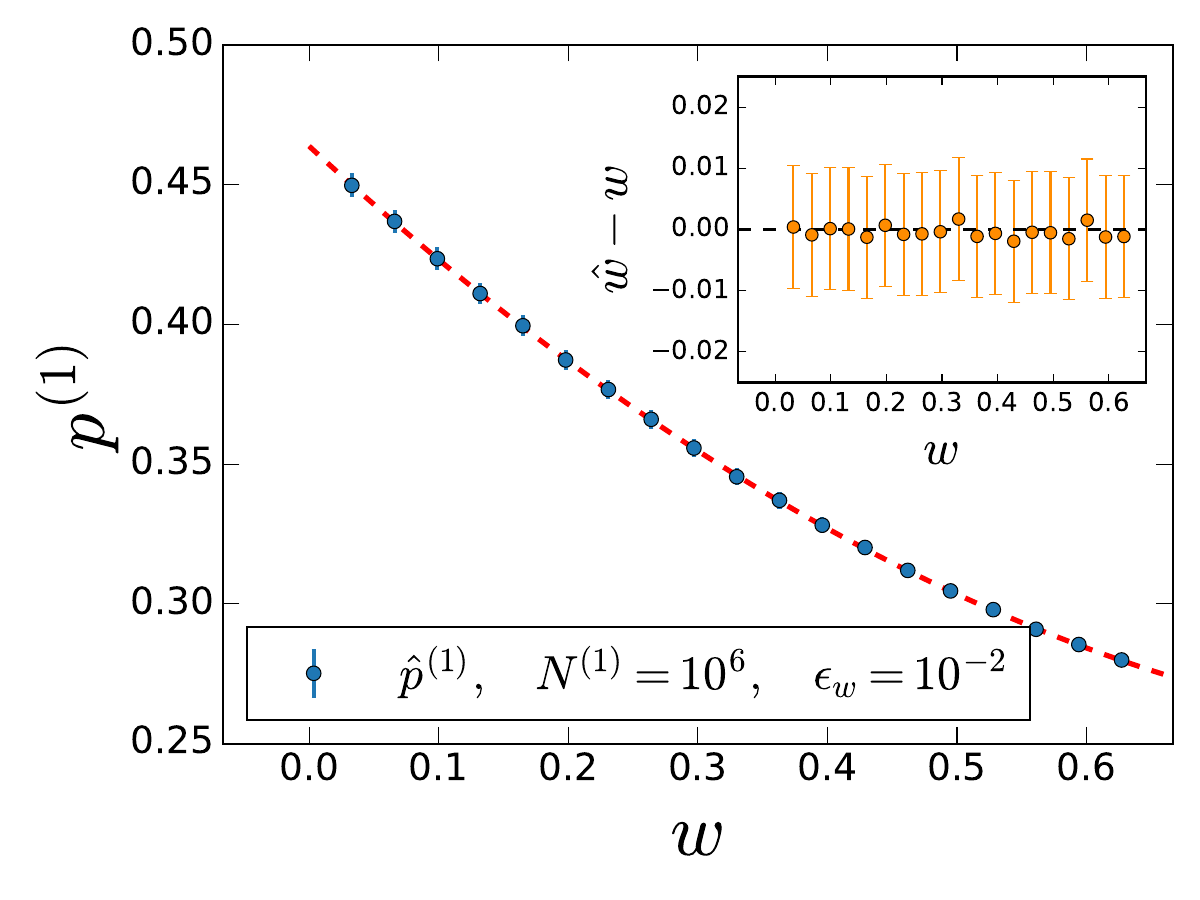}
        \textbf{(b)} \quad $N^{(1)} = 10^6$
    \end{minipage}%
    \quad
    \begin{minipage}{.32\textwidth}
        \centering
        \includegraphics[width=\textwidth]{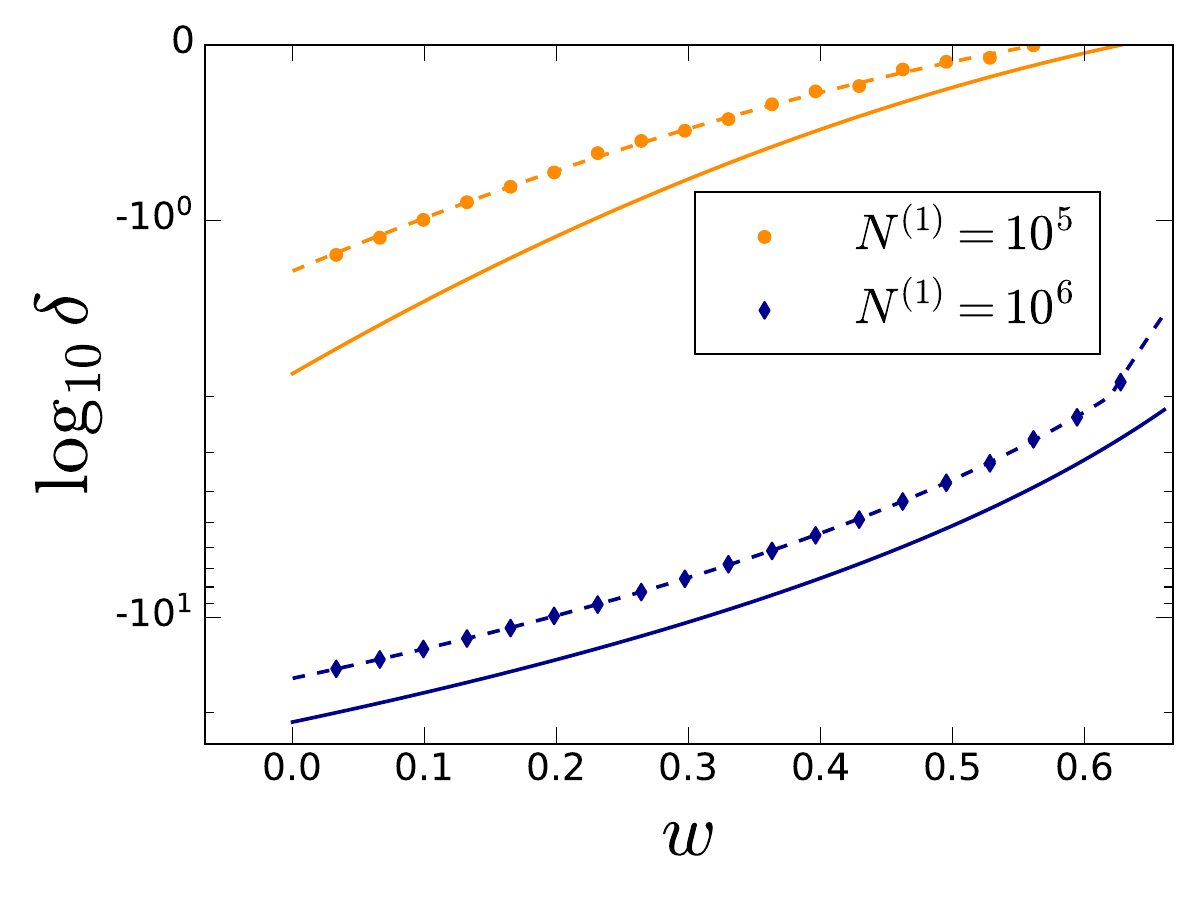}
        \textbf{(c)} \quad Failure probability bound $\delta$
    \end{minipage}%
    \caption{\textbf{Parameter estimation for Werner states in a simulated noisy distillation experiment}, with the following noise parameters: ${T_{A,B}^{\text{dpo},\text{dph}}}$ are the characteristic times for the depolarizing and dephasing channels acting on the first prepared state such that $\smash[t]{t/T_{A,B}^{\text{dpo},\text{dph}}=\Delta t_{\text{geom}}(p_g)/100}$, where $t_{\text{geom}}(p_g)$ is drawn from a geometric distribution with Bernoulli success probability equal to the $\rho_w$-generation rate of $p_g = 0.2$; $y_{A,B}=0.01$ are the CNOT depolarizing parameters; $\smash[t]{\eta_{A,B}^{Z}= 0.99}$ are the non-error probabilities of the measuring devices. We set the error bound as $\epsilon_w = 10^{-2}$. \textbf{(a)} Estimation with $N^{(1)} = 10^5$ $\rho_w^{\otimes2}$ Werner pairs; \textbf{(b)} with $N^{(1)} = 10^6$ pairs. For both cases, the empirical success probabilities $\hat{p}^{(1)}$ are shown alongside the red dashed curves representing the expected behavior. The inset plots show the deviation of the estimation from the true $w$. \textbf{(c)} The failure probability bound $\delta$ for each simulation. The solid curves indicate the expected bound given a noiseless distillation, while the dashed curves indicate the expected bound for a noisy distillation with the given noise parameters.}
    \label{fig:werner-estimation-combined}
\end{figure*}

We perform a numerical study of the estimation protocol given in Algorithm~\ref{alg:wernerParamEstimationI} to assess its performance in a realistic distillation experiment (see Fig.~\ref{fig:werner-estimation-combined}). We set our error bound for the Werner parameter to be $\epsilon_w= 10^{-2}$, and we take the following empirical parameters: $T_{A,B}^{\text{dpo},\text{dph}}$ are the characteristic times of the depolarization and dephasing channels in the preparation stage such that ${\Delta t/T_{A,B}^{\text{dpo},\text{dph}}=t_{\text{geom}}(p_g)/100}$, where $t_{\text{geom}}(p_g)$ is drawn from a geometric distribution with Bernoulli success probability equal to the $\rho_w$-generation rate of $p_g = 0.2$; $y_{A,B}= 0.01$ are the CNOT depolarizing parameters; $\eta_{A,B}^{Z}= 0.99$ are the non-error probabilities of the $Z$-measuring devices. We investigate the estimation performance of the state estimator for $N^{(1)} \in \{10^5, 10^6\}$ of the pairs $\rho_w^{\otimes 2}$. 

From the generated data, we observe that we can achieve a good estimation for $N^{(1)}=10^6$ for states far from the low-fidelity regime. This is supported by the $\delta$-plot in Fig.~\ref{fig:werner-estimation-combined}, where $\delta$ approaches zero exponentially fast if we instead approach $w$ close to zero. 
On the contrary, $N^{(1)}=10^5$ pairs are not enough to guarantee a reliable estimation, say, if we desire $\delta\leq 10^{-2}$. The simulation results agree with Fig.~\ref{fig:werner_samples}, where the resource cost for an estimation via a noisy distillation was larger than $10^5$ pairs if $\delta=10^{-2}$.

\textbf{Disti-Mator for Bell-diagonal states:} For estimating Bell-diagonal states $\overline{\rho}(\mathbf{q})$, we consider the measurement statistics from all three distillation protocols in Fig.~\ref{fig:distillation}. Here, multiple degrees of freedom can lead to several valid estimations prior to a gauge fixing. That is, if the inversion process is not restricted, we might obtain an estimation far from the true state that satisfies the same measurement statistics. We impose the constraint $q_1> 1/2$ which ensures that the Disti-Mator produces a unique estimation (see Methods-\ref{M_e}). This assumption is also consistent with recent experimental demonstration of fidelity up to $0.88$ between entangled trapped-ion qubits~\cite{krutyanskiy2023entanglement} separated by 230 meters.

We use three distillation protocols fully described by $\mathcal{D}_i$ ($i\in \{1,2,3\}$), each executed $N^{(i)}$ times. For our noise model, we find that $p^{(i)}$ remains a monotonically increasing function of a single intermediate variable $x_i$, similar to the noiseless case (see Methods-\ref{M_e}). Hence, the state estimator can independently estimate $\hat{x}_i$ from $\hat{p}^{(i)}$ via a bisection search. The estimation $\hat{\mathbf{q}}$ can then be determined once we obtain $\hat{\mathbf{x}}$. The overall failure probability bound $\delta$ incorporates the failure probability bounds associated with all three measurement statistics,
\begin{equation}
    \delta = 1 - \prod_{i=1}^3\left(1-\sum_{m=L,R}\exp[-2N^{(i)}(\epsilon_m^{(i)})^2]\right), 
\end{equation}
where $\epsilon_L^{(i)} := \hat{p}^{(i)}-\mathcal{D}_i(\hat{x}_i-\epsilon_i)$ and $\epsilon_R^{(i)} := \mathcal{D}_i(\hat{x}_i+\epsilon_i)-\hat{p}^{(i)}$. Thus, we expect the trace distance $D(\hat{\rho}(\hat{\mathbf{q}}),\overline{\rho}(\mathbf{q}))$ between the estimation and the true Bell-diagonal state to exceed $\epsilon_T =\sum_{i=1,2,3}\epsilon_i$ with probability no greater than $\delta$. The entire Bell-diagonal state estimation protocol is summarized in Algorithm~\ref{alg:bellParamEstimationI} (see Supplementary Material for more details).


\begin{figure}[htbp]
\begin{minipage}{\linewidth}
\begin{algorithm}[H]
{\small
\begin{algorithmic}[1]
\caption{\small Bell-diagonal estimation protocol}
\label{alg:bellParamEstimationI}
\Require 
\Statex Total number of unknown ${\rho}\otimes{\rho}$: $N = \sum_{i\in\{1,2,3\}} N^{(i)}$ 
\Statex Distillation experiments: $\mathcal{D}_i$ \Comment{Via a noise model}
\Statex Outcome statistics: $\hat{p}^{(i)} := {n^{(i)}}/{N^{(i)}}$
\Statex Error bound w.r.t. $x_i$: $\epsilon_i$ 
\Statex \Comment{Error bound on trace distance is $\epsilon_T:=\sum_{i\in\{1,2,3\}}\epsilon_i$}

\Ensure 
\Statex Estimated Bell-diagonal parameters $\hat{\mathbf{q}}:=(\hat{q}_1, \hat{q}_2, \hat{q}_3,\hat{q}_4)$
\Statex Failure probability: $\delta$

\For{$i=1$ to $3$}
    \State $a_0 \leftarrow 1/2$; $b_0\leftarrow 1$ \Comment{Initial endpoints of search}
    \State $\hat{x}_i \leftarrow$ \Call{BisectionMethod}{$\mathcal{D}_i$, $a_0$, $b_0$, $\hat{p}^{(i)}$}
    \Statex \Comment{see Supplementary Material}
    \State $\epsilon_L^{(i)} \leftarrow \hat{p}^{(i)}-\mathcal{D}_i(\hat{x}_i-\epsilon_i)$
    \State $\epsilon_R^{(i)} \leftarrow \mathcal{D}_i(\hat{x}_i+\epsilon_i)-\hat{p}^{(i)}$ \Comment{Monotonically increasing}
\EndFor 

\State $\hat{q}_1 \leftarrow (  -1 + \hat{x}_1 + \hat{x}_2 + \hat{x}_3 )/2$
\State $\hat{q}_2 \leftarrow (~~~1 + \hat{x}_1 - \hat{x}_2 - \hat{x}_3 )/2$
\State $\hat{q}_3 \leftarrow (~~~1 - \hat{x}_1 + \hat{x}_2 - \hat{x}_3 )/2$
\State $\hat{q}_4 \leftarrow (~~~1 - \hat{x}_1 - \hat{x}_2 + \hat{x}_3 )/2$ \Comment{Calculate Bell parameters}

\State $\delta \leftarrow 1 - \prod_{i=1}^3\left(1-\sum_{m=L,R}\exp[-2N^{(i)}(\epsilon_m^{(i)})^2]\right)$
\Statex \Comment{Calculate failure probability}
\end{algorithmic}
}
\end{algorithm}
\end{minipage}
\end{figure}

\begin{figure*}[t]
    \centering
    \begin{minipage}{.44\textwidth}
        \includegraphics[width=\textwidth]{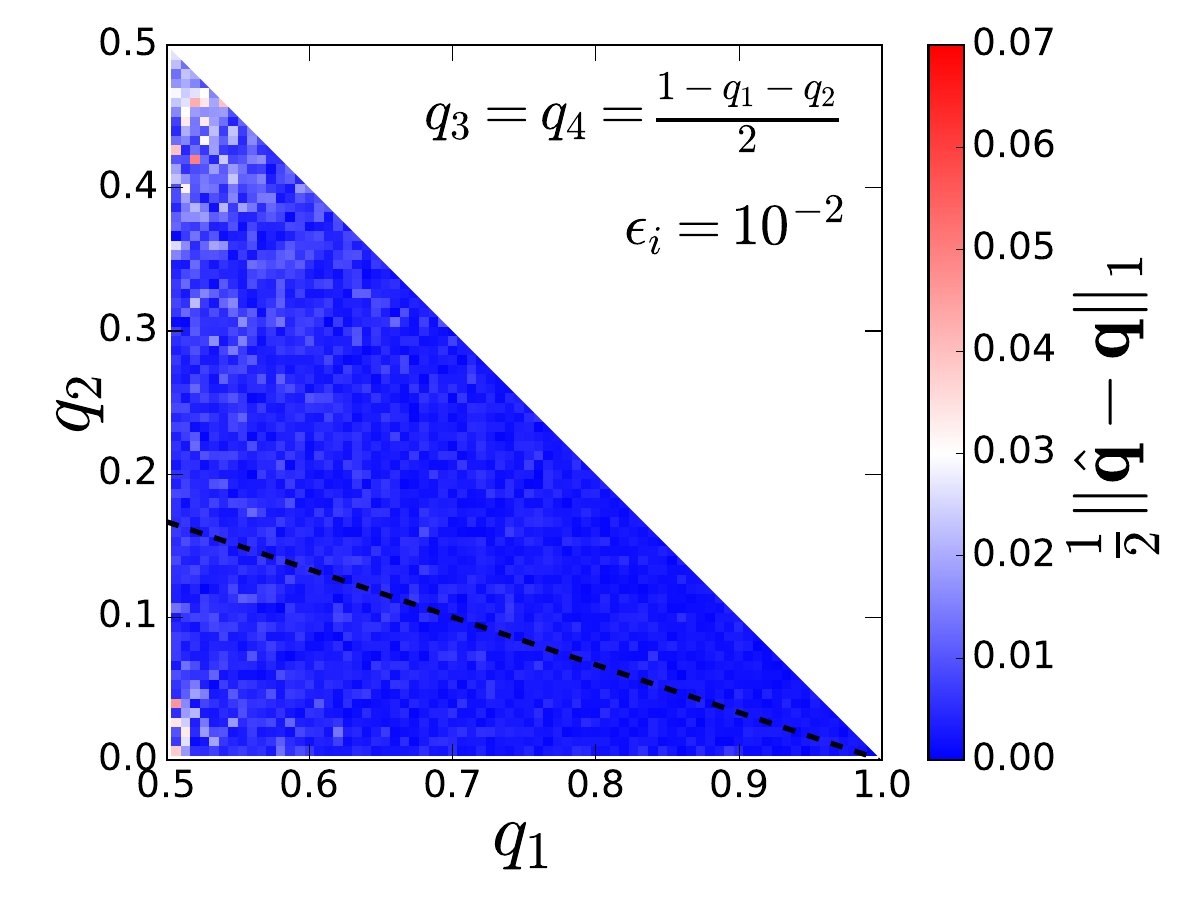}
        \textbf{(a)} Trace distance $D(\hat{\rho}(\hat{\mathbf{q}}),\overline{\rho}(\mathbf{q}))=\frac{1}{2}\|\hat{\mathbf{q}}-\mathbf{q}\|_1$
    \end{minipage}%
    \qquad
    \begin{minipage}{.44\textwidth}
        \includegraphics[width=\textwidth]{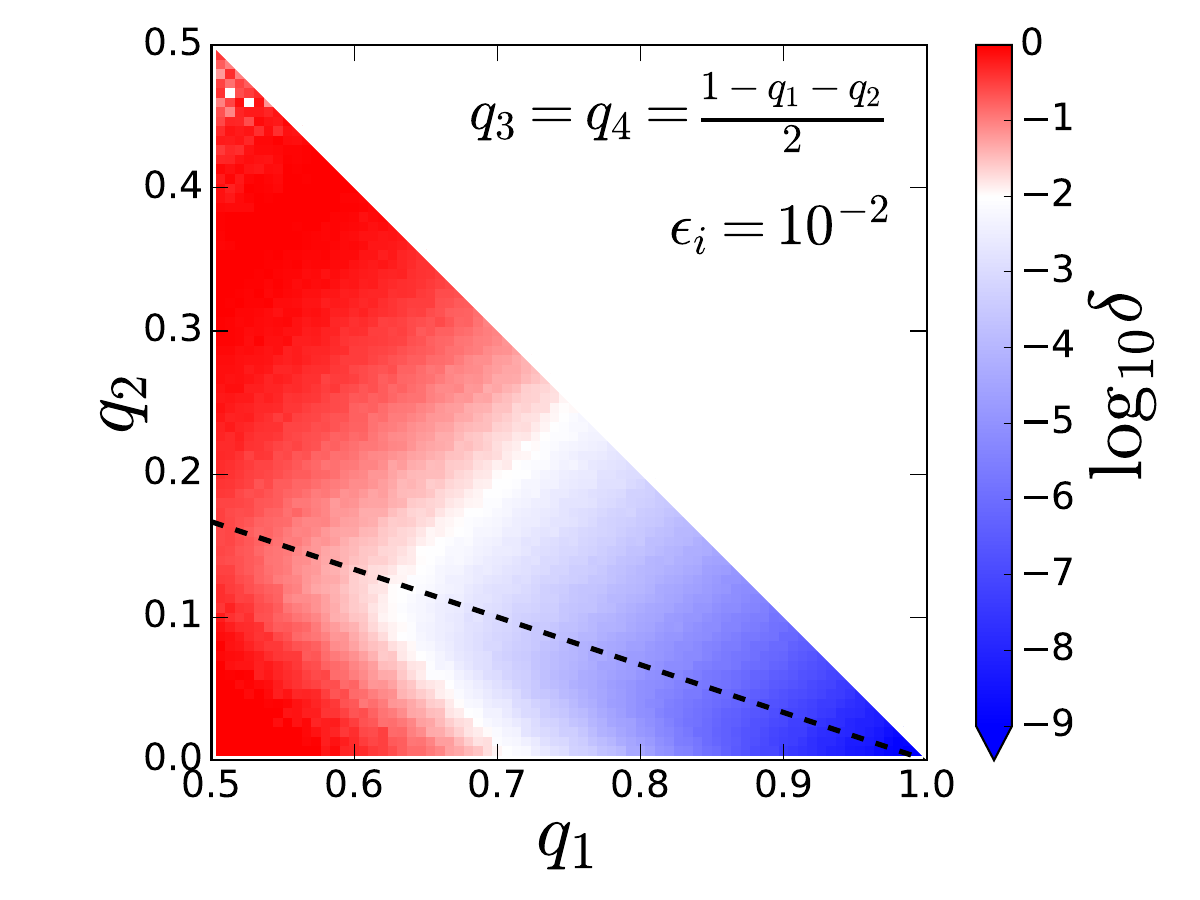}
        \textbf{(b)} Failure probability bound $\delta$ 
    \end{minipage}%
    \caption{\textbf{Parameter estimation for Bell-diagonal states in a simulated noisy distillation experiment,} with $N^{(i)} = 2\times 10^5$ the number of $\overline{\rho}^{\otimes2}$ pairs for each distillation protocol, and with the following noise parameters: characteristic times $\smash[t]{T_{A,B}^{\text{dpo},\text{dph}}}$ of depolarizing and dephasing channels acting on the first prepared state such that $\smash[t]{\Delta t/T_{A,B}^{\text{dpo},\text{dph}}=t_{\text{geom}}(p_g)/100}$, where $t_{\text{geom}}(p_g)$ is drawn from a geometric distribution with Bernoulli success probability equal to the $\overline{\rho}$-generation rate of $p_g = 0.2$; $m_{A,B}=0.01$ are the depolarizing parameters for the local $\pm\pi/2$ rotations; $y_{A,B}= 0.01$ are the CNOT depolarizing parameters; ${\eta_{A,B}^{Z,X}= 0.99}$ are the non-error probabilities of the measuring devices. We set $q_3=q_4=(1-q_1-q_2)/2$, and the error bound $\epsilon_i=10^{-2}$ for all $i$. The dashed lines describe the collection of Werner states. \textbf{(a)} Trace distance $D(\hat{\rho}(\hat{\mathbf{q}}),\overline{\rho}(\mathbf{q}))$ between the estimation $\hat{\rho}(\hat{\mathbf{q}})$ and the expected state in Bell-diagonal form $\overline{\rho}(\mathbf{q})$. We observe that the estimation is close to the expected $\mathbf{q}$ whenever $q_1$ is close to one. \textbf{(b)} The failure probability bound $\delta$ of the trace distance exceeding $\epsilon_T = 3\times10^{-2}$.}
    \label{fig:bell-estimation-combined}
\end{figure*}

We simulated distillation experiments in a realistic setting to assess the performance of the estimation protocol given in Algorithm~\ref{alg:bellParamEstimationI} (see Fig.~\ref{fig:bell-estimation-combined}). Here, we investigate over the collection of Bell-diagonal states with coefficients $\mathbf{q}=(q_1,q_2,(1-q_1-q_2)/2,(1-q_1-q_2)/2)$. We set the number of pairs $N^{(i)} = 2\times10^{5}$ and the error bounds $\epsilon_i = 10^{-2}$ for all $i$. This means that our desired trace distance $D(\hat{\rho}(\hat{\mathbf{q}}),\overline{\rho}(\mathbf{q}))$ must not exceed $\epsilon_T = 3\times10^{-2}$. We consider $m_{A,B}=0.01$ as the depolarizing parameters for the local $\pm\pi/2$ rotations and $\eta_{A,B}^{X}= 0.99$ as the non-error probabilities of the $X$-measuring devices. The remaining noise parameters are set to be the same as those in the Werner case.

From the generated data, we find that the failure probability decays exponentially fast as we approach $q_1 = 1$. Hence, high-fidelity states can be reliably estimated with the chosen $N^{(i)}$. If we want to guarantee $\delta\leq 10^{-2}$, we find that the estimator can estimate the true states within the error bound $\epsilon_T$ if $q_1\geq 0.6$. 


\section{Discussion}\label{chap:discussion}
In this work, we have shown that we can robustly and efficiently characterize entangled states using the measurement statistics obtained from probabilistic two-way entanglement distillation protocols. In particular, whenever distillation is a necessary step, our estimation method offers an efficient way to certify the Bell-diagonal parameters of an arbitrary state from the distillation statistics without consuming further quantum resources beyond what is needed for the distillation itself. This method allows us to bypass an additional step for noise estimation such as a tomographic protocol. Moreover, if the initial entangled states are of high fidelity, our method has a more favorable sample complexity than the conventional tomographic protocol. 

Further, adopting this theoretical framework, we have developed a state estimator---the `Disti-Mator'---that is robust even in the presence of SPAM errors and noisy distillation gates. 
We have also assessed the utility of our state estimator by conducting numerical simulations of noisy distillation experiments. We show that the estimator works well for Bell-diagonal states, and we also argue that this estimator can characterize the Bell-diagonal elements of an arbitrary state under the same realistic conditions. Our work will thus be useful for characterizing the entangled links generated in a quantum network in the near-term era. As a direct extension of our work, a Disti-Mator-like estimator can be constructed with more sophisticated distillation protocols (say, in Refs.~\cite{fujii2009entanglement,goodenough2024near}), which may reduce the sample complexity further. 

The importance of our work can also be justified from the perspective of learning Pauli noise, partially explored in Ref.~\cite{Maity23}. Recently, much interest has been devoted on efficiently estimating Pauli noise, which is crucial for practical information processing tasks \cite{Fujiwara03,Chiuri11,Ruppert12,Flammia20,Flammia21,Harper21,Chen23}. A direct extension for this work is to design a Disti-Mator-like estimator for a general Pauli noise model.  

We note that the high fidelities required for the distillation protocol to outperform the tomographic approach have been achieved experimentally over 230 meters with ion-trap qubits in Ref.~\cite{krutyanskiy2023entanglement}.
Experiments over several kilometers using telecom-wavelength photons~\cite{krutyanskiy2023telecom,knaut2024entanglement} have come remarkably close to this threshold, demonstrating that with near-future improvements to the hardware, noise estimation via distillation protocols will be more suitable at long distance.
Other systems such as solid-state spins~\cite{hermans2022qubit} have demonstrated higher-than-required fidelities with qubit separation of a few meters as well.

Distillation-based noise characterization opens a number of interesting possibilities when operating a quantum network.
Information about noise parameters can be extracted while delivering high-fidelity entangled pairs, provided that distillation succeeds, to the applications.
Distillation can potentially act as a real-time monitoring of physical links, allowing the quantum network to react dynamically to inevitable changes, as well as provide link updates useful in routing protocols.
Finally, Disti-Mator can serve as an efficient tool for quantum state verification of link-level as well as end-to-end entangled states.


\section{Methods}\label{chap:methods}

\subsection{Statistical tools used for parameter estimation}\label{M_a}
For the purpose of estimating state parameters, we use Hoeffding's inequality \cite{hoeffding1963prob}. Suppose $X_1,...,X_n$ are independent Bernoulli random variables, but not necessarily identically distributed, with $a \leq X_i \leq b$ for all $i$. Then, for any $t > 0$, 
\begin{equation}
    \Pr[\frac{1}{n}(\mathcal{S}-\mathbb{E}[\mathcal{S}])  \geq t] \leq  \exp(-\frac{2nt^2}{(b -a)^2}), 
\end{equation}
where $\mathbb{E}[\mathcal{S}]$ is the expectation value of $\mathcal{S} =\sum_{i=1}^n X_i$. 


\subsection{Distillation protocols}\label{sec:methods_distillation}

In this section, we discuss the probabilistic distillation protocols that are central to our estimator. We first consider the protocol proposed by Bennett {\it et al.} \cite{Bennett96(1)} (\texttt{Distillation-(a)} in Fig.~\ref{fig:distillation}). The undistilled copies are given in Eq.~\eqref{eqn:arbitrary-state}. The goal is to distill the copies towards the $\ket{\Phi^+}$ state. The protocol proceeds as follows.

1) Alice and Bob start with two copies of $\rho$.

2) Both apply a local CNOT operation on their respective halves, with the first copy as the control and the second as the target.

3) Each party measures their respective target qubit in the $Z$-basis. Their results are classically communicated to each other.

4) If they obtain correlated outcomes (i.e., both get ``up'' or both get ``down''), then they keep the unmeasured control copy. Otherwise, they discard this copy. 

This distillation protocol decreases the occurrence of $X$ errors while it increases that of $Z$ errors \cite{horodecki2009quantument}. For our estimator, we say that the success probability $p^{(1)}$ of both measurement outcomes being ``up'' is half the coincidence probability of the protocol, i.e., 
$p^{(1)}=((q_1+q_2)^2 + (q_3+q_4)^2)/2$. Whenever they keep the control copy, and provided that $q_1 > 1/2$, its fidelity with respect to the $\ket{\Phi^+}$ state increases to $(q_1^2 + q_2^2)/2p^{(1)}$.

We also consider other distillation protocols since this distillation protocol alone is not sufficient for a general estimation of the Bell-diagonal coefficients. For \texttt{Distillation-(b)} in Fig.~\ref{fig:distillation}, the control copy is measured in the $X$ basis instead of the target copy. The success probability of this new protocol is $p^{(2)} = ((q_1+q_3)^2 + (q_2+q_4)^2)/2$. However, this protocol increases the occurrence of $X$ errors. \texttt{Distillation-(c)} is similar to the first protocol, but with single-qubit $X$ rotations prior to the CNOT operations, thereby switching between $X$ and $Z$ errors \cite{Deutsch96,horodecki2009quantument}. For this protocol, the success probability is $p^{(3)} = ((q_1 + q_4)^2 + (q_2 + q_3)^2)/2$. We note that these two protocols successfully distill the unmeasured copy (with probability $2p^{(i)}$) whenever $q_1 > 1/2$.


\subsection{Estimating Werner states}\label{M_b}
Consider a scenario where Alice and Bob perform a noiseless \texttt{Distillation-(a)} on shared pairs of two-qubit i.i.d. Werner states. Upon Alice and Bob jointly measuring the second copy in the $Z$ basis, they obtain the measurement statistics $\hat{p}^{(1)}$. Using this, we can estimate the Werner parameter $\hat{w}$ by inversion, since $\hat{p}^{(1)} = \frac{1}{4}(2-2\hat{w}+ \hat{w}^2)$. However, since the number of samples used for the estimation is finite, $\hat{p}^{(1)}$ may differ from its mean $p^{(1)}$ and therefore the estimated $\hat{w}$ will also deviate from $w$ by at most $\epsilon_w$.

We now bound the difference between the estimated $\hat{w}$ and its empirical mean in detail. Taking $|\hat{w}- w| \geq \epsilon_w$ for some error bound $\epsilon_w$, and Eq.~\eqref{eqn:wfromp1}, we obtain 
\begin{align}
    &|\hat{w}- w| \geq \epsilon_w \nonumber \\
    &\Rightarrow \abs{\left(1 - \sqrt{4\hat{p}^{(1)}-1}\right) - \left(1 - \sqrt{4p^{(1)}-1}\right)} \geq \epsilon_w \nonumber \\
    &\Rightarrow ~\text{either}~~ 4p^{(1)} - 1 \geq \epsilon_w^{2} + 4\hat{p}^{(1)} - 1 + 2\epsilon_w\sqrt{4\hat{p}^{(1)} - 1} \nonumber\\
    &~~~~~~~\text{or}~~ 4p^{(1)} - 1 \leq \epsilon_w^{2} + 4\hat{p}^{(1)} - 1 - 2\epsilon_w\sqrt{4\hat{p}^{(1)} - 1} \nonumber \\
    &\Rightarrow ~\text{either}~~ -(\hat{p}^{(1)} - p^{(1)}) \geq \frac{1}{4} (\epsilon_w^{2} + 2\epsilon_w(1-\hat{w}))  \nonumber \\
    &~~~~~~~\text{or}~~ \hat{p}^{(1)} - p^{(1)} \geq \frac{1}{4} (-\epsilon_w^{2} + 2\epsilon_w(1-\hat{w})). 
\end{align}
Denoting $\epsilon^{(1)}_R(\hat{w}, \epsilon_w):= \frac{1}{4} (\epsilon_w^{2} + 2\epsilon_w(1-\hat{w}))~~\text{and}~~ \epsilon^{(1)}_L (\hat{w}, \epsilon_w):= \frac{1}{4} (-\epsilon_w^{2} + 2\epsilon_w(1-\hat{w}))$, we obtain a bound on the failure probability based on Hoeffding's inequality, 
\begin{equation}
    \text{Pr}(|\hat{w}- w| \geq \epsilon_w) \leq \sum_{m=L,R}\exp{-2 N^{(1)} (\epsilon^{(1)}_m(\hat{w}, \epsilon_w))^2}.\nonumber
\end{equation}

As a first step toward implementing these concepts in a realistic scenario, we consider a joint depolarization of the idle qubits. This depolarization is defined by Eq. \eqref{eqn:global_depolarizing} with the depolarizing parameter $\lambda (t)$ that varies over time. We attribute this to the memory decoherence experienced by the stored first copy while waiting for the second copy to be successfully distributed. As mentioned in the main text, the second copy is unaffected by the depolarizing noise. After some storage time $t$, the state in Eq.~\eqref{eqn:Werner_state} has the following form
\begin{align*}
    \Lambda(\rho_w) =& \,\,(1-\lambda(t))(1-w)\ketbra{\Phi^+}{\Phi^+} \\
    &\,\,+ \frac{(1-\lambda(t))w+\lambda(t)}{4}\mathbb{I}.
\end{align*}
We follow the same procedure discussed for the noiseless case to estimate the Werner parameter. In the presence of this depolarizing noise, the measurement statistics becomes
\begin{equation}
    p^{(1)}_{\lambda(t)} = \frac{1}{2}[(q'_1+q'_2)(q_1+q_2)+ (q'_3+q'_4)(q_3+q_4)] \nonumber
\end{equation}
where $q'_1(t) = (1-\lambda(t))(1-w)+ (1-\lambda(t))\frac{w}{4}+\frac{\lambda(t)}{4}$ and $q'_2(t)=q'_3(t)=q'_4(t)= (1-\lambda(t))\frac{w}{4}+\frac{\lambda(t)}{4}$. However, each instance $j$ in a real experiment will have a different $\lambda(t)=\lambda_j(t)$. Therefore, the estimate of $p_S^{(1)}$ can be stated as an arithmetic mean, 
\begin{equation}
    p_S^{(1)} = \frac{1}{N^{(1)}}\sum_{j=1}^{N^{(1)}} p^{(1)}_{\lambda_j(t)} = \frac{1}{4}(Sw^2-2S w+ S+1), \nonumber
\end{equation}
where $S= (1/N^{(1)})\sum_{j=1}^{N^{(1)}} (1-\lambda_{j}(t))$. Inverting the equation above, we obtain
\begin{equation}
    w = 1-\sqrt{1-\frac{S+1-4p_S^{(1)}}{S}}.\nonumber
\end{equation}
Assuming that $|\hat{w}- w|  \geq \epsilon_w$, the bounds on the difference between $p_S^{(i)}$ and $\hat{p}^{(i)}$ propagate as follows:
\begin{align}
    &|\hat{w}- w| \geq \epsilon_w \nonumber \\
    &\Rightarrow \abs{ \sqrt{1-\frac{S+1-4\hat{p}^{(1)}}{S}} - \sqrt{1-\frac{S+1-4p_S^{(1)}}{S}}} \geq \epsilon_w \nonumber \\
    &\Rightarrow ~\text{either,}~~ -(\hat{p}^{(1)} - p_S^{(1)}) \geq \frac{S}{4} (\epsilon_w^{2} + 2\epsilon_w(1-\hat{w}))  \nonumber \\
    &~~~~~~~\text{or}~~ \hat{p}^{(1)} - p_S^{(1)} \geq \frac{S}{4} (-\epsilon_w^{2} + 2\epsilon_w(1-\hat{w})). \nonumber
\end{align}
From the above analysis we obtain the bound
\begin{equation}
    \text{Pr}(|\hat{w}- w|\geq \epsilon_w) \leq \sum_{m=L,R}\exp{-2 N^{(1)} (\epsilon'^{(1)}_m(\hat{w}, \epsilon_w))^2},
\end{equation}
where $\epsilon'^{(1)}_R(\hat{w}, \epsilon_w) := \frac{S}{4} (\epsilon_w^{2} + 2\epsilon_w(1-\hat{w}))$ and $\epsilon'^{(1)}_L(\hat{w}, \epsilon_w):= \frac{S}{4} (-\epsilon_w^{2} + 2\epsilon_w(1-\hat{w}))$.

\textbf{Werner parameter estimation in tomography:}
We now compare these results to a tomographic estimation. We first rewrite the Werner state into a locally decomposable form $\rho_w = (\mathbb{I}\otimes\mathbb{I} + \sum_{\sigma=X,Y,Z} (1-w)\sigma\otimes\sigma)/4$. Now, Alice and Bob can measure their local qubits in $\lbrace X,Y,Z \rbrace$ basis and depending on the joint outcome probability of $X \otimes X$, or $Y \otimes Y$, or $Z \otimes Z$ they can estimate the Werner parameter. The probability of both Alice and Bob getting ``up'' outcomes is $p^{(1)}_{\text{tom}} = {(2-w)}/{4}$.

Now if $|\hat{w}- w| \geq \epsilon_w$, then we expect that $|\hat{p}^{(1)}_{\text{tom}} - p^{(1)}_{\text{tom}}|\geq {\epsilon_w}/{4}$. Using Hoeffding's inequality, 
\begin{equation}\label{eqn:tomography_Hoff}
    \text{Pr}(|\hat{w}- w| \geq \epsilon_w) \leq 2\exp{-\frac{1}{8}N_{\text{tom}}^{(1)}\epsilon_w^{2}}. 
\end{equation}
We note that for tomography the states can be measured as soon as they are generated, and therefore do not experience any memory decoherence.

\subsection{Estimating Bell-diagonal states}\label{M_c}
For estimating the parameters of a Bell-diagonal state (defined in Eq.~\eqref{eqn:Bell-diagonal}), we consider the measurement statistics $\lbrace \hat{p}^{(1)},\hat{p}^{(2)},\hat{p}^{(3)} \rbrace$ from \texttt{Distillation-(a)}, \texttt{-(b)}, and \texttt{-(c)}, respectively. We then invert these to obtain the state parameters of a Bell-diagonal state in the noiseless scenario, as modeled in Eq. \eqref{eqn:x_i}. The estimate $\hat{\mathbf{q}}$ can be calculated from the intermediate estimate $\hat{\mathbf{x}}$ with the following relations:
\begin{align}\label{q_i_est}
    \hat{q}_1 &= \frac{1}{2}(-1+\hat{x}_1+\hat{x}_2+\hat{x}_3), \nonumber \\
    \hat{q}_2 &= \frac{1}{2}(~~~1+\hat{x}_1-\hat{x}_2-\hat{x}_3), \nonumber \\
    \hat{q}_3 &= \frac{1}{2}(~~~1-\hat{x}_1+\hat{x}_2-\hat{x}_3), \nonumber \\
    \hat{q}_4 &= \frac{1}{2}(~~~1-\hat{x}_1-\hat{x}_2+\hat{x}_3),
\end{align}
where we have the estimators $\hat{x}_i:= \hat{q}_1+\hat{q}_{i+1} = \frac{1}{2}\left(1+\smash{\sqrt{4\hat{p}^{(i)}-1}} \right)$. Since $\hat{p}^{(i)}$ will differ from the expected value $p^{(i)}$, the estimated parameter $\hat{x}_i$ will consequently differ from $x_i$. Considering a failure condition $\abs{\hat{x}_i- x_i} \geq \epsilon_i$ for an error bound $\epsilon_i$, we obtain that
\begin{align}
    &\abs{\hat{x}_i- x_i} \geq \epsilon_i \nonumber \\
    &\Rightarrow \abs{\hat{x}_i - \frac{1}{2}\left(1 + \sqrt{4p^{(i)}-1}\right)} \geq \epsilon_i \nonumber \\
    &\Rightarrow ~\text{either}~ \hat{p}^{(i)} - p^{(i)} \geq   -\epsilon_i^{2} + \epsilon_i(2\hat{x}_i-1) =:\epsilon^{(i)}_L(\hat{x}_i,\epsilon_i) \nonumber \\
    &~~~~~~~\text{or}~~~~~~  p^{(i)} - \hat{p}^{(i)} \geq ~~~\epsilon_i^{2} + \epsilon_i(2\hat{x}_i-1)=:\epsilon^{(i)}_R(\hat{x}_i,\epsilon_i). \nonumber
\end{align}
Using Hoeffding's inequality, we find that the probability for each failure condition is bounded by
\begin{equation}\label{eqn:hoeffdingbound-individual-x}
    \Pr[\abs{\hat{x}_i- x_i} \geq \epsilon_i] \leq \sum_{m=L,R}\exp{-2 N^{(i)} (\epsilon^{(i)}_m(\hat{x}_i, \epsilon_i))^2}.
\end{equation}

We now consider the trace distance $D(\hat{\rho}(\hat{\mathbf{q}}),\overline{\rho}(\mathbf{q})) :=\frac{1}{2}\|\hat{\mathbf{q}}-\mathbf{q}\|_1$ between the expected state $\overline{\rho}(\mathbf{q})$ and its estimation $\hat{\rho}(\hat{\mathbf{q}})$. Notice that after a repeated use of the triangle inequality, we have
\begin{equation}
    D(\hat{\rho}(\hat{\mathbf{q}}),\overline{\rho}(\mathbf{q})) \leq \frac{1}{2}\sum_{k=1}^4 \frac{1}{2}\left(\sum_{i=1}^3\abs{\hat{x}_i-x_i}\right).
\end{equation}
For a successful estimation, we demand $\abs{\hat{x}_i- x_i} \leq \epsilon_i$, which implies that $D(\hat{\rho}(\hat{\mathbf{q}}),\overline{\rho}(\mathbf{q})) \leq \epsilon_T$, where $\epsilon_T := \sum_{i\in\{1,2,3\}}\epsilon_i$. This also results to an estimation on $\hat{\mathbf{q}}$ that satisfies $\abs{\hat{q}_i - q_i} \leq \epsilon_T/2$. We then find that
\begin{equation}\label{eqn:conc-bound-trace-distance-via-x}
    \Pr[\bigcap_{i=1}^{3}\{\abs{\hat{x}_i - x_i}\!\leq \epsilon_i\}] \leq \Pr[D(\hat{\rho}(\hat{\mathbf{q}}),\overline{\rho}(\mathbf{q}))\leq \epsilon_T].
\end{equation}
The expression on the left-hand side can be calculated using Eq.~\eqref{eqn:hoeffdingbound-individual-x}. Rearranging Eq.~\eqref{eqn:conc-bound-trace-distance-via-x} further, we obtain the concentration bound
\begin{equation}
    \Pr[D(\hat{\rho}(\hat{\mathbf{q}}),\overline{\rho}(\mathbf{q}))\geq \epsilon_T] \leq \delta(\hat{\mathbf{x}}, \{N^{(i)}\}, \{\epsilon_i\}),
\end{equation}
where 
\begin{align}
    &\delta(\hat{\mathbf{x}}, \{N^{(i)}\}, \{\epsilon_i\}) \nonumber\\
    &:= 1 - \prod_{i=1,2,3}\left[ 1-\sum_{m=L,R}\exp~(-2N^{(i)}\epsilon^{(i)}_m(\hat{x}_i,\epsilon_i)^2)\right].
\end{align}


To apply these concepts in a realistic scenario, we first introduce depolarizing noise to idling control qubits. Again, this noise is characterized by the depolarizing channel in Eq.~\eqref{eqn:global_depolarizing}. Because of the presence of depolarizing noise, the control Bell-diagonal state alters after a finite time $t$ into $\rho'=(1-\lambda(t))\overline{\rho} + \frac{\lambda(t)}{4}\mathbb{I}$,
where $\lambda(t)= 1- \exp(-{t}/{T^{\text{dpo}}})$. For \texttt{Distillation-(a)}, the success probability changes into 
\begin{align}
    p^{(1)}_{x(t)} &= \frac{1}{2}[(q'_1+q'_2)(q_1+q_2)+ (q'_3+q'_4)(q_3+q_4)], \nonumber
\end{align}
where we introduce the notation $q'_i(t) := (1-\lambda(t))q_i+ \frac{\lambda(t)}{4}$ for all $i$. However, $\lambda(t)=\lambda_j(t)$ may be different for each instance $j$ of the experiment. Hence, we have
\begin{equation}
    p^{(1)} = \frac{1}{N^{(1)}}\sum_{i=1}^{N^{(1)}} p^{(1)}_{x_i(t)} = S_1 x_1^2-S_1 x_1+ \frac{1}{4}(S_1+1), \nonumber
\end{equation}
where $S_i:= (1/N^{(i)})\sum_{j=1}^{N^{(i)}} (1-\lambda_{j}(t))$ is an arithmetic mean. From the equation above one may calculate $q_1+q_2$. In general, we have
\begin{equation}
    x_i=q_1+q_{i+1} = \frac{1}{2}\left(1+\sqrt{1-\frac{S_i+1-4p^{(i)}}{S_i}}\right). \nonumber
\end{equation}
One can now follow the same procedure as discussed earlier for the estimation of the Bell parameters in the noiseless scenario and obtain the solution
\begin{equation}
    \Pr[D(\hat{\rho}(\hat{\mathbf{q}}),\overline{\rho}(\mathbf{q}))\geq \epsilon_T] \leq \delta'(\hat{\mathbf{x}}, \{N^{(i)}\}, \{\epsilon_i\}).
\end{equation}
Here, $\delta'$ is similar to $\delta$ in the noiseless case, but with $\epsilon^{(i)}_R(\hat{x}_i,\epsilon_i) :=  S_i(\epsilon_i^{2} + \epsilon_i(2\hat{x}_i-1))$ and $\epsilon^{(i)}_L(\hat{x}_i,\epsilon_i):= S_i(-\epsilon_i^{2} + \epsilon_i(2\hat{x}_i-1))$.

\textbf{Parameter estimation of a Bell-diagonal state in tomography:} To facilitate quantum state tomography, we express the Bell-diagonal state in a locally decomposable form $\overline{\rho}({\mathbf{q}}) = (\mathbb{I}\otimes\mathbb{I} + \sum_{\sigma=X,Y,Z} t_{\sigma}({\mathbf{q}})\sigma\otimes\sigma)/4$, where $t_X({\mathbf{q}})=q_1-q_2+q_3-q_4$, $t_Y({\mathbf{q}}) = -q_1+q_2+q_3-q_4$, and $t_Z({\mathbf{q}})= q_1+q_2-q_3-q_4$. Alice and Bob locally measure on $\lbrace X,Y,Z \rbrace$ basis, and then estimate the state parameters depending on the joint measurement outcomes of $X \otimes X$, $Y \otimes Y$, $Z \otimes Z$. The probabilities $p_{\mathrm{tom}}^{(1)},p_{\mathrm{tom}}^{(2)},p_{\mathrm{tom}}^{(3)}$ of Alice and Bob getting both ``up'' outcomes when both measure in the $Z,X,Y$ basis, respectively, are
\begin{eqnarray}
    p_{\mathrm{tom}}^{(1)} &=& \frac{q_1+q_2}{2} = \frac{x_1}{2},\nonumber \\
    p_{\mathrm{tom}}^{(2)} &=& \frac{q_1+q_3}{2} = \frac{x_2}{2}, \nonumber \\
    p_{\mathrm{tom}}^{(3)} &=& \frac{q_2+q_3}{2} = \frac{1-x_3}{2}. 
\end{eqnarray}
Thus, if $\abs*{\hat{x}_i- x_i} \geq \epsilon_i$, then $\abs*{\hat{p}_{\mathrm{tom}}^{(i)}-{p}_{\mathrm{tom}}^{(i)}}\geq \epsilon_i/2 =: \epsilon^{(i)}$. This implies a concentration bound for the trace distance with an error threshold $\epsilon_T$,
\begin{eqnarray}
    & &\Pr[D(\hat{\rho}(\hat{\mathbf{q}}),\overline{\rho}(\mathbf{q}))\geq \epsilon_T]\leq \delta(\{N^{(i)}\},\{\epsilon_i\})\nonumber\\
    & &\quad= 1-\!\!\!\prod_{i=1,2,3}\left[1-2\exp(-2N_{\mathrm{tom}}^{(i)}(\epsilon^{(i)})^2)\right].
\end{eqnarray}
This again follows from a repeated application of the Hoeffding's inequality.


\subsection{Noise model for the Disti-Mator}\label{M_d}

In the following, we describe the noise model assumed by the Disti-Mator, and explain how this model affects the success probabilities of the distillation protocols. We first consider the depolarizing and dephasing effects on the first state stored in a quantum memory. The composite noise map reads
\begin{eqnarray}
    &\Delta_{\mathrm{dph},A}^{\zeta_A}&\!\!\circ\Delta_{\mathrm{dpo},A}^{\lambda_A}\circ\Delta_{\mathrm{dph},B}^{\zeta_B}\circ\Delta_{\mathrm{dpo},B}^{\lambda_B}(\overline{\rho}_{AB}) \nonumber\\
    &=&\!\!\!\!\!\! (1-\zeta_A)(1-\lambda_A)(1-\zeta_B)(1-\lambda_B)\overline{\rho}_{AB}\nonumber\\
    &+&\!\!\!\!\!\! (1-\zeta_A)(1-\lambda_A)\zeta_B(1-\lambda_B)\mathbb{I}_A\otimes Z_B\overline{\rho}_{AB}\mathbb{I}_A\otimes Z_B\nonumber\\
    &+&\!\!\!\!\!\! \zeta_A(1-\lambda_A)(1-\zeta_B)(1-\lambda_B)Z_A\otimes \mathbb{I}_B\overline{\rho}_{AB}Z_A\otimes \mathbb{I}_B\nonumber\\
    &+&\!\!\!\!\!\! \zeta_A(1-\lambda_A)\zeta_B(1-\lambda_B)Z_A\otimes Z_B\overline{\rho}_{AB}Z_A\otimes Z_B\nonumber\\
    &+&\!\!\!\!\!\! (\lambda_A+\lambda_B -\lambda_A\lambda_B)\frac{\mathbb{I}_{AB}}{4}.
\end{eqnarray}
Here, we have the channel parameters $\lambda_{A,B}(\Delta t) = 1-\exp\,({-{\Delta t}/{T_{A,B}^{\mathrm{dpo}}}})$ and $\zeta_{A,B}(\Delta t) = (1-\exp\,({-{\Delta t}/{T_{A,B}^{\mathrm{dph}}}}))/2$, with $T_{A,B}^{\mathrm{dpo}}$ and $T_{A,B}^{\mathrm{dph}}$ as their respective characteristic times. For \texttt{Distillation-(c)} in Fig.~\ref{fig:distillation}, we model the noisy local $\pm\pi/2$ rotations as
\begin{eqnarray}
    &\Lambda_{R_X}^{m_A,m_B}&\!\!\!\!(\overline{\rho}_{AB}) \nonumber\\
    &=&\!\!\!\! (1-m_A)(1-m_B)R_X\left(-\frac{\pi}{2}\right)\otimes R_X\left(+\frac{\pi}{2}\right)\nonumber\\
    & & \overline{\rho}_{AB}R_X\left(-\frac{\pi}{2}\right)^{\dagger}\otimes R_X\left(+\frac{\pi}{2}\right)^{\dagger}\nonumber\\
    & & +(m_A + m_B -m_Am_B)\frac{\mathbb{I}_{AB}}{4},
\end{eqnarray}
where $m_{A,B}$ are the associated depolarizing parameters. We model the noisy CNOT operations as
\begin{eqnarray}
    &\Lambda_{\mathrm{CNOT}}^{y_A,y_B}&\!\!\!\!(\overline{\rho}_{A_1B_1}\otimes\overline{\rho}_{A_2B_2}) \nonumber\\
    &=&\!\!\!\! (1-y_A)(1-y_B)(\mathrm{CNOT}_{A_1A_2}\otimes\mathrm{CNOT}_{B_1B_2})\nonumber\\
    & & (\overline{\rho}_{A_1B_1}\otimes\overline{\rho}_{A_2B_2})(\mathrm{CNOT}_{A_1A_2}\otimes\mathrm{CNOT}_{B_1B_2})\nonumber\\
    & & +(y_A + y_B -y_Ay_B)\frac{\mathbb{I}_{A_1B_1A_2B_2}}{16},
\end{eqnarray}
where $y_{A,B}$ are the CNOT depolarizing parameters. Finally, the ``up'' POVM elements read
\begin{eqnarray}
    \hat{M}_Z^{00} &=& (\eta_A^Z\ketbra{0}{0} + (1-\eta_A^Z)\ketbra{1}{1})\nonumber\\
    & &\otimes(\eta_B^Z\ketbra{0}{0} + (1-\eta_B^Z)\ketbra{1}{1}),\\
    \hat{M}_X^{++} &=& (\eta_A^X\ketbra{+}{+} + (1-\eta_A^X)\ketbra{-}{-})\nonumber\\
    & &\otimes(\eta_B^X\ketbra{+}{+} + (1-\eta_B^X)\ketbra{-}{-}),
\end{eqnarray}
where $\eta_{A,B}^{Z,X}$ are the non-error probabilities of the measuring devices.

We now discuss how $p^{(i)}$ varies with the above noise model. For \texttt{Distillation-(a)}, direct calculation gives
\begin{equation}
    p^{(1)} = f_1(\bm{\lambda},\bm{y},\bm{\eta}^Z)x_1^2 -f_1(\bm{\lambda},\bm{y},\bm{\eta}^Z)x_1 + C_1,
\end{equation}
where $f_1(\bm{\lambda},\bm{y},\bm{\eta}^Z) = \prod_{j=A,B}(1-\lambda_j)(1-y_j)(1-2\eta_j^Z)$, and $C_1$ is a constant written in terms of the noise parameters such that $C_1 = 1/2$ in the noiseless case. Moreover, we observe that the derivative $\partial_{x_1}p^{(1)}(x_1) = (2x_1 - 1)f_1(\bm{\lambda},\bm{y},\bm{\eta}^Z)$, which shows that $p^{(1)}$ is monotonically increasing with respect to $x_1$ over the interval $x_1\in[1/2,1]$. In fact, for \texttt{Distillation-(b)}, we get
\begin{equation}
    p^{(2)} = f_2(\bm{\lambda},\bm{\zeta},\bm{y},\bm{\eta}^X)x_2^2 -f_2(\bm{\lambda},\bm{\zeta},\bm{y},\bm{\eta}^X)x_2 + C_2,
\end{equation}
where $f_2(\bm{\lambda},\bm{\zeta},\bm{y},\bm{\eta}^X) = \prod_{j=A,B}(1-\lambda_j)(1-2\zeta_j)(1-y_j)(1-2\eta_j^X)$. We also see a similar result for \texttt{Distillation-(c)}. We summarize these results into the following Lemma.

{\bf Lemma 1:} For the given noise model, the success probability $p^{(i)}$ is a strictly monotonically increasing function with respect to the parameter $x_i$, with $x_i\in[1/2,1]$.


\subsection{Estimating general states using the Disti-Mator}\label{M_e}

We now discuss the Disti-Mator's ability to characterize general states by estimating their Bell-diagonal parameters $\mathbf{q}$. For this analysis, we calculate for the Bell-diagonal state $\overline{\rho}(\mathbf{q})$ that takes $\mathbf{q}$ as diagonal elements. Following the results from the previous section, we find that the success probability $p^{(i)}$ is a function of $x_i$, for $i\in \{1,2,3\}$. Unfortunately, due to the inherent gauge symmetry in $x_i$'s, a distillation protocol can be transformed (due to the overall effects of noise) in such a way that the measurement statistics remain invariant for two different Bell parameterizations. Hence, multiple valid $\hat{\mathbf{q}}$'s can be obtained from an estimation of $\hat{\mathbf{x}}$. Here, we confine ourselves to the region $\expval{\overline{\rho}(\mathbf{q})}{\Phi^+}= q_1> 1/2$, such that we successfully distill towards the $\ket{\Phi^+}$ state ($q_1=1$ acts as an attractor in this region \cite{Deutsch96}). The following lemma is a result of this gauge fixing.

{\bf Lemma 2:} If $\expval{\overline{\rho}(\mathbf{q})}{\Phi^+}> 1/2$, then the measurement statistics obtained from the three distillation protocols correspond to a unique $\overline{\rho}(\mathbf{q})$.
\begin{proof}
    Let the Bell parameterization $\mathbf{q}=(q_1, q_2, q_3, q_4)$ correspond to the three measurement statistics $p^{(1)}, p^{(2)}, p^{(3)}$ from the three distillation protocols, with $q_1>1/2$. Suppose that there exists another solution $\mathbf{q}'=(q'_1, q'_2, q'_3, q'_4)$ that yields the same measurement statistics, and that satisfies $q'_1>1/2$. Following Lemma 1, all $x_i$'s must be unique since the measurement statistics are strictly monotonically increasing. Thus, $x_1=q_1+q_2=q_1'+q_2'$, $x_2=q_1+q_3=q_1'+q_3'$, and $x_3=q_1+q_4=q_1'+q_4'$ must also be true. We can take $q'_1 = q_1 + \epsilon$, with $\epsilon$ being some positive or negative number. Since the measurement statistics remain invariant, then $q'_2 = q_2 - \epsilon$, $q'_3 = q_3 - \epsilon$, and $q'_4 = q_4 - \epsilon$. This leads to the contradiction since $q'_1+q'_2+q'_3+q'_4=1-2\epsilon \neq 1$. Therefore, $q_1$ must be unique. The above analysis implies we can also uniquely determine $q_2,q_3,q_4$, completing the proof.
\end{proof}

Let $\mathcal{D}_i$ describe the action of all quantum gates in all the $N^{(i)}$ repetitions of the distillation protocol $i$ (including noise effects) such that we can express the expected value as $p^{(i)} := \mathcal{D}_i(x_i)$. The inversion step is just an application of a bisection search, in which the solution $\hat{x}_i$ is unique via Lemma 2. Then, for an error bound $\epsilon_i$,
\begin{eqnarray}
    &\abs{\hat{x}_i - x_i}&\geq \epsilon_i \nonumber\\
    &\Longrightarrow& \text{either}~~~~ \hat{p}^{(i)} - p^{(i)} \geq \hat{p}^{(i)} - \mathcal{D}_i(\hat{x}_i -\epsilon_i)\nonumber\\
    & & \text{or}~~~~ \hat{p}^{(i)} - p^{(i)} \leq \hat{p}^{(i)} - \mathcal{D}_i(\hat{x}_i +\epsilon_i),
\end{eqnarray}
where the success probability is monotonically increasing over $x_i\in[1/2,1]$ following Lemma 1. Hence, the Hoeffding bound in terms of $x_i$ reads
\begin{eqnarray}
    &\Pr&\!\![\abs{\hat{x}_i - x_i}\geq \epsilon_i]\nonumber\\
    &\leq&  \sum_{m = \pm \epsilon_i} \exp[-2N^{(i)}(\hat{p}^{(i)} - \mathcal{D}_i(\hat{x}_i + m))^2].
\end{eqnarray}
Let $\hat{\rho}(\hat{\mathbf{q}})$ be our estimated state in Bell-diagonal form. Solving for the concentration bounds is similar to the analysis presented in Methods-\ref{M_c} that gave Eq.~\eqref{eqn:conc-bound-trace-distance-via-x}. Using this previously calculated Hoeffding bound results to the concentration bound for the trace distance:
\begin{eqnarray}
    &\Pr&\!\![D(\hat{\rho}(\hat{\mathbf{q}}),\overline{\rho}(\mathbf{q}))\geq \epsilon_T] \nonumber \\ 
    &\leq& 1-\prod_{i=1}^3\biggl(1-\!\!\!\sum_{m = \pm \epsilon_i}\!\!\exp[-2N^{(i)}(\hat{p}^{(i)} - \mathcal{D}_i(\hat{x}_i + m))^2] \biggr) \nonumber \\
    &=:& \delta(\hat{\mathbf{x}}, \{N^{(i)}\}, \{\epsilon_i\}). \label{eqn:methods_trdist_hoeffdingbound}
\end{eqnarray}
The above analysis applies even when estimating the Bell-diagonal elements of an arbitrary state. 


\bibliography{Bibliography.bib}

\begin{thebibliography}{76}%
\makeatletter
\providecommand \@ifxundefined [1]{%
 \@ifx{#1\undefined}
}%
\providecommand \@ifnum [1]{%
 \ifnum #1\expandafter \@firstoftwo
 \else \expandafter \@secondoftwo
 \fi
}%
\providecommand \@ifx [1]{%
 \ifx #1\expandafter \@firstoftwo
 \else \expandafter \@secondoftwo
 \fi
}%
\providecommand \natexlab [1]{#1}%
\providecommand \enquote  [1]{``#1''}%
\providecommand \bibnamefont  [1]{#1}%
\providecommand \bibfnamefont [1]{#1}%
\providecommand \citenamefont [1]{#1}%
\providecommand \href@noop [0]{\@secondoftwo}%
\providecommand \href [0]{\begingroup \@sanitize@url \@href}%
\providecommand \@href[1]{\@@startlink{#1}\@@href}%
\providecommand \@@href[1]{\endgroup#1\@@endlink}%
\providecommand \@sanitize@url [0]{\catcode `\\12\catcode `\$12\catcode
  `\&12\catcode `\#12\catcode `\^12\catcode `\_12\catcode `\%12\relax}%
\providecommand \@@startlink[1]{}%
\providecommand \@@endlink[0]{}%
\providecommand \url  [0]{\begingroup\@sanitize@url \@url }%
\providecommand \@url [1]{\endgroup\@href {#1}{\urlprefix }}%
\providecommand \urlprefix  [0]{URL }%
\providecommand \Eprint [0]{\href }%
\providecommand \doibase [0]{http://dx.doi.org/}%
\providecommand \selectlanguage [0]{\@gobble}%
\providecommand \bibinfo  [0]{\@secondoftwo}%
\providecommand \bibfield  [0]{\@secondoftwo}%
\providecommand \translation [1]{[#1]}%
\providecommand \BibitemOpen [0]{}%
\providecommand \bibitemStop [0]{}%
\providecommand \bibitemNoStop [0]{.\EOS\space}%
\providecommand \EOS [0]{\spacefactor3000\relax}%
\providecommand \BibitemShut  [1]{\csname bibitem#1\endcsname}%
\let\auto@bib@innerbib\@empty
\bibitem [{\citenamefont {Wehner}\ \emph {et~al.}(2018)\citenamefont {Wehner},
  \citenamefont {Elkouss},\ and\ \citenamefont {Hanson}}]{Wehner18}%
  \BibitemOpen
  \bibfield  {author} {\bibinfo {author} {\bibfnamefont {S.}~\bibnamefont
  {Wehner}}, \bibinfo {author} {\bibfnamefont {D.}~\bibnamefont {Elkouss}}, \
  and\ \bibinfo {author} {\bibfnamefont {R.}~\bibnamefont {Hanson}},\ }\href
  {\doibase 10.1126/science.aam9288} {\bibfield  {journal} {\bibinfo  {journal}
  {Science}\ }\textbf {\bibinfo {volume} {362}},\ \bibinfo {pages} {eaam9288}
  (\bibinfo {year} {2018})},\ \Eprint
  {http://arxiv.org/abs/https://www.science.org/doi/pdf/10.1126/science.aam9288}
  {https://www.science.org/doi/pdf/10.1126/science.aam9288} \BibitemShut
  {NoStop}%
\bibitem [{\citenamefont {{Van Meter}}\ \emph {et~al.}(2022)\citenamefont {{Van
  Meter}}, \citenamefont {Satoh}, \citenamefont {Benchasattabuse},
  \citenamefont {Teramoto}, \citenamefont {Matsuo}, \citenamefont {Hajdušek},
  \citenamefont {Satoh}, \citenamefont {Nagayama},\ and\ \citenamefont
  {Suzuki}}]{VanMeter22}%
  \BibitemOpen
  \bibfield  {author} {\bibinfo {author} {\bibfnamefont {R.}~\bibnamefont {{Van
  Meter}}}, \bibinfo {author} {\bibfnamefont {R.}~\bibnamefont {Satoh}},
  \bibinfo {author} {\bibfnamefont {N.}~\bibnamefont {Benchasattabuse}},
  \bibinfo {author} {\bibfnamefont {K.}~\bibnamefont {Teramoto}}, \bibinfo
  {author} {\bibfnamefont {T.}~\bibnamefont {Matsuo}}, \bibinfo {author}
  {\bibfnamefont {M.}~\bibnamefont {Hajdušek}}, \bibinfo {author}
  {\bibfnamefont {T.}~\bibnamefont {Satoh}}, \bibinfo {author} {\bibfnamefont
  {S.}~\bibnamefont {Nagayama}}, \ and\ \bibinfo {author} {\bibfnamefont
  {S.}~\bibnamefont {Suzuki}},\ }in\ \href {\doibase
  10.1109/QCE53715.2022.00055} {\emph {\bibinfo {booktitle} {2022 IEEE
  International Conference on Quantum Computing and Engineering (QCE)}}}\
  (\bibinfo {year} {2022})\ pp.\ \bibinfo {pages} {341--352}\BibitemShut
  {NoStop}%
\bibitem [{\citenamefont {Kozlowski}\ \emph {et~al.}(2023)\citenamefont
  {Kozlowski}, \citenamefont {Wehner}, \citenamefont {{Van Meter}},
  \citenamefont {Rijsman}, \citenamefont {Cacciapuoti}, \citenamefont
  {Caleffi},\ and\ \citenamefont {Nagayama}}]{rfc9340}%
  \BibitemOpen
  \bibfield  {author} {\bibinfo {author} {\bibfnamefont {W.}~\bibnamefont
  {Kozlowski}}, \bibinfo {author} {\bibfnamefont {S.}~\bibnamefont {Wehner}},
  \bibinfo {author} {\bibfnamefont {R.}~\bibnamefont {{Van Meter}}}, \bibinfo
  {author} {\bibfnamefont {B.}~\bibnamefont {Rijsman}}, \bibinfo {author}
  {\bibfnamefont {A.~S.}\ \bibnamefont {Cacciapuoti}}, \bibinfo {author}
  {\bibfnamefont {M.}~\bibnamefont {Caleffi}}, \ and\ \bibinfo {author}
  {\bibfnamefont {S.}~\bibnamefont {Nagayama}},\ }\href {\doibase
  10.17487/RFC9340} {\enquote {\bibinfo {title} {{Architectural Principles for
  a Quantum Internet}},}\ }\bibinfo {howpublished} {RFC 9340} (\bibinfo {year}
  {2023})\BibitemShut {NoStop}%
\bibitem [{\citenamefont {Bennett}\ and\ \citenamefont
  {Brassard}(2014)}]{Bennet84}%
  \BibitemOpen
  \bibfield  {author} {\bibinfo {author} {\bibfnamefont {C.~H.}\ \bibnamefont
  {Bennett}}\ and\ \bibinfo {author} {\bibfnamefont {G.}~\bibnamefont
  {Brassard}},\ }\href {\doibase https://doi.org/10.1016/j.tcs.2014.05.025}
  {\bibfield  {journal} {\bibinfo  {journal} {Theoretical Computer Science}\
  }\textbf {\bibinfo {volume} {560}},\ \bibinfo {pages} {7} (\bibinfo {year}
  {2014})},\ \bibinfo {note} {theoretical Aspects of Quantum Cryptography –
  celebrating 30 years of BB84}\BibitemShut {NoStop}%
\bibitem [{\citenamefont {Ekert}(1991)}]{Ekert91}%
  \BibitemOpen
  \bibfield  {author} {\bibinfo {author} {\bibfnamefont {A.~K.}\ \bibnamefont
  {Ekert}},\ }\href {\doibase 10.1103/PhysRevLett.67.661} {\bibfield  {journal}
  {\bibinfo  {journal} {Phys. Rev. Lett.}\ }\textbf {\bibinfo {volume} {67}},\
  \bibinfo {pages} {661} (\bibinfo {year} {1991})}\BibitemShut {NoStop}%
\bibitem [{\citenamefont {B\"{a}uml}\ \emph {et~al.}(2020)\citenamefont
  {B\"{a}uml}, \citenamefont {Azuma}, \citenamefont {Kato},\ and\ \citenamefont
  {Elkouss}}]{Bauml20}%
  \BibitemOpen
  \bibfield  {author} {\bibinfo {author} {\bibfnamefont {S.}~\bibnamefont
  {B\"{a}uml}}, \bibinfo {author} {\bibfnamefont {K.}~\bibnamefont {Azuma}},
  \bibinfo {author} {\bibfnamefont {G.}~\bibnamefont {Kato}}, \ and\ \bibinfo
  {author} {\bibfnamefont {D.}~\bibnamefont {Elkouss}},\ }\href {\doibase
  10.1038/s42005-020-0318-2} {\bibfield  {journal} {\bibinfo  {journal}
  {Communications Physics}\ }\textbf {\bibinfo {volume} {3}} (\bibinfo {year}
  {2020}),\ 10.1038/s42005-020-0318-2}\BibitemShut {NoStop}%
\bibitem [{\citenamefont {Gottesman}\ \emph {et~al.}(2012)\citenamefont
  {Gottesman}, \citenamefont {Jennewein},\ and\ \citenamefont
  {Croke}}]{Gottesman12}%
  \BibitemOpen
  \bibfield  {author} {\bibinfo {author} {\bibfnamefont {D.}~\bibnamefont
  {Gottesman}}, \bibinfo {author} {\bibfnamefont {T.}~\bibnamefont
  {Jennewein}}, \ and\ \bibinfo {author} {\bibfnamefont {S.}~\bibnamefont
  {Croke}},\ }\href {\doibase 10.1103/PhysRevLett.109.070503} {\bibfield
  {journal} {\bibinfo  {journal} {Phys. Rev. Lett.}\ }\textbf {\bibinfo
  {volume} {109}},\ \bibinfo {pages} {070503} (\bibinfo {year}
  {2012})}\BibitemShut {NoStop}%
\bibitem [{\citenamefont {K{\'{o}}m{\'{a}}r}\ \emph {et~al.}(2014)\citenamefont
  {K{\'{o}}m{\'{a}}r}, \citenamefont {Kessler}, \citenamefont {Bishof},
  \citenamefont {Jiang}, \citenamefont {S{\o}rensen}, \citenamefont {Ye},\ and\
  \citenamefont {Lukin}}]{Kmr14}%
  \BibitemOpen
  \bibfield  {author} {\bibinfo {author} {\bibfnamefont {P.}~\bibnamefont
  {K{\'{o}}m{\'{a}}r}}, \bibinfo {author} {\bibfnamefont {E.~M.}\ \bibnamefont
  {Kessler}}, \bibinfo {author} {\bibfnamefont {M.}~\bibnamefont {Bishof}},
  \bibinfo {author} {\bibfnamefont {L.}~\bibnamefont {Jiang}}, \bibinfo
  {author} {\bibfnamefont {A.~S.}\ \bibnamefont {S{\o}rensen}}, \bibinfo
  {author} {\bibfnamefont {J.}~\bibnamefont {Ye}}, \ and\ \bibinfo {author}
  {\bibfnamefont {M.~D.}\ \bibnamefont {Lukin}},\ }\href {\doibase
  10.1038/nphys3000} {\bibfield  {journal} {\bibinfo  {journal} {Nature
  Physics}\ }\textbf {\bibinfo {volume} {10}},\ \bibinfo {pages} {582}
  (\bibinfo {year} {2014})}\BibitemShut {NoStop}%
\bibitem [{\citenamefont {Ilo-Okeke}\ \emph {et~al.}(2018)\citenamefont
  {Ilo-Okeke}, \citenamefont {Tessler}, \citenamefont {Dowling},\ and\
  \citenamefont {Byrnes}}]{IloOkeke18}%
  \BibitemOpen
  \bibfield  {author} {\bibinfo {author} {\bibfnamefont {E.~O.}\ \bibnamefont
  {Ilo-Okeke}}, \bibinfo {author} {\bibfnamefont {L.}~\bibnamefont {Tessler}},
  \bibinfo {author} {\bibfnamefont {J.~P.}\ \bibnamefont {Dowling}}, \ and\
  \bibinfo {author} {\bibfnamefont {T.}~\bibnamefont {Byrnes}},\ }\href
  {\doibase 10.1038/s41534-018-0090-2} {\bibfield  {journal} {\bibinfo
  {journal} {npj Quantum Information}\ }\textbf {\bibinfo {volume} {4}}
  (\bibinfo {year} {2018}),\ 10.1038/s41534-018-0090-2}\BibitemShut {NoStop}%
\bibitem [{\citenamefont {Buhrman}\ and\ \citenamefont
  {R{\"o}hrig}(2003)}]{Buhrman03}%
  \BibitemOpen
  \bibfield  {author} {\bibinfo {author} {\bibfnamefont {H.}~\bibnamefont
  {Buhrman}}\ and\ \bibinfo {author} {\bibfnamefont {H.}~\bibnamefont
  {R{\"o}hrig}},\ }in\ \href@noop {} {\emph {\bibinfo {booktitle} {Mathematical
  Foundations of Computer Science 2003}}},\ \bibinfo {editor} {edited by\
  \bibinfo {editor} {\bibfnamefont {B.}~\bibnamefont {Rovan}}\ and\ \bibinfo
  {editor} {\bibfnamefont {P.}~\bibnamefont {Vojt{\'a}{\v{s}}}}}\ (\bibinfo
  {publisher} {Springer Berlin Heidelberg},\ \bibinfo {address} {Berlin,
  Heidelberg},\ \bibinfo {year} {2003})\ pp.\ \bibinfo {pages}
  {1--20}\BibitemShut {NoStop}%
\bibitem [{\citenamefont {Broadbent}\ \emph {et~al.}(2009)\citenamefont
  {Broadbent}, \citenamefont {Fitzsimons},\ and\ \citenamefont
  {Kashefi}}]{Broadbent09}%
  \BibitemOpen
  \bibfield  {author} {\bibinfo {author} {\bibfnamefont {A.}~\bibnamefont
  {Broadbent}}, \bibinfo {author} {\bibfnamefont {J.}~\bibnamefont
  {Fitzsimons}}, \ and\ \bibinfo {author} {\bibfnamefont {E.}~\bibnamefont
  {Kashefi}},\ }in\ \href {\doibase 10.1109/FOCS.2009.36} {\emph {\bibinfo
  {booktitle} {2009 50th Annual IEEE Symposium on Foundations of Computer
  Science}}}\ (\bibinfo {year} {2009})\ pp.\ \bibinfo {pages}
  {517--526}\BibitemShut {NoStop}%
\bibitem [{\citenamefont {Fitzsimons}(2017)}]{Fitzsimons17}%
  \BibitemOpen
  \bibfield  {author} {\bibinfo {author} {\bibfnamefont {J.~F.}\ \bibnamefont
  {Fitzsimons}},\ }\href {\doibase 10.1038/s41534-017-0025-3} {\bibfield
  {journal} {\bibinfo  {journal} {npj Quantum Information}\ }\textbf {\bibinfo
  {volume} {3}} (\bibinfo {year} {2017}),\
  10.1038/s41534-017-0025-3}\BibitemShut {NoStop}%
\bibitem [{\citenamefont {Cuomo}\ \emph {et~al.}(2020)\citenamefont {Cuomo},
  \citenamefont {Caleffi},\ and\ \citenamefont {Cacciapuoti}}]{Cuomo20}%
  \BibitemOpen
  \bibfield  {author} {\bibinfo {author} {\bibfnamefont {D.}~\bibnamefont
  {Cuomo}}, \bibinfo {author} {\bibfnamefont {M.}~\bibnamefont {Caleffi}}, \
  and\ \bibinfo {author} {\bibfnamefont {A.~S.}\ \bibnamefont {Cacciapuoti}},\
  }\href {\doibase 10.1049/iet-qtc.2020.0002} {\bibfield  {journal} {\bibinfo
  {journal} {{IET} Quantum Communication}\ }\textbf {\bibinfo {volume} {1}},\
  \bibinfo {pages} {3} (\bibinfo {year} {2020})}\BibitemShut {NoStop}%
\bibitem [{\citenamefont {Chakraborty}\ \emph {et~al.}(2020)\citenamefont
  {Chakraborty}, \citenamefont {Elkouss}, \citenamefont {Rijsman},\ and\
  \citenamefont {Wehner}}]{Chakraborty20}%
  \BibitemOpen
  \bibfield  {author} {\bibinfo {author} {\bibfnamefont {K.}~\bibnamefont
  {Chakraborty}}, \bibinfo {author} {\bibfnamefont {D.}~\bibnamefont
  {Elkouss}}, \bibinfo {author} {\bibfnamefont {B.}~\bibnamefont {Rijsman}}, \
  and\ \bibinfo {author} {\bibfnamefont {S.}~\bibnamefont {Wehner}},\ }\href
  {\doibase 10.1109/TQE.2020.3028172} {\bibfield  {journal} {\bibinfo
  {journal} {IEEE Transactions on Quantum Engineering}\ }\textbf {\bibinfo
  {volume} {1}},\ \bibinfo {pages} {1} (\bibinfo {year} {2020})}\BibitemShut
  {NoStop}%
\bibitem [{\citenamefont {Azuma}\ \emph {et~al.}(2021)\citenamefont {Azuma},
  \citenamefont {Bäuml}, \citenamefont {Coopmans}, \citenamefont {Elkouss},\
  and\ \citenamefont {Li}}]{Azuma21}%
  \BibitemOpen
  \bibfield  {author} {\bibinfo {author} {\bibfnamefont {K.}~\bibnamefont
  {Azuma}}, \bibinfo {author} {\bibfnamefont {S.}~\bibnamefont {Bäuml}},
  \bibinfo {author} {\bibfnamefont {T.}~\bibnamefont {Coopmans}}, \bibinfo
  {author} {\bibfnamefont {D.}~\bibnamefont {Elkouss}}, \ and\ \bibinfo
  {author} {\bibfnamefont {B.}~\bibnamefont {Li}},\ }\href {\doibase
  10.1116/5.0024062} {\bibfield  {journal} {\bibinfo  {journal} {AVS Quantum
  Science}\ }\textbf {\bibinfo {volume} {3}},\ \bibinfo {pages} {014101}
  (\bibinfo {year} {2021})},\ \Eprint
  {http://arxiv.org/abs/https://doi.org/10.1116/5.0024062}
  {https://doi.org/10.1116/5.0024062} \BibitemShut {NoStop}%
\bibitem [{\citenamefont {Coopmans}\ \emph {et~al.}(2021)\citenamefont
  {Coopmans}, \citenamefont {Knegjens}, \citenamefont {Dahlberg}, \citenamefont
  {Maier}, \citenamefont {Nijsten}, \citenamefont {de~Oliveira~Filho},
  \citenamefont {Papendrecht}, \citenamefont {Rabbie}, \citenamefont
  {Rozp{\k{e}}dek}, \citenamefont {Skrzypczyk}, \citenamefont {Wubben},
  \citenamefont {de~Jong}, \citenamefont {Podareanu}, \citenamefont
  {Torres-Knoop}, \citenamefont {Elkouss},\ and\ \citenamefont
  {Wehner}}]{Coopmans21}%
  \BibitemOpen
  \bibfield  {author} {\bibinfo {author} {\bibfnamefont {T.}~\bibnamefont
  {Coopmans}}, \bibinfo {author} {\bibfnamefont {R.}~\bibnamefont {Knegjens}},
  \bibinfo {author} {\bibfnamefont {A.}~\bibnamefont {Dahlberg}}, \bibinfo
  {author} {\bibfnamefont {D.}~\bibnamefont {Maier}}, \bibinfo {author}
  {\bibfnamefont {L.}~\bibnamefont {Nijsten}}, \bibinfo {author} {\bibfnamefont
  {J.}~\bibnamefont {de~Oliveira~Filho}}, \bibinfo {author} {\bibfnamefont
  {M.}~\bibnamefont {Papendrecht}}, \bibinfo {author} {\bibfnamefont
  {J.}~\bibnamefont {Rabbie}}, \bibinfo {author} {\bibfnamefont
  {F.}~\bibnamefont {Rozp{\k{e}}dek}}, \bibinfo {author} {\bibfnamefont
  {M.}~\bibnamefont {Skrzypczyk}}, \bibinfo {author} {\bibfnamefont
  {L.}~\bibnamefont {Wubben}}, \bibinfo {author} {\bibfnamefont
  {W.}~\bibnamefont {de~Jong}}, \bibinfo {author} {\bibfnamefont
  {D.}~\bibnamefont {Podareanu}}, \bibinfo {author} {\bibfnamefont
  {A.}~\bibnamefont {Torres-Knoop}}, \bibinfo {author} {\bibfnamefont
  {D.}~\bibnamefont {Elkouss}}, \ and\ \bibinfo {author} {\bibfnamefont
  {S.}~\bibnamefont {Wehner}},\ }\href {\doibase 10.1038/s42005-021-00647-8}
  {\bibfield  {journal} {\bibinfo  {journal} {Communications Physics}\ }\textbf
  {\bibinfo {volume} {4}} (\bibinfo {year} {2021}),\
  10.1038/s42005-021-00647-8}\BibitemShut {NoStop}%
\bibitem [{\citenamefont {Satoh}\ \emph {et~al.}(2022)\citenamefont {Satoh},
  \citenamefont {Hajdušek}, \citenamefont {Benchasattabuse}, \citenamefont
  {Nagayama}, \citenamefont {Teramoto}, \citenamefont {Matsuo}, \citenamefont
  {Metwalli}, \citenamefont {Pathumsoot}, \citenamefont {Satoh}, \citenamefont
  {Suzuki},\ and\ \citenamefont {Van~Meter}}]{Satoh22}%
  \BibitemOpen
  \bibfield  {author} {\bibinfo {author} {\bibfnamefont {R.}~\bibnamefont
  {Satoh}}, \bibinfo {author} {\bibfnamefont {M.}~\bibnamefont {Hajdušek}},
  \bibinfo {author} {\bibfnamefont {N.}~\bibnamefont {Benchasattabuse}},
  \bibinfo {author} {\bibfnamefont {S.}~\bibnamefont {Nagayama}}, \bibinfo
  {author} {\bibfnamefont {K.}~\bibnamefont {Teramoto}}, \bibinfo {author}
  {\bibfnamefont {T.}~\bibnamefont {Matsuo}}, \bibinfo {author} {\bibfnamefont
  {S.~A.}\ \bibnamefont {Metwalli}}, \bibinfo {author} {\bibfnamefont
  {P.}~\bibnamefont {Pathumsoot}}, \bibinfo {author} {\bibfnamefont
  {T.}~\bibnamefont {Satoh}}, \bibinfo {author} {\bibfnamefont
  {S.}~\bibnamefont {Suzuki}}, \ and\ \bibinfo {author} {\bibfnamefont
  {R.}~\bibnamefont {Van~Meter}},\ }in\ \href {\doibase
  10.1109/QCE53715.2022.00056} {\emph {\bibinfo {booktitle} {2022 IEEE
  International Conference on Quantum Computing and Engineering (QCE)}}}\
  (\bibinfo {year} {2022})\ pp.\ \bibinfo {pages} {353--364}\BibitemShut
  {NoStop}%
\bibitem [{\citenamefont {Dieks}(1982)}]{dieks1982communication}%
  \BibitemOpen
  \bibfield  {author} {\bibinfo {author} {\bibfnamefont {D.}~\bibnamefont
  {Dieks}},\ }\href
  {https://www.sciencedirect.com/science/article/abs/pii/0375960182900846}
  {\bibfield  {journal} {\bibinfo  {journal} {Physics Letters A}\ }\textbf
  {\bibinfo {volume} {92}},\ \bibinfo {pages} {271} (\bibinfo {year}
  {1982})}\BibitemShut {NoStop}%
\bibitem [{\citenamefont {Wootters}\ and\ \citenamefont
  {Zurek}(1982)}]{wootters1982single}%
  \BibitemOpen
  \bibfield  {author} {\bibinfo {author} {\bibfnamefont {W.~K.}\ \bibnamefont
  {Wootters}}\ and\ \bibinfo {author} {\bibfnamefont {W.~H.}\ \bibnamefont
  {Zurek}},\ }\href@noop {} {\bibfield  {journal} {\bibinfo  {journal}
  {Nature}\ }\textbf {\bibinfo {volume} {299}},\ \bibinfo {pages} {802}
  (\bibinfo {year} {1982})}\BibitemShut {NoStop}%
\bibitem [{\citenamefont {Azuma}\ \emph {et~al.}(2023)\citenamefont {Azuma},
  \citenamefont {Economou}, \citenamefont {Elkouss}, \citenamefont {Hilaire},
  \citenamefont {Jiang}, \citenamefont {Lo},\ and\ \citenamefont
  {Tzitrin}}]{azuma2023quantum}%
  \BibitemOpen
  \bibfield  {author} {\bibinfo {author} {\bibfnamefont {K.}~\bibnamefont
  {Azuma}}, \bibinfo {author} {\bibfnamefont {S.~E.}\ \bibnamefont {Economou}},
  \bibinfo {author} {\bibfnamefont {D.}~\bibnamefont {Elkouss}}, \bibinfo
  {author} {\bibfnamefont {P.}~\bibnamefont {Hilaire}}, \bibinfo {author}
  {\bibfnamefont {L.}~\bibnamefont {Jiang}}, \bibinfo {author} {\bibfnamefont
  {H.-K.}\ \bibnamefont {Lo}}, \ and\ \bibinfo {author} {\bibfnamefont
  {I.}~\bibnamefont {Tzitrin}},\ }\href {\doibase 10.1103/RevModPhys.95.045006}
  {\bibfield  {journal} {\bibinfo  {journal} {Reviews of Modern Physics}\
  }\textbf {\bibinfo {volume} {95}},\ \bibinfo {pages} {045006} (\bibinfo
  {year} {2023})}\BibitemShut {NoStop}%
\bibitem [{\citenamefont {Goodenough}\ \emph {et~al.}(2021)\citenamefont
  {Goodenough}, \citenamefont {Elkouss},\ and\ \citenamefont
  {Wehner}}]{Goodenough21}%
  \BibitemOpen
  \bibfield  {author} {\bibinfo {author} {\bibfnamefont {K.}~\bibnamefont
  {Goodenough}}, \bibinfo {author} {\bibfnamefont {D.}~\bibnamefont {Elkouss}},
  \ and\ \bibinfo {author} {\bibfnamefont {S.}~\bibnamefont {Wehner}},\ }\href
  {\doibase 10.1103/PhysRevA.103.032610} {\bibfield  {journal} {\bibinfo
  {journal} {Phys. Rev. A}\ }\textbf {\bibinfo {volume} {103}},\ \bibinfo
  {pages} {032610} (\bibinfo {year} {2021})}\BibitemShut {NoStop}%
\bibitem [{\citenamefont {Bennett}\ \emph
  {et~al.}(1996{\natexlab{a}})\citenamefont {Bennett}, \citenamefont
  {Brassard}, \citenamefont {Popescu}, \citenamefont {Schumacher},
  \citenamefont {Smolin},\ and\ \citenamefont {Wootters}}]{Bennett96(1)}%
  \BibitemOpen
  \bibfield  {author} {\bibinfo {author} {\bibfnamefont {C.~H.}\ \bibnamefont
  {Bennett}}, \bibinfo {author} {\bibfnamefont {G.}~\bibnamefont {Brassard}},
  \bibinfo {author} {\bibfnamefont {S.}~\bibnamefont {Popescu}}, \bibinfo
  {author} {\bibfnamefont {B.}~\bibnamefont {Schumacher}}, \bibinfo {author}
  {\bibfnamefont {J.~A.}\ \bibnamefont {Smolin}}, \ and\ \bibinfo {author}
  {\bibfnamefont {W.~K.}\ \bibnamefont {Wootters}},\ }\href {\doibase
  10.1103/PhysRevLett.76.722} {\bibfield  {journal} {\bibinfo  {journal} {Phys.
  Rev. Lett.}\ }\textbf {\bibinfo {volume} {76}},\ \bibinfo {pages} {722}
  (\bibinfo {year} {1996}{\natexlab{a}})}\BibitemShut {NoStop}%
\bibitem [{\citenamefont {Bennett}\ \emph
  {et~al.}(1996{\natexlab{b}})\citenamefont {Bennett}, \citenamefont
  {DiVincenzo}, \citenamefont {Smolin},\ and\ \citenamefont
  {Wootters}}]{Bennett96(2)}%
  \BibitemOpen
  \bibfield  {author} {\bibinfo {author} {\bibfnamefont {C.~H.}\ \bibnamefont
  {Bennett}}, \bibinfo {author} {\bibfnamefont {D.~P.}\ \bibnamefont
  {DiVincenzo}}, \bibinfo {author} {\bibfnamefont {J.~A.}\ \bibnamefont
  {Smolin}}, \ and\ \bibinfo {author} {\bibfnamefont {W.~K.}\ \bibnamefont
  {Wootters}},\ }\href {\doibase 10.1103/PhysRevA.54.3824} {\bibfield
  {journal} {\bibinfo  {journal} {Phys. Rev. A}\ }\textbf {\bibinfo {volume}
  {54}},\ \bibinfo {pages} {3824} (\bibinfo {year}
  {1996}{\natexlab{b}})}\BibitemShut {NoStop}%
\bibitem [{\citenamefont {Deutsch}\ \emph {et~al.}(1996)\citenamefont
  {Deutsch}, \citenamefont {Ekert}, \citenamefont {Jozsa}, \citenamefont
  {Macchiavello}, \citenamefont {Popescu},\ and\ \citenamefont
  {Sanpera}}]{Deutsch96}%
  \BibitemOpen
  \bibfield  {author} {\bibinfo {author} {\bibfnamefont {D.}~\bibnamefont
  {Deutsch}}, \bibinfo {author} {\bibfnamefont {A.}~\bibnamefont {Ekert}},
  \bibinfo {author} {\bibfnamefont {R.}~\bibnamefont {Jozsa}}, \bibinfo
  {author} {\bibfnamefont {C.}~\bibnamefont {Macchiavello}}, \bibinfo {author}
  {\bibfnamefont {S.}~\bibnamefont {Popescu}}, \ and\ \bibinfo {author}
  {\bibfnamefont {A.}~\bibnamefont {Sanpera}},\ }\href {\doibase
  10.1103/PhysRevLett.77.2818} {\bibfield  {journal} {\bibinfo  {journal}
  {Phys. Rev. Lett.}\ }\textbf {\bibinfo {volume} {77}},\ \bibinfo {pages}
  {2818} (\bibinfo {year} {1996})}\BibitemShut {NoStop}%
\bibitem [{\citenamefont {Yamamoto}\ \emph {et~al.}(2001)\citenamefont
  {Yamamoto}, \citenamefont {Koashi},\ and\ \citenamefont
  {Imoto}}]{Yamamoto01}%
  \BibitemOpen
  \bibfield  {author} {\bibinfo {author} {\bibfnamefont {T.}~\bibnamefont
  {Yamamoto}}, \bibinfo {author} {\bibfnamefont {M.}~\bibnamefont {Koashi}}, \
  and\ \bibinfo {author} {\bibfnamefont {N.}~\bibnamefont {Imoto}},\ }\href
  {\doibase 10.1103/PhysRevA.64.012304} {\bibfield  {journal} {\bibinfo
  {journal} {Phys. Rev. A}\ }\textbf {\bibinfo {volume} {64}},\ \bibinfo
  {pages} {012304} (\bibinfo {year} {2001})}\BibitemShut {NoStop}%
\bibitem [{\citenamefont {Pan}\ \emph {et~al.}(2001)\citenamefont {Pan},
  \citenamefont {Simon}, \citenamefont {Brukner},\ and\ \citenamefont
  {Zeilinger}}]{Pan01}%
  \BibitemOpen
  \bibfield  {author} {\bibinfo {author} {\bibfnamefont {J.-W.}\ \bibnamefont
  {Pan}}, \bibinfo {author} {\bibfnamefont {C.}~\bibnamefont {Simon}}, \bibinfo
  {author} {\bibfnamefont {{\v{C}}.}~\bibnamefont {Brukner}}, \ and\ \bibinfo
  {author} {\bibfnamefont {A.}~\bibnamefont {Zeilinger}},\ }\href {\doibase
  10.1038/35074041} {\bibfield  {journal} {\bibinfo  {journal} {Nature}\
  }\textbf {\bibinfo {volume} {410}},\ \bibinfo {pages} {1067} (\bibinfo {year}
  {2001})}\BibitemShut {NoStop}%
\bibitem [{\citenamefont {Nielsen}\ and\ \citenamefont
  {Chuang}(2010)}]{nielsen_chuang_2010}%
  \BibitemOpen
  \bibfield  {author} {\bibinfo {author} {\bibfnamefont {M.~A.}\ \bibnamefont
  {Nielsen}}\ and\ \bibinfo {author} {\bibfnamefont {I.~L.}\ \bibnamefont
  {Chuang}},\ }\href {\doibase 10.1017/CBO9780511976667} {\emph {\bibinfo
  {title} {Quantum Computation and Quantum Information: 10th Anniversary
  Edition}}}\ (\bibinfo  {publisher} {Cambridge University Press},\ \bibinfo
  {year} {2010})\BibitemShut {NoStop}%
\bibitem [{\citenamefont {Altepeter}\ \emph {et~al.}(2003)\citenamefont
  {Altepeter}, \citenamefont {Branning}, \citenamefont {Jeffrey}, \citenamefont
  {Wei}, \citenamefont {Kwiat}, \citenamefont {Thew}, \citenamefont {O'Brien},
  \citenamefont {Nielsen},\ and\ \citenamefont {White}}]{Altepeter03}%
  \BibitemOpen
  \bibfield  {author} {\bibinfo {author} {\bibfnamefont {J.~B.}\ \bibnamefont
  {Altepeter}}, \bibinfo {author} {\bibfnamefont {D.}~\bibnamefont {Branning}},
  \bibinfo {author} {\bibfnamefont {E.}~\bibnamefont {Jeffrey}}, \bibinfo
  {author} {\bibfnamefont {T.~C.}\ \bibnamefont {Wei}}, \bibinfo {author}
  {\bibfnamefont {P.~G.}\ \bibnamefont {Kwiat}}, \bibinfo {author}
  {\bibfnamefont {R.~T.}\ \bibnamefont {Thew}}, \bibinfo {author}
  {\bibfnamefont {J.~L.}\ \bibnamefont {O'Brien}}, \bibinfo {author}
  {\bibfnamefont {M.~A.}\ \bibnamefont {Nielsen}}, \ and\ \bibinfo {author}
  {\bibfnamefont {A.~G.}\ \bibnamefont {White}},\ }\href {\doibase
  10.1103/PhysRevLett.90.193601} {\bibfield  {journal} {\bibinfo  {journal}
  {Phys. Rev. Lett.}\ }\textbf {\bibinfo {volume} {90}},\ \bibinfo {pages}
  {193601} (\bibinfo {year} {2003})}\BibitemShut {NoStop}%
\bibitem [{\citenamefont {Poyatos}\ \emph {et~al.}(1997)\citenamefont
  {Poyatos}, \citenamefont {Cirac},\ and\ \citenamefont {Zoller}}]{Poyatos97}%
  \BibitemOpen
  \bibfield  {author} {\bibinfo {author} {\bibfnamefont {J.~F.}\ \bibnamefont
  {Poyatos}}, \bibinfo {author} {\bibfnamefont {J.~I.}\ \bibnamefont {Cirac}},
  \ and\ \bibinfo {author} {\bibfnamefont {P.}~\bibnamefont {Zoller}},\ }\href
  {\doibase 10.1103/PhysRevLett.78.390} {\bibfield  {journal} {\bibinfo
  {journal} {Phys. Rev. Lett.}\ }\textbf {\bibinfo {volume} {78}},\ \bibinfo
  {pages} {390} (\bibinfo {year} {1997})}\BibitemShut {NoStop}%
\bibitem [{\citenamefont {Chuang}\ and\ \citenamefont
  {Nielsen}(1997)}]{Chuang97}%
  \BibitemOpen
  \bibfield  {author} {\bibinfo {author} {\bibfnamefont {I.~L.}\ \bibnamefont
  {Chuang}}\ and\ \bibinfo {author} {\bibfnamefont {M.~A.}\ \bibnamefont
  {Nielsen}},\ }\href {\doibase 10.1080/09500349708231894} {\bibfield
  {journal} {\bibinfo  {journal} {Journal of Modern Optics}\ }\textbf {\bibinfo
  {volume} {44}},\ \bibinfo {pages} {2455} (\bibinfo {year}
  {1997})}\BibitemShut {NoStop}%
\bibitem [{\citenamefont {Fujiwara}(2001)}]{Fujiwara01}%
  \BibitemOpen
  \bibfield  {author} {\bibinfo {author} {\bibfnamefont {A.}~\bibnamefont
  {Fujiwara}},\ }\href {\doibase 10.1103/PhysRevA.63.042304} {\bibfield
  {journal} {\bibinfo  {journal} {Phys. Rev. A}\ }\textbf {\bibinfo {volume}
  {63}},\ \bibinfo {pages} {042304} (\bibinfo {year} {2001})}\BibitemShut
  {NoStop}%
\bibitem [{\citenamefont {O'Brien}\ \emph {et~al.}(2004)\citenamefont
  {O'Brien}, \citenamefont {Pryde}, \citenamefont {Gilchrist}, \citenamefont
  {James}, \citenamefont {Langford}, \citenamefont {Ralph},\ and\ \citenamefont
  {White}}]{OBrien04}%
  \BibitemOpen
  \bibfield  {author} {\bibinfo {author} {\bibfnamefont {J.~L.}\ \bibnamefont
  {O'Brien}}, \bibinfo {author} {\bibfnamefont {G.~J.}\ \bibnamefont {Pryde}},
  \bibinfo {author} {\bibfnamefont {A.}~\bibnamefont {Gilchrist}}, \bibinfo
  {author} {\bibfnamefont {D.~F.~V.}\ \bibnamefont {James}}, \bibinfo {author}
  {\bibfnamefont {N.~K.}\ \bibnamefont {Langford}}, \bibinfo {author}
  {\bibfnamefont {T.~C.}\ \bibnamefont {Ralph}}, \ and\ \bibinfo {author}
  {\bibfnamefont {A.~G.}\ \bibnamefont {White}},\ }\href {\doibase
  10.1103/PhysRevLett.93.080502} {\bibfield  {journal} {\bibinfo  {journal}
  {Phys. Rev. Lett.}\ }\textbf {\bibinfo {volume} {93}},\ \bibinfo {pages}
  {080502} (\bibinfo {year} {2004})}\BibitemShut {NoStop}%
\bibitem [{\citenamefont {Emerson}\ \emph {et~al.}(2005)\citenamefont
  {Emerson}, \citenamefont {Alicki},\ and\ \citenamefont
  {Życzkowski}}]{Emerson05}%
  \BibitemOpen
  \bibfield  {author} {\bibinfo {author} {\bibfnamefont {J.}~\bibnamefont
  {Emerson}}, \bibinfo {author} {\bibfnamefont {R.}~\bibnamefont {Alicki}}, \
  and\ \bibinfo {author} {\bibfnamefont {K.}~\bibnamefont {Życzkowski}},\
  }\href {\doibase 10.1088/1464-4266/7/10/021} {\bibfield  {journal} {\bibinfo
  {journal} {Journal of Optics B: Quantum and Semiclassical Optics}\ }\textbf
  {\bibinfo {volume} {7}},\ \bibinfo {pages} {S347} (\bibinfo {year}
  {2005})}\BibitemShut {NoStop}%
\bibitem [{\citenamefont {Knill}\ \emph {et~al.}(2008)\citenamefont {Knill},
  \citenamefont {Leibfried}, \citenamefont {Reichle}, \citenamefont {Britton},
  \citenamefont {Blakestad}, \citenamefont {Jost}, \citenamefont {Langer},
  \citenamefont {Ozeri}, \citenamefont {Seidelin},\ and\ \citenamefont
  {Wineland}}]{Knill08}%
  \BibitemOpen
  \bibfield  {author} {\bibinfo {author} {\bibfnamefont {E.}~\bibnamefont
  {Knill}}, \bibinfo {author} {\bibfnamefont {D.}~\bibnamefont {Leibfried}},
  \bibinfo {author} {\bibfnamefont {R.}~\bibnamefont {Reichle}}, \bibinfo
  {author} {\bibfnamefont {J.}~\bibnamefont {Britton}}, \bibinfo {author}
  {\bibfnamefont {R.~B.}\ \bibnamefont {Blakestad}}, \bibinfo {author}
  {\bibfnamefont {J.~D.}\ \bibnamefont {Jost}}, \bibinfo {author}
  {\bibfnamefont {C.}~\bibnamefont {Langer}}, \bibinfo {author} {\bibfnamefont
  {R.}~\bibnamefont {Ozeri}}, \bibinfo {author} {\bibfnamefont
  {S.}~\bibnamefont {Seidelin}}, \ and\ \bibinfo {author} {\bibfnamefont
  {D.~J.}\ \bibnamefont {Wineland}},\ }\href {\doibase
  10.1103/PhysRevA.77.012307} {\bibfield  {journal} {\bibinfo  {journal} {Phys.
  Rev. A}\ }\textbf {\bibinfo {volume} {77}},\ \bibinfo {pages} {012307}
  (\bibinfo {year} {2008})}\BibitemShut {NoStop}%
\bibitem [{\citenamefont {Dankert}\ \emph {et~al.}(2009)\citenamefont
  {Dankert}, \citenamefont {Cleve}, \citenamefont {Emerson},\ and\
  \citenamefont {Livine}}]{Dankert09}%
  \BibitemOpen
  \bibfield  {author} {\bibinfo {author} {\bibfnamefont {C.}~\bibnamefont
  {Dankert}}, \bibinfo {author} {\bibfnamefont {R.}~\bibnamefont {Cleve}},
  \bibinfo {author} {\bibfnamefont {J.}~\bibnamefont {Emerson}}, \ and\
  \bibinfo {author} {\bibfnamefont {E.}~\bibnamefont {Livine}},\ }\href
  {\doibase 10.1103/PhysRevA.80.012304} {\bibfield  {journal} {\bibinfo
  {journal} {Phys. Rev. A}\ }\textbf {\bibinfo {volume} {80}},\ \bibinfo
  {pages} {012304} (\bibinfo {year} {2009})}\BibitemShut {NoStop}%
\bibitem [{\citenamefont {Magesan}\ \emph {et~al.}(2011)\citenamefont
  {Magesan}, \citenamefont {Gambetta},\ and\ \citenamefont
  {Emerson}}]{Magesan11}%
  \BibitemOpen
  \bibfield  {author} {\bibinfo {author} {\bibfnamefont {E.}~\bibnamefont
  {Magesan}}, \bibinfo {author} {\bibfnamefont {J.~M.}\ \bibnamefont
  {Gambetta}}, \ and\ \bibinfo {author} {\bibfnamefont {J.}~\bibnamefont
  {Emerson}},\ }\href {\doibase 10.1103/PhysRevLett.106.180504} {\bibfield
  {journal} {\bibinfo  {journal} {Phys. Rev. Lett.}\ }\textbf {\bibinfo
  {volume} {106}},\ \bibinfo {pages} {180504} (\bibinfo {year}
  {2011})}\BibitemShut {NoStop}%
\bibitem [{\citenamefont {Magesan}\ \emph {et~al.}(2012)\citenamefont
  {Magesan}, \citenamefont {Gambetta},\ and\ \citenamefont
  {Emerson}}]{Magesan12}%
  \BibitemOpen
  \bibfield  {author} {\bibinfo {author} {\bibfnamefont {E.}~\bibnamefont
  {Magesan}}, \bibinfo {author} {\bibfnamefont {J.~M.}\ \bibnamefont
  {Gambetta}}, \ and\ \bibinfo {author} {\bibfnamefont {J.}~\bibnamefont
  {Emerson}},\ }\href {\doibase 10.1103/PhysRevA.85.042311} {\bibfield
  {journal} {\bibinfo  {journal} {Phys. Rev. A}\ }\textbf {\bibinfo {volume}
  {85}},\ \bibinfo {pages} {042311} (\bibinfo {year} {2012})}\BibitemShut
  {NoStop}%
\bibitem [{\citenamefont {Erhard}\ \emph {et~al.}(2019)\citenamefont {Erhard},
  \citenamefont {Wallman}, \citenamefont {Postler}, \citenamefont {Meth},
  \citenamefont {Stricker}, \citenamefont {Martinez}, \citenamefont
  {Schindler}, \citenamefont {Monz}, \citenamefont {Emerson},\ and\
  \citenamefont {Blatt}}]{Erhard19}%
  \BibitemOpen
  \bibfield  {author} {\bibinfo {author} {\bibfnamefont {A.}~\bibnamefont
  {Erhard}}, \bibinfo {author} {\bibfnamefont {J.~J.}\ \bibnamefont {Wallman}},
  \bibinfo {author} {\bibfnamefont {L.}~\bibnamefont {Postler}}, \bibinfo
  {author} {\bibfnamefont {M.}~\bibnamefont {Meth}}, \bibinfo {author}
  {\bibfnamefont {R.}~\bibnamefont {Stricker}}, \bibinfo {author}
  {\bibfnamefont {E.~A.}\ \bibnamefont {Martinez}}, \bibinfo {author}
  {\bibfnamefont {P.}~\bibnamefont {Schindler}}, \bibinfo {author}
  {\bibfnamefont {T.}~\bibnamefont {Monz}}, \bibinfo {author} {\bibfnamefont
  {J.}~\bibnamefont {Emerson}}, \ and\ \bibinfo {author} {\bibfnamefont
  {R.}~\bibnamefont {Blatt}},\ }\href {\doibase 10.1038/s41467-019-13068-7}
  {\bibfield  {journal} {\bibinfo  {journal} {Nature Communications}\ }\textbf
  {\bibinfo {volume} {10}} (\bibinfo {year} {2019}),\
  10.1038/s41467-019-13068-7}\BibitemShut {NoStop}%
\bibitem [{\citenamefont {Helsen}\ \emph {et~al.}(2022)\citenamefont {Helsen},
  \citenamefont {Roth}, \citenamefont {Onorati}, \citenamefont {Werner},\ and\
  \citenamefont {Eisert}}]{Helsen22}%
  \BibitemOpen
  \bibfield  {author} {\bibinfo {author} {\bibfnamefont {J.}~\bibnamefont
  {Helsen}}, \bibinfo {author} {\bibfnamefont {I.}~\bibnamefont {Roth}},
  \bibinfo {author} {\bibfnamefont {E.}~\bibnamefont {Onorati}}, \bibinfo
  {author} {\bibfnamefont {A.}~\bibnamefont {Werner}}, \ and\ \bibinfo {author}
  {\bibfnamefont {J.}~\bibnamefont {Eisert}},\ }\href {\doibase
  10.1103/PRXQuantum.3.020357} {\bibfield  {journal} {\bibinfo  {journal} {PRX
  Quantum}\ }\textbf {\bibinfo {volume} {3}},\ \bibinfo {pages} {020357}
  (\bibinfo {year} {2022})}\BibitemShut {NoStop}%
\bibitem [{\citenamefont {{\v{S}}upi{\'{c}}}\ and\ \citenamefont
  {Bowles}(2020)}]{Supic2020}%
  \BibitemOpen
  \bibfield  {author} {\bibinfo {author} {\bibfnamefont {I.}~\bibnamefont
  {{\v{S}}upi{\'{c}}}}\ and\ \bibinfo {author} {\bibfnamefont {J.}~\bibnamefont
  {Bowles}},\ }\href {\doibase 10.22331/q-2020-09-30-337} {\bibfield  {journal}
  {\bibinfo  {journal} {{Quantum}}\ }\textbf {\bibinfo {volume} {4}},\ \bibinfo
  {pages} {337} (\bibinfo {year} {2020})}\BibitemShut {NoStop}%
\bibitem [{\citenamefont {Merkel}\ \emph {et~al.}(2013)\citenamefont {Merkel},
  \citenamefont {Gambetta}, \citenamefont {Smolin}, \citenamefont {Poletto},
  \citenamefont {C\'orcoles}, \citenamefont {Johnson}, \citenamefont {Ryan},\
  and\ \citenamefont {Steffen}}]{Merkel13}%
  \BibitemOpen
  \bibfield  {author} {\bibinfo {author} {\bibfnamefont {S.~T.}\ \bibnamefont
  {Merkel}}, \bibinfo {author} {\bibfnamefont {J.~M.}\ \bibnamefont
  {Gambetta}}, \bibinfo {author} {\bibfnamefont {J.~A.}\ \bibnamefont
  {Smolin}}, \bibinfo {author} {\bibfnamefont {S.}~\bibnamefont {Poletto}},
  \bibinfo {author} {\bibfnamefont {A.~D.}\ \bibnamefont {C\'orcoles}},
  \bibinfo {author} {\bibfnamefont {B.~R.}\ \bibnamefont {Johnson}}, \bibinfo
  {author} {\bibfnamefont {C.~A.}\ \bibnamefont {Ryan}}, \ and\ \bibinfo
  {author} {\bibfnamefont {M.}~\bibnamefont {Steffen}},\ }\href {\doibase
  10.1103/PhysRevA.87.062119} {\bibfield  {journal} {\bibinfo  {journal} {Phys.
  Rev. A}\ }\textbf {\bibinfo {volume} {87}},\ \bibinfo {pages} {062119}
  (\bibinfo {year} {2013})}\BibitemShut {NoStop}%
\bibitem [{\citenamefont {Nielsen}\ \emph {et~al.}(2021)\citenamefont
  {Nielsen}, \citenamefont {Gamble}, \citenamefont {Rudinger}, \citenamefont
  {Scholten}, \citenamefont {Young},\ and\ \citenamefont
  {Blume-Kohout}}]{Nielsen21}%
  \BibitemOpen
  \bibfield  {author} {\bibinfo {author} {\bibfnamefont {E.}~\bibnamefont
  {Nielsen}}, \bibinfo {author} {\bibfnamefont {J.~K.}\ \bibnamefont {Gamble}},
  \bibinfo {author} {\bibfnamefont {K.}~\bibnamefont {Rudinger}}, \bibinfo
  {author} {\bibfnamefont {T.}~\bibnamefont {Scholten}}, \bibinfo {author}
  {\bibfnamefont {K.}~\bibnamefont {Young}}, \ and\ \bibinfo {author}
  {\bibfnamefont {R.}~\bibnamefont {Blume-Kohout}},\ }\href {\doibase
  10.22331/q-2021-10-05-557} {\bibfield  {journal} {\bibinfo  {journal}
  {{Quantum}}\ }\textbf {\bibinfo {volume} {5}},\ \bibinfo {pages} {557}
  (\bibinfo {year} {2021})}\BibitemShut {NoStop}%
\bibitem [{\citenamefont {Yuan}\ and\ \citenamefont {Fung}(2017)}]{Yuan2017}%
  \BibitemOpen
  \bibfield  {author} {\bibinfo {author} {\bibfnamefont {H.}~\bibnamefont
  {Yuan}}\ and\ \bibinfo {author} {\bibfnamefont {C.-H.~F.}\ \bibnamefont
  {Fung}},\ }\href {\doibase 10.1038/s41534-017-0014-6} {\bibfield  {journal}
  {\bibinfo  {journal} {npj Quantum Information}\ }\textbf {\bibinfo {volume}
  {3}} (\bibinfo {year} {2017}),\ 10.1038/s41534-017-0014-6}\BibitemShut
  {NoStop}%
\bibitem [{\citenamefont {Qi}\ \emph {et~al.}(2017)\citenamefont {Qi},
  \citenamefont {Hou}, \citenamefont {Wang}, \citenamefont {Dong},
  \citenamefont {Zhong}, \citenamefont {Li}, \citenamefont {Xiang},
  \citenamefont {Wiseman}, \citenamefont {Li},\ and\ \citenamefont
  {Guo}}]{Qi2017}%
  \BibitemOpen
  \bibfield  {author} {\bibinfo {author} {\bibfnamefont {B.}~\bibnamefont
  {Qi}}, \bibinfo {author} {\bibfnamefont {Z.}~\bibnamefont {Hou}}, \bibinfo
  {author} {\bibfnamefont {Y.}~\bibnamefont {Wang}}, \bibinfo {author}
  {\bibfnamefont {D.}~\bibnamefont {Dong}}, \bibinfo {author} {\bibfnamefont
  {H.-S.}\ \bibnamefont {Zhong}}, \bibinfo {author} {\bibfnamefont
  {L.}~\bibnamefont {Li}}, \bibinfo {author} {\bibfnamefont {G.-Y.}\
  \bibnamefont {Xiang}}, \bibinfo {author} {\bibfnamefont {H.~M.}\ \bibnamefont
  {Wiseman}}, \bibinfo {author} {\bibfnamefont {C.-F.}\ \bibnamefont {Li}}, \
  and\ \bibinfo {author} {\bibfnamefont {G.-C.}\ \bibnamefont {Guo}},\ }\href
  {\doibase 10.1038/s41534-017-0016-4} {\bibfield  {journal} {\bibinfo
  {journal} {npj Quantum Information}\ }\textbf {\bibinfo {volume} {3}}
  (\bibinfo {year} {2017}),\ 10.1038/s41534-017-0016-4}\BibitemShut {NoStop}%
\bibitem [{\citenamefont {Yu}\ \emph {et~al.}(2022)\citenamefont {Yu},
  \citenamefont {Liu}, \citenamefont {Yang}, \citenamefont {Gong},
  \citenamefont {Cao}, \citenamefont {Zhang}, \citenamefont {Liu},
  \citenamefont {Heyl}, \citenamefont {Ozawa}, \citenamefont {Goldman},\ and\
  \citenamefont {Cai}}]{Yu2022}%
  \BibitemOpen
  \bibfield  {author} {\bibinfo {author} {\bibfnamefont {M.}~\bibnamefont
  {Yu}}, \bibinfo {author} {\bibfnamefont {Y.}~\bibnamefont {Liu}}, \bibinfo
  {author} {\bibfnamefont {P.}~\bibnamefont {Yang}}, \bibinfo {author}
  {\bibfnamefont {M.}~\bibnamefont {Gong}}, \bibinfo {author} {\bibfnamefont
  {Q.}~\bibnamefont {Cao}}, \bibinfo {author} {\bibfnamefont {S.}~\bibnamefont
  {Zhang}}, \bibinfo {author} {\bibfnamefont {H.}~\bibnamefont {Liu}}, \bibinfo
  {author} {\bibfnamefont {M.}~\bibnamefont {Heyl}}, \bibinfo {author}
  {\bibfnamefont {T.}~\bibnamefont {Ozawa}}, \bibinfo {author} {\bibfnamefont
  {N.}~\bibnamefont {Goldman}}, \ and\ \bibinfo {author} {\bibfnamefont
  {J.}~\bibnamefont {Cai}},\ }\href {\doibase 10.1038/s41534-022-00547-x}
  {\bibfield  {journal} {\bibinfo  {journal} {npj Quantum Information}\
  }\textbf {\bibinfo {volume} {8}} (\bibinfo {year} {2022}),\
  10.1038/s41534-022-00547-x}\BibitemShut {NoStop}%
\bibitem [{\citenamefont {Xu}\ \emph {et~al.}(2019)\citenamefont {Xu},
  \citenamefont {Li}, \citenamefont {Liu}, \citenamefont {Wang}, \citenamefont
  {Yuan},\ and\ \citenamefont {Wang}}]{Xu2019}%
  \BibitemOpen
  \bibfield  {author} {\bibinfo {author} {\bibfnamefont {H.}~\bibnamefont
  {Xu}}, \bibinfo {author} {\bibfnamefont {J.}~\bibnamefont {Li}}, \bibinfo
  {author} {\bibfnamefont {L.}~\bibnamefont {Liu}}, \bibinfo {author}
  {\bibfnamefont {Y.}~\bibnamefont {Wang}}, \bibinfo {author} {\bibfnamefont
  {H.}~\bibnamefont {Yuan}}, \ and\ \bibinfo {author} {\bibfnamefont
  {X.}~\bibnamefont {Wang}},\ }\href {\doibase 10.1038/s41534-019-0198-z}
  {\bibfield  {journal} {\bibinfo  {journal} {npj Quantum Information}\
  }\textbf {\bibinfo {volume} {5}} (\bibinfo {year} {2019}),\
  10.1038/s41534-019-0198-z}\BibitemShut {NoStop}%
\bibitem [{\citenamefont {Meyer}\ \emph {et~al.}(2021)\citenamefont {Meyer},
  \citenamefont {Borregaard},\ and\ \citenamefont {Eisert}}]{Meyer2021}%
  \BibitemOpen
  \bibfield  {author} {\bibinfo {author} {\bibfnamefont {J.~J.}\ \bibnamefont
  {Meyer}}, \bibinfo {author} {\bibfnamefont {J.}~\bibnamefont {Borregaard}}, \
  and\ \bibinfo {author} {\bibfnamefont {J.}~\bibnamefont {Eisert}},\ }\href
  {\doibase 10.1038/s41534-021-00425-y} {\bibfield  {journal} {\bibinfo
  {journal} {npj Quantum Information}\ }\textbf {\bibinfo {volume} {7}}
  (\bibinfo {year} {2021}),\ 10.1038/s41534-021-00425-y}\BibitemShut {NoStop}%
\bibitem [{\citenamefont {Pereira}\ \emph {et~al.}(2022)\citenamefont
  {Pereira}, \citenamefont {Zambrano},\ and\ \citenamefont
  {Delgado}}]{Pereira2022}%
  \BibitemOpen
  \bibfield  {author} {\bibinfo {author} {\bibfnamefont {L.}~\bibnamefont
  {Pereira}}, \bibinfo {author} {\bibfnamefont {L.}~\bibnamefont {Zambrano}}, \
  and\ \bibinfo {author} {\bibfnamefont {A.}~\bibnamefont {Delgado}},\ }\href
  {\doibase 10.1038/s41534-022-00565-9} {\bibfield  {journal} {\bibinfo
  {journal} {npj Quantum Information}\ }\textbf {\bibinfo {volume} {8}}
  (\bibinfo {year} {2022}),\ 10.1038/s41534-022-00565-9}\BibitemShut {NoStop}%
\bibitem [{\citenamefont {Xiao}\ \emph {et~al.}(2022)\citenamefont {Xiao},
  \citenamefont {Fan},\ and\ \citenamefont {Zeng}}]{Xiao2022}%
  \BibitemOpen
  \bibfield  {author} {\bibinfo {author} {\bibfnamefont {T.}~\bibnamefont
  {Xiao}}, \bibinfo {author} {\bibfnamefont {J.}~\bibnamefont {Fan}}, \ and\
  \bibinfo {author} {\bibfnamefont {G.}~\bibnamefont {Zeng}},\ }\href {\doibase
  10.1038/s41534-021-00513-z} {\bibfield  {journal} {\bibinfo  {journal} {npj
  Quantum Information}\ }\textbf {\bibinfo {volume} {8}} (\bibinfo {year}
  {2022}),\ 10.1038/s41534-021-00513-z}\BibitemShut {NoStop}%
\bibitem [{\citenamefont {Nolan}\ \emph {et~al.}(2021)\citenamefont {Nolan},
  \citenamefont {Smerzi},\ and\ \citenamefont {Pezz{\`{e}}}}]{Nolan2021}%
  \BibitemOpen
  \bibfield  {author} {\bibinfo {author} {\bibfnamefont {S.}~\bibnamefont
  {Nolan}}, \bibinfo {author} {\bibfnamefont {A.}~\bibnamefont {Smerzi}}, \
  and\ \bibinfo {author} {\bibfnamefont {L.}~\bibnamefont {Pezz{\`{e}}}},\
  }\href {\doibase 10.1038/s41534-021-00497-w} {\bibfield  {journal} {\bibinfo
  {journal} {npj Quantum Information}\ }\textbf {\bibinfo {volume} {7}}
  (\bibinfo {year} {2021}),\ 10.1038/s41534-021-00497-w}\BibitemShut {NoStop}%
\bibitem [{\citenamefont {Conlon}\ \emph {et~al.}(2021)\citenamefont {Conlon},
  \citenamefont {Suzuki}, \citenamefont {Lam},\ and\ \citenamefont
  {Assad}}]{Conlon2021}%
  \BibitemOpen
  \bibfield  {author} {\bibinfo {author} {\bibfnamefont {L.~O.}\ \bibnamefont
  {Conlon}}, \bibinfo {author} {\bibfnamefont {J.}~\bibnamefont {Suzuki}},
  \bibinfo {author} {\bibfnamefont {P.~K.}\ \bibnamefont {Lam}}, \ and\
  \bibinfo {author} {\bibfnamefont {S.~M.}\ \bibnamefont {Assad}},\ }\href
  {\doibase 10.1038/s41534-021-00414-1} {\bibfield  {journal} {\bibinfo
  {journal} {npj Quantum Information}\ }\textbf {\bibinfo {volume} {7}}
  (\bibinfo {year} {2021}),\ 10.1038/s41534-021-00414-1}\BibitemShut {NoStop}%
\bibitem [{\citenamefont {Hou}\ \emph {et~al.}(2016)\citenamefont {Hou},
  \citenamefont {Zhu}, \citenamefont {Xiang}, \citenamefont {Li},\ and\
  \citenamefont {Guo}}]{Hou2016}%
  \BibitemOpen
  \bibfield  {author} {\bibinfo {author} {\bibfnamefont {Z.}~\bibnamefont
  {Hou}}, \bibinfo {author} {\bibfnamefont {H.}~\bibnamefont {Zhu}}, \bibinfo
  {author} {\bibfnamefont {G.-Y.}\ \bibnamefont {Xiang}}, \bibinfo {author}
  {\bibfnamefont {C.-F.}\ \bibnamefont {Li}}, \ and\ \bibinfo {author}
  {\bibfnamefont {G.-C.}\ \bibnamefont {Guo}},\ }\href {\doibase
  10.1038/npjqi.2016.1} {\bibfield  {journal} {\bibinfo  {journal} {npj Quantum
  Information}\ }\textbf {\bibinfo {volume} {2}} (\bibinfo {year} {2016}),\
  10.1038/npjqi.2016.1}\BibitemShut {NoStop}%
\bibitem [{\citenamefont {Grinko}\ \emph {et~al.}(2021)\citenamefont {Grinko},
  \citenamefont {Gacon}, \citenamefont {Zoufal},\ and\ \citenamefont
  {Woerner}}]{Grinko2021}%
  \BibitemOpen
  \bibfield  {author} {\bibinfo {author} {\bibfnamefont {D.}~\bibnamefont
  {Grinko}}, \bibinfo {author} {\bibfnamefont {J.}~\bibnamefont {Gacon}},
  \bibinfo {author} {\bibfnamefont {C.}~\bibnamefont {Zoufal}}, \ and\ \bibinfo
  {author} {\bibfnamefont {S.}~\bibnamefont {Woerner}},\ }\href {\doibase
  10.1038/s41534-021-00379-1} {\bibfield  {journal} {\bibinfo  {journal} {npj
  Quantum Information}\ }\textbf {\bibinfo {volume} {7}} (\bibinfo {year}
  {2021}),\ 10.1038/s41534-021-00379-1}\BibitemShut {NoStop}%
\bibitem [{\citenamefont {Helsen}\ and\ \citenamefont
  {Wehner}(2023)}]{Helsen2023}%
  \BibitemOpen
  \bibfield  {author} {\bibinfo {author} {\bibfnamefont {J.}~\bibnamefont
  {Helsen}}\ and\ \bibinfo {author} {\bibfnamefont {S.}~\bibnamefont
  {Wehner}},\ }\href {\doibase 10.1038/s41534-022-00628-x} {\bibfield
  {journal} {\bibinfo  {journal} {npj Quantum Information}\ }\textbf {\bibinfo
  {volume} {9}} (\bibinfo {year} {2023}),\
  10.1038/s41534-022-00628-x}\BibitemShut {NoStop}%
\bibitem [{\citenamefont {De~Andrade}\ \emph {et~al.}(2022)\citenamefont
  {De~Andrade}, \citenamefont {Diaz}, \citenamefont {Navas}, \citenamefont
  {Guha}, \citenamefont {Montaño}, \citenamefont {Smith}, \citenamefont
  {Raymer},\ and\ \citenamefont {Towsley}}]{Andrade22}%
  \BibitemOpen
  \bibfield  {author} {\bibinfo {author} {\bibfnamefont {M.~G.}\ \bibnamefont
  {De~Andrade}}, \bibinfo {author} {\bibfnamefont {J.}~\bibnamefont {Diaz}},
  \bibinfo {author} {\bibfnamefont {J.}~\bibnamefont {Navas}}, \bibinfo
  {author} {\bibfnamefont {S.}~\bibnamefont {Guha}}, \bibinfo {author}
  {\bibfnamefont {I.}~\bibnamefont {Montaño}}, \bibinfo {author}
  {\bibfnamefont {B.}~\bibnamefont {Smith}}, \bibinfo {author} {\bibfnamefont
  {M.}~\bibnamefont {Raymer}}, \ and\ \bibinfo {author} {\bibfnamefont
  {D.}~\bibnamefont {Towsley}},\ }in\ \href {\doibase
  10.1109/QCE53715.2022.00061} {\emph {\bibinfo {booktitle} {2022 IEEE
  International Conference on Quantum Computing and Engineering (QCE)}}}\
  (\bibinfo {year} {2022})\ pp.\ \bibinfo {pages} {400--409}\BibitemShut
  {NoStop}%
\bibitem [{\citenamefont {De~Andrade}\ \emph {et~al.}(2023)\citenamefont
  {De~Andrade}, \citenamefont {Navas}, \citenamefont {Montaño},\ and\
  \citenamefont {Towsley}}]{Andrade23}%
  \BibitemOpen
  \bibfield  {author} {\bibinfo {author} {\bibfnamefont {M.~G.}\ \bibnamefont
  {De~Andrade}}, \bibinfo {author} {\bibfnamefont {J.}~\bibnamefont {Navas}},
  \bibinfo {author} {\bibfnamefont {I.}~\bibnamefont {Montaño}}, \ and\
  \bibinfo {author} {\bibfnamefont {D.}~\bibnamefont {Towsley}},\ }in\ \href
  {\doibase 10.1109/QCE57702.2023.00142} {\emph {\bibinfo {booktitle} {2023
  IEEE International Conference on Quantum Computing and Engineering (QCE)}}},\
  Vol.~\bibinfo {volume} {01}\ (\bibinfo {year} {2023})\ pp.\ \bibinfo {pages}
  {1260--1270}\BibitemShut {NoStop}%
\bibitem [{\citenamefont {De~Andrade}\ \emph {et~al.}(2024)\citenamefont
  {De~Andrade}, \citenamefont {Navas}, \citenamefont {Guha}, \citenamefont
  {Montaño}, \citenamefont {Raymer}, \citenamefont {Smith},\ and\
  \citenamefont {Towsley}}]{Andrade24}%
  \BibitemOpen
  \bibfield  {author} {\bibinfo {author} {\bibfnamefont {M.~G.}\ \bibnamefont
  {De~Andrade}}, \bibinfo {author} {\bibfnamefont {J.}~\bibnamefont {Navas}},
  \bibinfo {author} {\bibfnamefont {S.}~\bibnamefont {Guha}}, \bibinfo {author}
  {\bibfnamefont {I.}~\bibnamefont {Montaño}}, \bibinfo {author}
  {\bibfnamefont {M.}~\bibnamefont {Raymer}}, \bibinfo {author} {\bibfnamefont
  {B.}~\bibnamefont {Smith}}, \ and\ \bibinfo {author} {\bibfnamefont
  {D.}~\bibnamefont {Towsley}},\ }\href {\doibase 10.1109/MNET.2024.3403805}
  {\bibfield  {journal} {\bibinfo  {journal} {IEEE Network}\ ,\ \bibinfo
  {pages} {1}} (\bibinfo {year} {2024})}\BibitemShut {NoStop}%
\bibitem [{\citenamefont {Eisert}\ \emph {et~al.}(2020)\citenamefont {Eisert},
  \citenamefont {Hangleiter}, \citenamefont {Walk}, \citenamefont {Roth},
  \citenamefont {Markham}, \citenamefont {Parekh}, \citenamefont {Chabaud},\
  and\ \citenamefont {Kashefi}}]{Eisert2020}%
  \BibitemOpen
  \bibfield  {author} {\bibinfo {author} {\bibfnamefont {J.}~\bibnamefont
  {Eisert}}, \bibinfo {author} {\bibfnamefont {D.}~\bibnamefont {Hangleiter}},
  \bibinfo {author} {\bibfnamefont {N.}~\bibnamefont {Walk}}, \bibinfo {author}
  {\bibfnamefont {I.}~\bibnamefont {Roth}}, \bibinfo {author} {\bibfnamefont
  {D.}~\bibnamefont {Markham}}, \bibinfo {author} {\bibfnamefont
  {R.}~\bibnamefont {Parekh}}, \bibinfo {author} {\bibfnamefont
  {U.}~\bibnamefont {Chabaud}}, \ and\ \bibinfo {author} {\bibfnamefont
  {E.}~\bibnamefont {Kashefi}},\ }\href {\doibase 10.1038/s42254-020-0186-4}
  {\bibfield  {journal} {\bibinfo  {journal} {Nature Reviews Physics}\ }\textbf
  {\bibinfo {volume} {2}},\ \bibinfo {pages} {382} (\bibinfo {year}
  {2020})}\BibitemShut {NoStop}%
\bibitem [{\citenamefont {Wagner}\ \emph {et~al.}(2021)\citenamefont {Wagner},
  \citenamefont {Kampermann}, \citenamefont {Bru\ss{}},\ and\ \citenamefont
  {Kliesch}}]{Wagner21}%
  \BibitemOpen
  \bibfield  {author} {\bibinfo {author} {\bibfnamefont {T.}~\bibnamefont
  {Wagner}}, \bibinfo {author} {\bibfnamefont {H.}~\bibnamefont {Kampermann}},
  \bibinfo {author} {\bibfnamefont {D.}~\bibnamefont {Bru\ss{}}}, \ and\
  \bibinfo {author} {\bibfnamefont {M.}~\bibnamefont {Kliesch}},\ }\href
  {\doibase 10.1103/PhysRevResearch.3.013292} {\bibfield  {journal} {\bibinfo
  {journal} {Phys. Rev. Res.}\ }\textbf {\bibinfo {volume} {3}},\ \bibinfo
  {pages} {013292} (\bibinfo {year} {2021})}\BibitemShut {NoStop}%
\bibitem [{\citenamefont {Wagner}\ \emph {et~al.}(2022)\citenamefont {Wagner},
  \citenamefont {Kampermann}, \citenamefont {Bru{\ss{}}},\ and\ \citenamefont
  {Kliesch}}]{Wagner22}%
  \BibitemOpen
  \bibfield  {author} {\bibinfo {author} {\bibfnamefont {T.}~\bibnamefont
  {Wagner}}, \bibinfo {author} {\bibfnamefont {H.}~\bibnamefont {Kampermann}},
  \bibinfo {author} {\bibfnamefont {D.}~\bibnamefont {Bru{\ss{}}}}, \ and\
  \bibinfo {author} {\bibfnamefont {M.}~\bibnamefont {Kliesch}},\ }\href
  {\doibase 10.22331/q-2022-09-19-809} {\bibfield  {journal} {\bibinfo
  {journal} {{Quantum}}\ }\textbf {\bibinfo {volume} {6}},\ \bibinfo {pages}
  {809} (\bibinfo {year} {2022})}\BibitemShut {NoStop}%
\bibitem [{\citenamefont {Maity}\ \emph {et~al.}(2023)\citenamefont {Maity},
  \citenamefont {Casapao}, \citenamefont {Benchasattabuse}, \citenamefont
  {Hajdusek}, \citenamefont {{Van Meter}},\ and\ \citenamefont
  {Elkouss}}]{Maity23}%
  \BibitemOpen
  \bibfield  {author} {\bibinfo {author} {\bibfnamefont {A.~G.}\ \bibnamefont
  {Maity}}, \bibinfo {author} {\bibfnamefont {J.~C.~A.}\ \bibnamefont
  {Casapao}}, \bibinfo {author} {\bibfnamefont {N.}~\bibnamefont
  {Benchasattabuse}}, \bibinfo {author} {\bibfnamefont {M.}~\bibnamefont
  {Hajdusek}}, \bibinfo {author} {\bibfnamefont {R.}~\bibnamefont {{Van
  Meter}}}, \ and\ \bibinfo {author} {\bibfnamefont {D.}~\bibnamefont
  {Elkouss}},\ }\href {\doibase 10.1145/3626570.3626594} {\bibfield  {journal}
  {\bibinfo  {journal} {SIGMETRICS Perform. Eval. Rev.}\ }\textbf {\bibinfo
  {volume} {51}},\ \bibinfo {pages} {66–68} (\bibinfo {year}
  {2023})}\BibitemShut {NoStop}%
\bibitem [{\citenamefont {Hoeffding}(1963)}]{hoeffding1963prob}%
  \BibitemOpen
  \bibfield  {author} {\bibinfo {author} {\bibfnamefont {W.}~\bibnamefont
  {Hoeffding}},\ }\href {\doibase 10.1080/01621459.1963.10500830} {\bibfield
  {journal} {\bibinfo  {journal} {Journal of the American Statistical
  Association}\ }\textbf {\bibinfo {volume} {58}},\ \bibinfo {pages} {13}
  (\bibinfo {year} {1963})}\BibitemShut {NoStop}%
\bibitem [{\citenamefont {Krutyanskiy}\ \emph
  {et~al.}(2023{\natexlab{a}})\citenamefont {Krutyanskiy}, \citenamefont
  {Galli}, \citenamefont {Krcmarsky}, \citenamefont {Baier}, \citenamefont
  {Fioretto}, \citenamefont {Pu}, \citenamefont {Mazloom}, \citenamefont
  {Sekatski}, \citenamefont {Canteri}, \citenamefont {Teller}, \citenamefont
  {Schupp}, \citenamefont {Bate}, \citenamefont {Meraner}, \citenamefont
  {Sangouard}, \citenamefont {Lanyon},\ and\ \citenamefont
  {Northup}}]{krutyanskiy2023entanglement}%
  \BibitemOpen
  \bibfield  {author} {\bibinfo {author} {\bibfnamefont {V.}~\bibnamefont
  {Krutyanskiy}}, \bibinfo {author} {\bibfnamefont {M.}~\bibnamefont {Galli}},
  \bibinfo {author} {\bibfnamefont {V.}~\bibnamefont {Krcmarsky}}, \bibinfo
  {author} {\bibfnamefont {S.}~\bibnamefont {Baier}}, \bibinfo {author}
  {\bibfnamefont {D.~A.}\ \bibnamefont {Fioretto}}, \bibinfo {author}
  {\bibfnamefont {Y.}~\bibnamefont {Pu}}, \bibinfo {author} {\bibfnamefont
  {A.}~\bibnamefont {Mazloom}}, \bibinfo {author} {\bibfnamefont
  {P.}~\bibnamefont {Sekatski}}, \bibinfo {author} {\bibfnamefont
  {M.}~\bibnamefont {Canteri}}, \bibinfo {author} {\bibfnamefont
  {M.}~\bibnamefont {Teller}}, \bibinfo {author} {\bibfnamefont
  {J.}~\bibnamefont {Schupp}}, \bibinfo {author} {\bibfnamefont
  {J.}~\bibnamefont {Bate}}, \bibinfo {author} {\bibfnamefont {M.}~\bibnamefont
  {Meraner}}, \bibinfo {author} {\bibfnamefont {N.}~\bibnamefont {Sangouard}},
  \bibinfo {author} {\bibfnamefont {B.~P.}\ \bibnamefont {Lanyon}}, \ and\
  \bibinfo {author} {\bibfnamefont {T.~E.}\ \bibnamefont {Northup}},\ }\href
  {\doibase 10.1103/PhysRevLett.130.050803} {\bibfield  {journal} {\bibinfo
  {journal} {Phys. Rev. Lett.}\ }\textbf {\bibinfo {volume} {130}},\ \bibinfo
  {pages} {050803} (\bibinfo {year} {2023}{\natexlab{a}})}\BibitemShut
  {NoStop}%
\bibitem [{\citenamefont {Fujii}\ and\ \citenamefont
  {Yamamoto}(2009)}]{fujii2009entanglement}%
  \BibitemOpen
  \bibfield  {author} {\bibinfo {author} {\bibfnamefont {K.}~\bibnamefont
  {Fujii}}\ and\ \bibinfo {author} {\bibfnamefont {K.}~\bibnamefont
  {Yamamoto}},\ }\href {\doibase 10.1103/PhysRevA.80.042308} {\bibfield
  {journal} {\bibinfo  {journal} {Physical Review A}\ }\textbf {\bibinfo
  {volume} {80}},\ \bibinfo {pages} {042308} (\bibinfo {year}
  {2009})}\BibitemShut {NoStop}%
\bibitem [{\citenamefont {Goodenough}\ \emph {et~al.}(2024)\citenamefont
  {Goodenough}, \citenamefont {De~Bone}, \citenamefont {Addala}, \citenamefont
  {Krastanov}, \citenamefont {Jansen}, \citenamefont {Gijswijt},\ and\
  \citenamefont {Elkouss}}]{goodenough2024near}%
  \BibitemOpen
  \bibfield  {author} {\bibinfo {author} {\bibfnamefont {K.}~\bibnamefont
  {Goodenough}}, \bibinfo {author} {\bibfnamefont {S.}~\bibnamefont {De~Bone}},
  \bibinfo {author} {\bibfnamefont {V.}~\bibnamefont {Addala}}, \bibinfo
  {author} {\bibfnamefont {S.}~\bibnamefont {Krastanov}}, \bibinfo {author}
  {\bibfnamefont {S.}~\bibnamefont {Jansen}}, \bibinfo {author} {\bibfnamefont
  {D.}~\bibnamefont {Gijswijt}}, \ and\ \bibinfo {author} {\bibfnamefont
  {D.}~\bibnamefont {Elkouss}},\ }\href {\doibase 10.1109/JSAC.2024.3380094}
  {\bibfield  {journal} {\bibinfo  {journal} {IEEE Journal on Selected Areas in
  Communications}\ } (\bibinfo {year} {2024}),\
  10.1109/JSAC.2024.3380094}\BibitemShut {NoStop}%
\bibitem [{\citenamefont {Fujiwara}\ and\ \citenamefont
  {Imai}(2003)}]{Fujiwara03}%
  \BibitemOpen
  \bibfield  {author} {\bibinfo {author} {\bibfnamefont {A.}~\bibnamefont
  {Fujiwara}}\ and\ \bibinfo {author} {\bibfnamefont {H.}~\bibnamefont
  {Imai}},\ }\href {\doibase 10.1088/0305-4470/36/29/314} {\bibfield  {journal}
  {\bibinfo  {journal} {Journal of Physics A: Mathematical and General}\
  }\textbf {\bibinfo {volume} {36}},\ \bibinfo {pages} {8093} (\bibinfo {year}
  {2003})}\BibitemShut {NoStop}%
\bibitem [{\citenamefont {Chiuri}\ \emph {et~al.}(2011)\citenamefont {Chiuri},
  \citenamefont {Rosati}, \citenamefont {Vallone}, \citenamefont {P\'adua},
  \citenamefont {Imai}, \citenamefont {Giacomini}, \citenamefont
  {Macchiavello},\ and\ \citenamefont {Mataloni}}]{Chiuri11}%
  \BibitemOpen
  \bibfield  {author} {\bibinfo {author} {\bibfnamefont {A.}~\bibnamefont
  {Chiuri}}, \bibinfo {author} {\bibfnamefont {V.}~\bibnamefont {Rosati}},
  \bibinfo {author} {\bibfnamefont {G.}~\bibnamefont {Vallone}}, \bibinfo
  {author} {\bibfnamefont {S.}~\bibnamefont {P\'adua}}, \bibinfo {author}
  {\bibfnamefont {H.}~\bibnamefont {Imai}}, \bibinfo {author} {\bibfnamefont
  {S.}~\bibnamefont {Giacomini}}, \bibinfo {author} {\bibfnamefont
  {C.}~\bibnamefont {Macchiavello}}, \ and\ \bibinfo {author} {\bibfnamefont
  {P.}~\bibnamefont {Mataloni}},\ }\href {\doibase
  10.1103/PhysRevLett.107.253602} {\bibfield  {journal} {\bibinfo  {journal}
  {Phys. Rev. Lett.}\ }\textbf {\bibinfo {volume} {107}},\ \bibinfo {pages}
  {253602} (\bibinfo {year} {2011})}\BibitemShut {NoStop}%
\bibitem [{\citenamefont {Ruppert}\ \emph {et~al.}(2012)\citenamefont
  {Ruppert}, \citenamefont {Virosztek},\ and\ \citenamefont
  {Hangos}}]{Ruppert12}%
  \BibitemOpen
  \bibfield  {author} {\bibinfo {author} {\bibfnamefont {L.}~\bibnamefont
  {Ruppert}}, \bibinfo {author} {\bibfnamefont {D.}~\bibnamefont {Virosztek}},
  \ and\ \bibinfo {author} {\bibfnamefont {K.}~\bibnamefont {Hangos}},\ }\href
  {\doibase 10.1088/1751-8113/45/26/265305} {\bibfield  {journal} {\bibinfo
  {journal} {Journal of Physics A: Mathematical and Theoretical}\ }\textbf
  {\bibinfo {volume} {45}},\ \bibinfo {pages} {265305} (\bibinfo {year}
  {2012})}\BibitemShut {NoStop}%
\bibitem [{\citenamefont {Flammia}\ and\ \citenamefont
  {Wallman}(2020)}]{Flammia20}%
  \BibitemOpen
  \bibfield  {author} {\bibinfo {author} {\bibfnamefont {S.~T.}\ \bibnamefont
  {Flammia}}\ and\ \bibinfo {author} {\bibfnamefont {J.~J.}\ \bibnamefont
  {Wallman}},\ }\href {\doibase 10.1145/3408039} {\bibfield  {journal}
  {\bibinfo  {journal} {ACM Transactions on Quantum Computing}\ }\textbf
  {\bibinfo {volume} {1}} (\bibinfo {year} {2020}),\
  10.1145/3408039}\BibitemShut {NoStop}%
\bibitem [{\citenamefont {Flammia}\ and\ \citenamefont
  {O'Donnell}(2021)}]{Flammia21}%
  \BibitemOpen
  \bibfield  {author} {\bibinfo {author} {\bibfnamefont {S.~T.}\ \bibnamefont
  {Flammia}}\ and\ \bibinfo {author} {\bibfnamefont {R.}~\bibnamefont
  {O'Donnell}},\ }\href {\doibase 10.22331/q-2021-09-23-549} {\bibfield
  {journal} {\bibinfo  {journal} {{Quantum}}\ }\textbf {\bibinfo {volume}
  {5}},\ \bibinfo {pages} {549} (\bibinfo {year} {2021})}\BibitemShut {NoStop}%
\bibitem [{\citenamefont {Harper}\ \emph {et~al.}(2021)\citenamefont {Harper},
  \citenamefont {Yu},\ and\ \citenamefont {Flammia}}]{Harper21}%
  \BibitemOpen
  \bibfield  {author} {\bibinfo {author} {\bibfnamefont {R.}~\bibnamefont
  {Harper}}, \bibinfo {author} {\bibfnamefont {W.}~\bibnamefont {Yu}}, \ and\
  \bibinfo {author} {\bibfnamefont {S.~T.}\ \bibnamefont {Flammia}},\ }\href
  {\doibase 10.1103/PRXQuantum.2.010322} {\bibfield  {journal} {\bibinfo
  {journal} {PRX Quantum}\ }\textbf {\bibinfo {volume} {2}},\ \bibinfo {pages}
  {010322} (\bibinfo {year} {2021})}\BibitemShut {NoStop}%
\bibitem [{\citenamefont {Chen}\ \emph {et~al.}(2023)\citenamefont {Chen},
  \citenamefont {Liu}, \citenamefont {Otten}, \citenamefont {Seif},
  \citenamefont {Fefferman},\ and\ \citenamefont {Jiang}}]{Chen23}%
  \BibitemOpen
  \bibfield  {author} {\bibinfo {author} {\bibfnamefont {S.}~\bibnamefont
  {Chen}}, \bibinfo {author} {\bibfnamefont {Y.}~\bibnamefont {Liu}}, \bibinfo
  {author} {\bibfnamefont {M.}~\bibnamefont {Otten}}, \bibinfo {author}
  {\bibfnamefont {A.}~\bibnamefont {Seif}}, \bibinfo {author} {\bibfnamefont
  {B.}~\bibnamefont {Fefferman}}, \ and\ \bibinfo {author} {\bibfnamefont
  {L.}~\bibnamefont {Jiang}},\ }\href {\doibase 10.1038/s41467-022-35759-4}
  {\bibfield  {journal} {\bibinfo  {journal} {Nature Communications}\ }\textbf
  {\bibinfo {volume} {14}} (\bibinfo {year} {2023}),\
  10.1038/s41467-022-35759-4}\BibitemShut {NoStop}%
\bibitem [{\citenamefont {Krutyanskiy}\ \emph
  {et~al.}(2023{\natexlab{b}})\citenamefont {Krutyanskiy}, \citenamefont
  {Canteri}, \citenamefont {Meraner}, \citenamefont {Bate}, \citenamefont
  {Krcmarsky}, \citenamefont {Schupp}, \citenamefont {Sangouard},\ and\
  \citenamefont {Lanyon}}]{krutyanskiy2023telecom}%
  \BibitemOpen
  \bibfield  {author} {\bibinfo {author} {\bibfnamefont {V.}~\bibnamefont
  {Krutyanskiy}}, \bibinfo {author} {\bibfnamefont {M.}~\bibnamefont
  {Canteri}}, \bibinfo {author} {\bibfnamefont {M.}~\bibnamefont {Meraner}},
  \bibinfo {author} {\bibfnamefont {J.}~\bibnamefont {Bate}}, \bibinfo {author}
  {\bibfnamefont {V.}~\bibnamefont {Krcmarsky}}, \bibinfo {author}
  {\bibfnamefont {J.}~\bibnamefont {Schupp}}, \bibinfo {author} {\bibfnamefont
  {N.}~\bibnamefont {Sangouard}}, \ and\ \bibinfo {author} {\bibfnamefont
  {B.~P.}\ \bibnamefont {Lanyon}},\ }\href {\doibase
  10.1103/PhysRevLett.130.213601} {\bibfield  {journal} {\bibinfo  {journal}
  {Phys. Rev. Lett.}\ }\textbf {\bibinfo {volume} {130}},\ \bibinfo {pages}
  {213601} (\bibinfo {year} {2023}{\natexlab{b}})}\BibitemShut {NoStop}%
\bibitem [{\citenamefont {Knaut}\ \emph {et~al.}(2024)\citenamefont {Knaut},
  \citenamefont {Suleymanzade}, \citenamefont {Wei}, \citenamefont {Assumpcao},
  \citenamefont {Stas}, \citenamefont {Huan}, \citenamefont {Machielse},
  \citenamefont {Knall}, \citenamefont {Sutula}, \citenamefont {Baranes},
  \citenamefont {Sinclair}, \citenamefont {De-Eknamkul}, \citenamefont
  {Levonian}, \citenamefont {Bhaskar}, \citenamefont {Park}, \citenamefont
  {Lon{\v c}ar},\ and\ \citenamefont {Lukin}}]{knaut2024entanglement}%
  \BibitemOpen
  \bibfield  {author} {\bibinfo {author} {\bibfnamefont {C.~M.}\ \bibnamefont
  {Knaut}}, \bibinfo {author} {\bibfnamefont {A.}~\bibnamefont {Suleymanzade}},
  \bibinfo {author} {\bibfnamefont {Y.~C.}\ \bibnamefont {Wei}}, \bibinfo
  {author} {\bibfnamefont {D.~R.}\ \bibnamefont {Assumpcao}}, \bibinfo {author}
  {\bibfnamefont {P.~J.}\ \bibnamefont {Stas}}, \bibinfo {author}
  {\bibfnamefont {Y.~Q.}\ \bibnamefont {Huan}}, \bibinfo {author}
  {\bibfnamefont {B.}~\bibnamefont {Machielse}}, \bibinfo {author}
  {\bibfnamefont {E.~N.}\ \bibnamefont {Knall}}, \bibinfo {author}
  {\bibfnamefont {M.}~\bibnamefont {Sutula}}, \bibinfo {author} {\bibfnamefont
  {G.}~\bibnamefont {Baranes}}, \bibinfo {author} {\bibfnamefont
  {N.}~\bibnamefont {Sinclair}}, \bibinfo {author} {\bibfnamefont
  {C.}~\bibnamefont {De-Eknamkul}}, \bibinfo {author} {\bibfnamefont {D.~S.}\
  \bibnamefont {Levonian}}, \bibinfo {author} {\bibfnamefont {M.~K.}\
  \bibnamefont {Bhaskar}}, \bibinfo {author} {\bibfnamefont {H.}~\bibnamefont
  {Park}}, \bibinfo {author} {\bibfnamefont {M.}~\bibnamefont {Lon{\v c}ar}}, \
  and\ \bibinfo {author} {\bibfnamefont {M.~D.}\ \bibnamefont {Lukin}},\ }\href
  {\doibase 10.1038/s41586-024-07252-z} {\bibfield  {journal} {\bibinfo
  {journal} {Nature}\ }\textbf {\bibinfo {volume} {629}},\ \bibinfo {pages}
  {573} (\bibinfo {year} {2024})}\BibitemShut {NoStop}%
\bibitem [{\citenamefont {Hermans}\ \emph {et~al.}(2022)\citenamefont
  {Hermans}, \citenamefont {Pompili}, \citenamefont {Beukers}, \citenamefont
  {Baier}, \citenamefont {Borregaard},\ and\ \citenamefont
  {Hanson}}]{hermans2022qubit}%
  \BibitemOpen
  \bibfield  {author} {\bibinfo {author} {\bibfnamefont {S.~L.~N.}\
  \bibnamefont {Hermans}}, \bibinfo {author} {\bibfnamefont {M.}~\bibnamefont
  {Pompili}}, \bibinfo {author} {\bibfnamefont {H.~K.~C.}\ \bibnamefont
  {Beukers}}, \bibinfo {author} {\bibfnamefont {S.}~\bibnamefont {Baier}},
  \bibinfo {author} {\bibfnamefont {J.}~\bibnamefont {Borregaard}}, \ and\
  \bibinfo {author} {\bibfnamefont {R.}~\bibnamefont {Hanson}},\ }\href
  {\doibase 10.1038/s41586-022-04697-y} {\bibfield  {journal} {\bibinfo
  {journal} {Nature}\ }\textbf {\bibinfo {volume} {605}},\ \bibinfo {pages}
  {663–668} (\bibinfo {year} {2022})}\BibitemShut {NoStop}%
\bibitem [{\citenamefont {Horodecki}\ \emph {et~al.}(2009)\citenamefont
  {Horodecki}, \citenamefont {Horodecki}, \citenamefont {Horodecki},\ and\
  \citenamefont {Horodecki}}]{horodecki2009quantument}%
  \BibitemOpen
  \bibfield  {author} {\bibinfo {author} {\bibfnamefont {R.}~\bibnamefont
  {Horodecki}}, \bibinfo {author} {\bibfnamefont {P.}~\bibnamefont
  {Horodecki}}, \bibinfo {author} {\bibfnamefont {M.}~\bibnamefont
  {Horodecki}}, \ and\ \bibinfo {author} {\bibfnamefont {K.}~\bibnamefont
  {Horodecki}},\ }\href {\doibase 10.1103/RevModPhys.81.865} {\bibfield
  {journal} {\bibinfo  {journal} {Rev. Mod. Phys.}\ }\textbf {\bibinfo {volume}
  {81}},\ \bibinfo {pages} {865} (\bibinfo {year} {2009})}\BibitemShut
  {NoStop}%
\end{thebibliography}%

\section{Acknowledgements}
We thank Kenneth Goodenough, Shota Nagayama, Akihito Soeda, and Zherui Wang for fruitful discussions. This work was supported  by the JST Moonshot R\&D program under Grants JPMJMS226C.


\section*{Data and Code Availability}
The data generated via the numerical experiments are available at the GitHub repository \href{https://github.com/netq-oist/distimator}{https://github.com/netq-oist/distimator}. The Disti-Mator simulation toolbox itself and other codes that are used for the numerical experiment are available as GitHub repository at \href{https://github.com/netq-oist/distimator}{https://github.com/netq-oist/distimator}.

\section*{Author Contributions}
JCAC and AGM wrote the first version of the manuscript. JCAC and AGM wrote the numerical package with input from NB. MH, RDV, and DE supervised the project. All the authors contributed in deriving and analyzing the results, as well as in writing the final manuscript.

\section*{Competing Interests}
The authors declare no competing interests.


\newpage

\onecolumngrid

\appendix

\begin{center}
    \large{\textbf{Supplementary Material}} 
\end{center}

\section*{Sample complexity}

Here we discuss the sample complexity of the proposed estimation scheme, i.e., the required number of input pairs needed for a successful estimation. For estimating the Werner parameter, we have shown that the failure probability in the noiseless case is bounded by $\delta$, where
\begin{equation}
    \text{Pr}(|\hat{w}- w| \geq \epsilon_w) \leq \sum_{m=L,R}\exp{-2 N^{(1)} (\epsilon^{(1)}_m(\hat{w}, \epsilon_w))^2} =: \delta.
\end{equation}
Since $0\leq \hat{w} \leq 2/3$, we have
\begin{align}
    \delta &= \exp[-2 N^{(1)} \left(\frac{1}{4} (-\epsilon_w^{2} + 2\epsilon_w(1-\hat{w}))\right)^2] + \exp[-2 N^{(1)} \left(\frac{1}{4} (\epsilon_w^{2} + 2\epsilon_w(1-\hat{w}))\right)^2],\nonumber\\
    &\leq 2 \exp[-\frac{N^{(1)}}{8}\epsilon_w^2 \left(\frac{2}{3}-\epsilon_w\right)^2].
\end{align}
Hence, as $\epsilon_w \rightarrow 0^+$, we have the following upper bound:
\begin{equation}
    N^{(1)} \leq \frac{8\log(2/\delta)}{\epsilon_w^2 \left(\frac{2}{3}-\epsilon_w\right)^2} = \mathcal{O}\left(\frac{\log(1/\delta)}{\epsilon_w^2}\right).
\end{equation}

For estimating the Bell-diagonal elements in a noiseless scenario, we have shown in the main text that the estimator has a failure probability bounded by
\begin{equation}
    \Pr[D(\hat{\rho}(\hat{\mathbf{q}}),\overline{\rho}(\mathbf{q}))\geq \epsilon_T] \leq \delta(\hat{\mathbf{x}}, \{N^{(i)}\},\{\epsilon_i\}),
\end{equation}
where 
\begin{equation}
    \delta(\hat{\mathbf{x}}, \{N^{(i)}\},\{\epsilon_i\}) := 1 - \prod_{i\in\{1,2,3\}}\left[ 1-\sum_{m=L,R}\exp~(-2N^{(i)}\epsilon^{(i)}_m(\hat{x}_i,\epsilon_i)^2)\right] =: 1 - \prod_{i\in\{1,2,3\}}(1-\delta_i),
\end{equation}
and $N^{(i)}$ are the number of samples used for the $i$-th distillation protocol. Suppose that $N^{(i)} = \widetilde{N}$ and $\epsilon_i = \epsilon$ for all $i$. Then,
\begin{align}
    \delta(\hat{\mathbf{x}},\widetilde{N},\epsilon)
    &\leq \delta_1 + \delta_2 + \delta_3 + \delta_1\delta_2\delta_3, \\
    &\leq 4\max_i \delta_i,\\
    &= 4\max_i\left(\exp[-2\widetilde{N}(-\epsilon^2 + \epsilon(2\hat{x}_i-1))^2]+\exp[-2\widetilde{N}(\epsilon^2 + \epsilon(2\hat{x}_i-1))^2]\right),\\
    &\leq 8 \exp[-2\widetilde{N}(-\epsilon^2 + \epsilon(2\max_i\hat{x}_i-1))^2]
\end{align}
where $0\leq\delta_i\leq 1$, and $1/2 \leq \hat{x}_i \leq 1$. For fixed $\hat{\mathbf{x}}$, the expression inside the exponential is dominated by $\mathcal{O}(\epsilon^2)$ as $\epsilon\rightarrow 0^+$. Thus, with $\epsilon_T = 3\epsilon$, we have the following upper bound on $\widetilde{N}$:
\begin{equation}
    \widetilde{N} \leq \frac{\log(8/\delta)}{2(-(\epsilon_T/3)^2 + (\epsilon_T/3)(2\max_i\hat{x}_i-1))^2} = \mathcal{O}\left(\frac{\log(1/\delta)}{\epsilon_T^{2}}\right).
\end{equation}

\section*{Numerical simulations}

Here we discuss the relevant mathematical framework in building the  simulation program for the Disti-Mator. We first describe a simplified method for tracking the effects of noise in state preparation and measurement and in the distillation protocols by leveraging the underlying symmetries in Bell-diagonal states. We then discuss the strategies to invert the set of empirical probabilities $\hat{\mathbf{p}}$ to the intermediate estimation $\hat{\mathbf{x}}$ and finally to the desired estimation $\hat{\mathbf{q}}$. 

\subsection{Bell vectorization}

Standard quantum theory  elucidates that the noisy evolution of undistilled states can be computed through a series of matrix operations. However, executing many matrix operations to simulate a distillation experiment can be time-consuming. For our analysis, it is convenient to treat the states as a set of ``Bell blocks'', and subsequently associate this set to a vector that is easier to manipulate numerically (hence the idea for a Bell vectorization).

\subsubsection{Bell-diagonal states}

We start with the Bell-diagonal state, 
\begin{equation}
    \overline{\rho}(\mathbf{q})=  q_1\ketbra{\Phi^+}{\Phi^+} + q_2\ketbra{\Phi^-}{ \Phi^-}+ q_3\ketbra{\Psi^+}{ \Psi^+} + q_4\ketbra{\Psi^-}{\Psi^-},
\end{equation}
where $\ket{\Phi^{\pm}}  = (\ket{00}\pm\ket{11})/\sqrt{2}$, $\ket{\Psi^{\pm}}  = (\ket{01}\pm\ket{10})/\sqrt{2}$, and $q_1+q_2+q_3+q_4 = 1$. We naturally assign $\overline{\rho}(\mathbf{q}) \mapsto \mathbf{q} = [q_1;q_2;q_3;q_4]$ (we use semicolons for clarity). Thus, for a control-target pair $\rho_{A_1B_1}(\mathbf{q})\otimes \rho_{A_2B_2}(\tilde{\mathbf{q}})$ defined over the composite space $\mathcal{H}_{A_1}\otimes\mathcal{H}_{B_1}\otimes\mathcal{H}_{A_2}\otimes\mathcal{H}_{B_2}$, we assign a Bell vector $\mathbf{q}\otimes\tilde{\mathbf{q}}$, defined in $\mathbb{R}^{16}$, where
\begin{equation}
    \mathbf{q}\otimes\tilde{\mathbf{q}} := [q_1;q_2;q_3;q_4]\otimes[\tilde{q}_1;\tilde{q}_2;\tilde{q}_3;\tilde{q}_4].
\end{equation}
Here, we keep the order of the entries based on how the tensor product is evaluated. This approach can be extended to general states, focusing solely on tracking their Bell-diagonal elements.

\subsubsection{Noise contribution in the state preparation stage}

For each distillation round, we require $N$ control-target pairs $\rho_{A_1B_1}\otimes\rho_{A_2B_2}$. We assume that these pairs are not generated simultaneously. The target pair is generated at time $t$ subsequent to the control pair. Thus, the control pair is subjected to noise while the target pair is being distributed, which we model as a combination of depolarizing and dephasing channels on each qubit. Let $\lambda(t) = 1-e^{-{t}/{T_1}}$ be the depolarizing parameter, with characteristic time $T_1$, and let $\zeta(t) = \frac{1}{2}(1-e^{-{t}/{T_2}})$ be the dephasing parameter, with characteristic time $T_2$. Then, evolving the state gives
\begin{equation}
    \rho_{A_1B_1}\otimes\rho_{A_2B_2} \mapsto [(\Delta_{\mathrm{dph},A_1}^{\zeta_A}\circ \Delta_{\mathrm{dpo},A_1}^{\lambda_A})\otimes(\Delta_{\mathrm{dph},B_1}^{\zeta_B}\circ \Delta_{\mathrm{dpo},B_1}^{\lambda_B})](\rho_{A_1 B_1}) \otimes \rho_{A_2B_2} = \rho_{A_1B_1}^*\otimes \rho_{A_2B_2}.
\end{equation}
Here,
\begin{equation}
    \Delta_{\mathrm{dpo},B_1}^{\lambda_B} (\rho_{A_1B_1}) = (1-\lambda_B)\rho_{A_1B_1} + {\lambda_B}\mathrm{Tr}_{B_1}(\rho_{A_1B_1})\otimes\frac{I_{B_1}}{2} = (1-\lambda_B)\rho_{A_1B_1} +\lambda_B\frac{I_{A_1B_1}}{4}=: \rho_{A_1B_1}',
\end{equation}
followed by
\begin{align}
    \Delta_{\mathrm{dph},B_1}^{\zeta_B} (\rho_{A_1B_1}') &= (1-\zeta_B)\rho_{A_1B_1}' + \zeta_B I_{A_1}\otimes Z_{B_1}\rho_{A_1B_1}' I_{A_1}\otimes Z_{B_1}=:\rho_{A_1B_1}''.
\end{align}
Then,
\begin{equation}
    \Delta_{\mathrm{dpo},A_1}^{\lambda_A} (\rho_{A_1B_1}'') = (1-\lambda_A)\rho_{A_1B_1}'' + {\lambda_A}\frac{I_{A_1}}{2}\otimes\mathrm{Tr}_{A_1}(\rho_{A_1B_1}'')= (1-\lambda_A)\rho_{A_1B_1} +\lambda_A\frac{I_{A_1B_1}}{4}=:\rho_{A_1B_1}''',
\end{equation}
and finally,
\begin{equation}
    \Delta_{\mathrm{dph},A_1}^{\zeta_B} (\rho_{A_1B_1}''') = (1-\zeta_A)\rho_{A_1B_1}''' + \zeta_A Z_{A_1}\otimes I_{B_1}\rho_{A_1B_1}''' Z_{A_1}\otimes I_{B_1} =: \rho_{A_1 B_1}^*.
\end{equation}
In terms of Bell vectors,
\begin{align}
    \Delta_{\mathrm{dpo}}^{\lambda_B} (\rho_{AB}) &\mapsto \mathbf{q}':=\left[\frac{\lambda_B}{4}+(1-\lambda_B)q_1;\frac{\lambda_B}{4}+(1-\lambda_B)q_2;\frac{\lambda_B}{4}+(1-\lambda_B)q_3;\frac{\lambda_B}{4}+(1-\lambda_B)q_4\right],\\
    \Delta_{\mathrm{dph}}^{\zeta_B} (\rho_{AB}) &\mapsto \mathbf{q}':=[(1- \zeta_B)q_1 + \zeta_B q_2 ;(1-\zeta_B)q_2 + \zeta_B q_1 ;(1 - \zeta_B)q_3 + \zeta_B q_4 ;(1-\zeta_B)q_4 + \zeta_B q_3],
\end{align}
and so on. 

\subsubsection{Noisy local rotations}

For the Deutsch {\it et al.} protocol \cite{Deutsch96}, we require local rotations prior to local controlled gates. Given $\rho_{AB}$, we describe the action of a noisy local rotations as
\begin{align}
    \Lambda_{R_x}^{m_A,m_B}(\rho_{AB}) =&\quad (1-m_A)(1-m_B)R_x\left(-\frac{\pi}{2}\right)\otimes R_x\left(+\frac{\pi}{2}\right)\rho_{AB}R_x\left(-\frac{\pi}{2}\right)^{\dagger}\otimes R_x\left(+\frac{\pi}{2}\right)^{\dagger}\nonumber\\
    &+\left(m_A + m_B - m_A m_B\right)\frac{I_{AB}}{4},
\end{align}
for depolarizing parameters $m_i$, where $0\leq m_i \leq 1$. Let $\mathbf{q}=[q_1;q_2;q_3;q_4]$ be the associated Bell vector to $\rho_{AB}$. Then,
\begin{equation}
    \Lambda_{R_x}^{m_A,m_B}(\rho_{AB}) \mapsto \mathbf{q}' := (1-m_A)(1-m_B)[q_1;q_4;q_3;q_2]+\frac{m_A + m_B - m_A m_B}{4}[1;1;1;1].
\end{equation}

\subsubsection{Noisy CNOT operations}

In all distillation protocols, Alice and Bob perform a bilocal CNOT operation on their respective halves. We describe the action of both noisy CNOT operations on $\rho_{A_1B_1}(\mathbf{q})\otimes\rho_{A_2B_2}(\tilde{\mathbf{q}})$ in terms of the depolarizing parameters $y_A$ and $y_B$, where $0\leq y_i\leq 1$,
\begin{align}
    &\Lambda_{\text{CNOT}}^{y_A,y_B}(\rho_{A_1B_1}\otimes\rho_{A_2B_2})\nonumber\\
    &=(1-y_A)(1-y_B) \,(\text{CNOT}_{A_1 A_2}\otimes \text{CNOT}_{B_1 B_2}) \,  \rho_{A_1B_1}\otimes\rho_{A_2B_2} (\,\text{CNOT}_{A_1 A_2} \otimes \text{CNOT}_{B_1 B_2})\nonumber\\
    &\quad+ [y_A + y_B - y_A y_B]\frac{I_{A_1 A_2 B_1 B_2}}{16}.
\end{align}
For noisy CNOT operations, we take the following operation on the Bell vector:
\begin{align}
    &(\text{CNOT}_{A_1 A_2}\otimes \text{CNOT}_{B_1 B_2}) \,  \rho_{A_1B_1}\otimes\rho_{A_2B_2} (\,\text{CNOT}_{A_1 A_2} \otimes \text{CNOT}_{B_1 B_2})\nonumber\\
    &\mapsto [q_1\tilde{q}_1;q_2\tilde{q}_2;q_1\tilde{q}_3;q_2\tilde{q}_4;q_2\tilde{q}_1;q_1\tilde{q}_2;q_2\tilde{q}_3;q_1\tilde{q}_4;q_3\tilde{q}_3;q_4\tilde{q}_4;q_3\tilde{q}_1;q_4\tilde{q}_2;q_4\tilde{q}_3;q_3\tilde{q}_4;q_4\tilde{q}_1;q_3\tilde{q}_2] =: \mathbf{c}.
\end{align}
Thus, the vectorized version is
\begin{equation}\label{eqn:vector_after_CNOT}
    \Lambda_{\text{CNOT}}^{y_A,y_B}(\rho_{A_1B_1}\otimes\rho_{A_2B_2}) \mapsto \mathbf{f} := (1-y_A)(1-y_B)\mathbf{c} + \frac{y_A + y_B - y_A y_B}{16}[1;1;1;1]^{\otimes 2}.
\end{equation}
Notice that $\mathbf{f} \neq \mathbf{q}'\otimes\tilde{\mathbf{q}}'$ since the CNOT operations are entangling.

\subsubsection{Noisy measurements}

Finally, we consider imperfect measurement devices. We first take a noisy $Z$-measurement case. We describe the measurement devices with the stochastic map $\ket{0} \mapsto \ket{0}$ and $\ket{1} \mapsto \ket{1}$ with probability $\eta_z$, and $\ket{0} \mapsto \ket{1}$ and $\ket{1} \mapsto \ket{0}$ with probability $1-\eta_z$. So, the relevant POVM element when measuring the target pair is
\begin{align}
    I_{A_1B_1}\otimes M_{A_2 B_2}^{00}(\eta_z^A,\eta_z^B) &= I_{A_1B_1}\otimes\left(\eta_z^A\eta_z^B\ketbra{00}{00} + (1-\eta_z^A)(1-\eta_z^B)\ketbra{11}{11}\right. \nonumber\\
    &\quad+ \left.\eta_z^A(1-\eta_z^B)\ketbra{01}{01} + (1-\eta_z^A)\eta_z^B\ketbra{10}{10}\right).
\end{align}
Similarly, for the noisy $X$-measurement case, we take the stochastic map $\ket{\pm} \mapsto \ket{\pm}$ with probability $\eta_x$, and $\ket{\pm} \mapsto \ket{\mp}$ with probability $1-\eta_x$. Thus, the relevant POVM element when measuring the control pair is
\begin{align}
    M_{A_1 B_1}^{++}(\eta_x^A,\eta_x^B)\otimes I_{A_2B_2} &= \left(\eta_x^A\eta_x^B \ketbra{++}{++} + (1-\eta_x^A)(1-\eta_x^B)\ketbra{--}{--} \right.\nonumber\\
    &\quad+ \left.\eta_x^A(1-\eta_x^B)\ketbra{+-}{+-} + (1-\eta_x^A)\eta_x^B\ketbra{-+}{-+}\right)\otimes I_{A_2B_2}.
\end{align}
Then, for a noisy $Z$-measurement on the target pair, the success probability is
\begin{align}
    \mathrm{Tr}(I_{A_1B_1}\otimes M_{A_2 B_2}^{00}(\eta_z^A,\eta_z^B))&= \frac{1}{2}(1-\eta_z^B-\eta_z^A(1-2\eta_z^B))(f_1+f_2+f_5+f_6+f_9+f_{10}+f_{13}+f_{14})\nonumber\\
    &\quad+ \frac{1}{2}(\eta_z^B+\eta_z^A(1-2\eta_z^B))(f_3 + f_4 + f_7 + f_8 + f_{11}+f_{12}+f_{15}+f_{16}).
\end{align}
On the other hand, for a noisy $X$-measurement on the control pair, the success probability is
\begin{align}
    \mathrm{Tr}(M_{A_1 B_1}^{++}(\eta_x^A,\eta_x^B)\otimes I_{A_2B_2})&= \frac{1}{2}(1-\eta_x^B - \eta_x^A(1-2\eta_x^B))(f_1+f_2+f_3+f_4+f_9+f_{10}+f_{11}+f_{12})\nonumber\\
    &\quad+ \frac{1}{2}(\eta_x^B +\eta_x^A(1-2\eta_x^B))(f_5 + f_6 + f_7 + f_8 + f_{13}+f_{14}+f_{15}+f_{16}),
\end{align}
where $f_i$ is the $i$-th component of $\mathbf{f}$ in Eq.~\eqref{eqn:vector_after_CNOT}. 

\subsection{Inversion strategies}

Here, we outline the strategy to determine the Bell-diagonal elements from the empirical probabilities $\hat{\mathbf{p}} = [\hat{p}^{(1)},\hat{p}^{(2)},\hat{p}^{(3)}]$. We split the discussion into two, the Werner version, and the general Bell-diagonal version, for Algorithms 1 and 2 in the main text, respectively.

\subsubsection{Werner state estimation}
As a special case, we consider a Werner state, where $q_1=1-\frac{3w}{4}$ and $q_2= q_3=q_4=\frac{w}{4}$. Our goal is to estimate $\hat{w}$ using the first distillation protocol, with error bound $\epsilon_w$. We only consider the range $0\leq w < 2/3$, which ensures a successful distillation with some probability.
We assume that both the control and the target pairs are the same immediately after their generation.

The inversion strategy that we adopt involves a bisection search algorithm. We summarize the bisection search as a general function in Algorithm~\ref{alg:bisectionsearch}. This method can be applied since the success probability is continuous as well as monotonically decreasing with $w$. Here we deviate from the main text by introducing the statistical error ($\epsilon_w$ for the Werner case) as another input on the search function to justify our choice for the bisection search tolerance $\epsilon_{\mathrm{bis}}$.

Given $\epsilon_w$, the number of iterations needed is bounded by
\begin{equation}
    n_{1/2} \leq \left\lceil \log_2\left(\frac{b_0-a_0}{\epsilon_{\mathrm{bis}}}\right) \right\rceil = \left\lceil \log_2\left(\frac{2}{3\epsilon_w^3}\right) \right\rceil,
\end{equation}
where we have chosen $a_0=0$ and $b_0 = 2/3$ as the initial endpoints of the search. Referring back to the case of estimating Werner states via a noiseless distillation protocol, we find that $\epsilon_{L,R}^{(1)}$ have $\mathcal{O}(\epsilon_w^2)$ subleading terms. Hence, we choose the bisection search tolerance to be $\epsilon_{\mathrm{bis}} = o(\epsilon_w^2)$ as $\epsilon_w\rightarrow 0^+$ so that the tolerance is qualitatively smaller than these subleading terms. Here we simply set $\epsilon_{\mathrm{bis}} = \epsilon_w^3$. For our analysis, each inversion step calculates $p^{(1)}$ by using a Bell vectorization-based numerics. Thus, we guarantee an estimate $\hat{w}$ for a given $\hat{p}^{(1)}$. We then back-propagate $\hat{w}\pm\epsilon_w$ to determine the error bounds $\epsilon_{L,R}^{(1)}$.


\begin{figure}[htbp]
\begin{minipage}{\linewidth}
\begin{algorithm}[H]
{\small
\begin{algorithmic}[1]
\caption{\small Bisection search algorithm}
\label{alg:bisectionsearch}
\Require 
\Statex Continuous and monotonic function: $F$
\Statex Search endpoints: $a_0,b_0$, where $a_0<b_0$
\Statex `Zero' of the search: $q$ 
\Statex Statistical error associated to $q$: $\epsilon$ \Comment{Error defined outside of the bisection search}

\Ensure 
\Statex `Root'/Solution: $x_M$ \Comment{Such that $q= F(x_M)$}

\State \textbf{function} \Call{BisectionMethod}{$F$, $a_0$, $b_0$, $q$, $\epsilon$} \Comment{Inversion via bisection search}
\State $\epsilon_{\mathrm{bis}} \leftarrow o(\epsilon^2)$ \Comment{Search tolerance}
\State $n_{1/2} \leftarrow \lceil \log_2((b_0 - a_0)/\epsilon_{\mathrm{bis}}) \rceil$ \Comment{Maximum number of iterations}
\State $n\leftarrow 1$ \Comment{Initialize iteration number}
\While{$n\leq n_{1/2}$}
\State $x_{M} \leftarrow (b_0+a_0)/2$
\If{$F(x_{M})=q$ or $(b_0 - a_0)/2 \leq \epsilon_{\mathrm{bis}}$}
\State \Return $x_{M}$ \Comment{With absolute error $\epsilon_{\mathrm{bis}}$}
\State \textbf{break}
\EndIf
\State $n \leftarrow n+1$
\If{$\mathrm{sgn}(F(x_M) - q) = \mathrm{sgn}(a_0 - q)$} \Comment{Defining the new interval}
\State $a_0 \leftarrow x_M $
\Else
\State $b_0 \leftarrow x_M $
\EndIf

\EndWhile
\State \textbf{end function}

\end{algorithmic}
}
\end{algorithm}
\end{minipage}
\end{figure}


\subsubsection{Bell-diagonal estimation}

We now extend the inversion strategy to the general Bell-diagonal scenario. Given that we have $p^{(i)} = p^{(i)}(x_i)$, we can independently do a bisection search to estimate $\hat{x}_i$ given the empirical probability $\hat{p}^{(i)}$. Similar to the Werner case, we note that the success probabilities are continuous and monotonic over $x_i\in[1/2,1]$. From $\hat{\mathbf{x}}$, we can then determine the estimated vector $\hat{\mathbf{q}} = [\hat{q}_1;\hat{q}_2;\hat{q}_3;\hat{q}_4]$ using the relationships
\begin{align}
    q_1 &= \frac{1}{2}(-1 + x_1 + x_2 + x_3),\nonumber\\
    q_2 &= \frac{1}{2}(~~~1 + x_1 - x_2 - x_3),\nonumber\\
    q_3 &= \frac{1}{2}(~~~1 - x_1 + x_2 - x_3),\nonumber\\
    q_4 &= \frac{1}{2}(~~~1 - x_1 - x_2 + x_3).
\end{align}
We set the statistical error bound to be $\epsilon_i$, i.e., $\abs{x_i - \hat{x}_i} \leq \epsilon_i$. Taking $\epsilon_{\mathrm{bis},i}$ the bisection search tolerance for $\hat{x}_i$, the number of bisections needed, $n_{1/2, i}$, is bounded by 
\begin{equation}
    n_{1/2, i} \leq \left\lceil \log_2\left(\frac{b_0-a_0}{\epsilon_{\mathrm{bis},i}}\right) \right\rceil = \left\lceil \log_2\left(\frac{1}{2\epsilon_i^3}\right) \right\rceil.
\end{equation}
Here we have chosen $a_0=1/2$ and $b_0 = 1$ as the initial endpoints of the search, and $\epsilon_{\mathrm{bis},i} = \epsilon_i^3$. Because the $\hat{x}_i$'s can be searched independently, we can parameterize the Bell vectors with $[x_1;0;1-x_1;0]$, $[x_2;1-x_2;0;0]$, and $[x_3;1-x_3;0;0]$ for the first, second, and third distillation protocols, respectively. Finally, we back-propagate $\hat{x}_i\pm\epsilon_i$ to determine the error bounds $\epsilon_{L,R}^{(i)}$.


\end{document}